\documentclass[11pt,twoside,nohyper]{pallis}
\usepackage{epsfig}
\usepackage{latexsym}
\usepackage{amssymb}
\usepackage{fancyheda}
\usepackage{times}
\usepackage[varg]{txfonts}
\usepackage{caption}
\newcommand{\ftn}{\footnotesize}

\newcommand{\ssz}{\scriptsize}

\newcommand{\TeV}{{\mbox{\rm TeV}}}
\newcommand{\MeV}{{\mbox{\rm MeV}}}
\newcommand{\GeV}{{\mbox{\rm GeV}}}

\newcommand{\vH}{{\ensuremath{\bar H}}}

\newcommand{\vHi}{{\ensuremath{\bar H_{{\rm I}}}}}
\newcommand{\vrho}{{\ensuremath{\bar\rho}}}
\newcommand{\vn}{{\ensuremath{\bar n}}}
\newcommand{\vq}{{\ensuremath{\bar q}}}
\newcommand{\vqi}{{\ensuremath{\bar q_{\rm I}}}}

\newcommand{\vV}{{\ensuremath{\bar V}}}
\newcommand{\vVo}{{\ensuremath{\bar V_0}}}

\newcommand{\vTi}{{\ensuremath{T_{{\rm I}}}}}
\newcommand{\vti}{{\ensuremath{\vtau_{{\rm I}}}}}

\newcommand{\vtns}{{\ensuremath{\vtau_{{\rm BBN}}}}}
\newcommand{\vtkr}{{\ensuremath{\vtau_{{\rm KR}}}}}
\newcommand{\vtp}{{\ensuremath{\vtau_{{\rm ext}}}}}
\newcommand{\vtx}{{\ensuremath{\vtau_{{\chi}}}}}

\def\openep{\leavevmode\hbox{\normalsize$\iota$\kern-3.8pt$^$-}}
\newcommand{\vtauf}{\ensuremath{\tauup}}
\newcommand{\vtau}{\ensuremath{\tauup}}
\newcommand{\btauf}{\ensuremath{\tilde\tauup}}
\newcommand{\btau}{\ensuremath{\tilde\tauup}}
\newcommand{\vGamma}{{\mbox{$\bar \Gamma$}}}

\newcommand{\btrhp}{\ensuremath{\btau_{{\rm RH}}}}
\newcommand{\btf}{\ensuremath{\btau_{{\rm F}}}}
\newcommand{\vtf}{{\ensuremath{\vtau_{{\rm F}}}}}
\newcommand{\tf}{\ensuremath{\tau_{{\rm F}}}}

\newcommand{\brhofi}{{\ensuremath{\bar\rho_{\phi_{\rm I}}}}}
\newcommand{\brhoqi}{{\ensuremath{\bar\rho_{q_{\rm I}}}}}

\newcommand{\brhof}{{\ensuremath{\bar\rho_{\phi}}}}
\newcommand{\brhoq}{{\ensuremath{\bar\rho_{q}}}}
\newcommand{\brhoR}{{\ensuremath{\bar\rho_{\rm R}}}}

\newcommand{\brhoRi}{{\ensuremath{\vrho_{{\rm RI}}}}}

\def\beq{\begin{equation}}
\def\eeq{\end{equation}}
\def\bea{\begin{eqnarray}}
\def\eea{\end{eqnarray}}

\newcommand{\Ti}{\ensuremath{T_{\rm I}}}
\newcommand{\Tkr}{\ensuremath{T_{\rm KR}}}
\newcommand{\Trh}{\ensuremath{T_{\rm RH}}}

\newcommand{\Omx}{\ensuremath{\Omega_\chi h^2}}
\newcommand{\Domx}{\ensuremath{\Delta\Omega_{\chi}}}
\newcommand{\Omxsc}{\ensuremath{\left.\Omx\right|_{\rm SC}}}

\newcommand{\Omqns}{\ensuremath{\Omega_q^{\rm BBN}}}

\newcommand{\vrhoR}{{\ensuremath{\bar\rho_{{\rm R}}}}}

\newcommand\sigv{\ensuremath{\langle \sigma v\rangle}}
\newcommand{\vsv}{\ensuremath{\overline{\sigv}}}

\newcommand{\Nx}{\ensuremath{N_\chi}}

\newcommand\Gm[1]{\Gamma_{#1}}
\newcommand\vGm[1]{\bar\Gamma_{#1}}

\newcommand{\ps}{\ensuremath{e^+}}
\newcommand{\el}{\ensuremath{e^-}}
\newcommand{\mchi}{\ensuremath{\left.\chiup^2\right|_{\rm min}}}

\newcommand{\nt}{\ensuremath{\chi}}

\newcommand{\cxf}{\ensuremath{c_{\chi\phi}}}
\newcommand{\mff}{\ensuremath{m_\phi}}

\newcommand{\mx}{{\mbox{$m_\chi$}}}
\newcommand{\Gr}{\ensuremath{\widetilde{G}}}

\newcommand{\sFref}[2]{Fig.~\ref{#1}-{\small\sf ({#2})}}
\newcommand{\sEref}[2]{Eq.~(\ref{#1}{\small\sf {#2}})}

\newcommand{\Eref}[1]{Eq.~(\ref{#1})}
\newcommand{\Sref}[1]{Sec.~\ref{#1}}
\newcommand{\Fref}[1]{Fig.~\ref{#1}}
\newcommand{\Tref}[1]{Table~\ref{#1}}
\newcommand{\cref}[1]{Ref.~\cite{#1}}
\newcommand{\etal}{{\it et al.\/}}
\renewcommand{\email}[1]{{{\sl e-mail address:}~{\tt #1}}}
\newcommand{\dof}{{\mbox{\rm d.o.f}}}
\newcommand{\mP}{\ensuremath{m_{\rm P}}}

\newcommand\eqs[2]{Eqs.~(\ref{#1}) and (\ref{#2})}
\newcommand\eqss[3]{Eqs.~(\ref{#1}), (\ref{#2}) and (\ref{#3})}

\newcommand\ssFref[3]{Fig.~\ref{#1}-{\small\sf ({#2})} and
{\small\sf ({#3})}}

\title{\huge \bfseries\scshape Cold Dark Matter in non-Standard Cosmologies, \\ PAMELA, ATIC and Fermi LAT}

\author{\Large \bfseries\scshape C. Pallis\\
Department of Physics,  University of Patras,
\\ GR-265 00 Patras, GREECE \\ \vspace{3pt}
\email{kpallis@auth.gr}}

\abstract{We consider two non-standard cosmological scenaria
according to which the universe is reheated to a low reheating
temperature after the late decay of a scalar field or is dominated
by the kinetic energy of a quintessence field in the context of a
tracking quintessential model. In both cases, we calculate the
relic density of the \emph{Weakly Interacting Massive Particles}
(WIMPs) and show that it can be enhanced with respect to its value
in the standard cosmology. By adjusting the parameters of the low
reheating or the quintessential scenario, the cold dark matter
abundance in the universe can become compatible with large values
for the annihilation cross section times the velocity of the
WIMPs. Using these values and assuming that the WIMPs annihilate
predominantly to $e^+e^-$, $\mu^+\mu^-$ or $\tau^+\tau^-$, we
calculate the induced fluxes of $e^\pm$ cosmic rays and fit the
current data of PAMELA and ATIC or Fermi LAT. We achieve rather
good fits especially to PAMELA and Fermi-LAT data in conjunction
with a marginal fulfillment of the restriction arising from cosmic
microwave background, provided that the WIMPs annihilate
predominantly to $\mu^+\mu^-$. In both non-standard scenaria the
required transition temperature to the conventional radiation
dominated era turns out to be lower than about $0.7~\GeV$. In the
case of the low reheating, an appreciable non-thermal contribution
to the WIMP relic density is also necessitated.

{\sl\bfseries Published in} {\sl Nucl. Phys. }{\bf
B831}, 217 (2010)}

\begin{document}

\setcounter{page}{1} \pagestyle{fancyplain}

\addtolength{\headheight}{.5cm}

\rhead[\fancyplain{}{ \bf \thepage}]{\fancyplain{}{\scshape CDM in
non-Standard Cosmologies, PAMELA, ATIC and Fermi LAT}}
\lhead[\fancyplain{}{\sc \leftmark}]{\fancyplain{}{\bf \thepage}}
\cfoot{}

\section{Introduction}\label{intro}

The accurate determination of cosmological parameters by
up–-to–-date observations, most notably by the \emph{Wilkinson
Microwave Anisotropy Probe} (WMAP) \cite{wmap, wmapl}, establishes
a quite extensive and convincing evidence for the constitution of
the present universe by an enigmatic component called \emph{Cold
Dark Matter} (CDM) with abundance, $\Omega_{\rm CDM}h^2$, in the
following range:
\beq \Omega_{\rm CDM}h^2=0.1099\pm0.0124 \label{cdmba}\eeq
at $95\%$ \emph{confidence level} (c.l.). Natural candidates
\cite{candidates} to account for the CDM are \cite{jungman} the
\emph{weakly interacting massive particles} (WIMPs, hereafter
denoted as $\chi$) with prominent representative (for other WIMPs,
see \cref{lkk, wimps}) the lightest neutralino \cite{goldberg}
which turns out to be the \emph{lightest supersymmetric (SUSY)
particle} (LSP) in a sizeable fraction of the parameter space of
the SUSY models and therefore, stable under the assumption of the
$R$-parity conservation. In view of \Eref{cdmba}, the relic
density of $\chi$'s, $\Omega_{\chi}h^2$, is to satisfy a very
narrow range of values:
\beq \mbox{\sf\small (a)}~~0.097\lesssim \Omx~~\mbox{and}~~
\mbox{\sf \small (b)}~~\Omx\lesssim0.12,\label{cdmb}\eeq
with the lower bound being valid under the assumption that CDM is
entirely composed by $\chi$'s.

The calculation of $\Omx$ crucially depends \cite{Kam} on the
adopted assumption about the dominant component of the universe
during the decoupling of WIMPs. The usual assumption is that this
occurs during the \emph{radiation dominated} (RD) epoch which
commences after the primordial inflation. However, our ignorance
about the universal history before \emph{Big Bang nucleosynthesis}
(BBN) allows for other possibilities. E.g., the presence of a
scalar field, which dominates the budget of the universal energy
density through its potential \cite{Kam, moroi, riotto, rhodos,
gondolo, nonthermal} or kinetic \cite{Kam, salati, prof, jcapa}
energy density, can enhance significantly $\Omx$ \emph{with
respect to} (w.r.t) its value in the \emph{standard cosmology}
(SC). In the first case, the scalar field can generate an episode
of low reheating which can be accompanied by thermal and/or
non-thermal production of $\chi$'s. In the second case, a
\emph{kination dominated} (KD) epoch \cite{kination}, which may be
embedded \cite{jcapa, dimopoulos, chung, mas, patra} in a
quintessential framework, can arise. As a bonus, in the latter
case, the problem of the second major component of the present
universe, called \emph{Dark Energy} (DE) can be addressed -- for
reviews see, e.g., \cref{der}.

The aforementioned enhancements of $\Omx$ have attracted much
attention recently \cite{khalil} since they assist us to
interpret, through WIMP annihilation in the galaxy and
consistently with \Eref{cdmba}, the reported \cite{pamela, atic}
excess on the positron ($e^+$) and/or electron ($\el$)
\emph{cosmic-ray} (CR) flux, without invoking any pole effect
\cite{pole}, ad-hoc boost factor \cite{sommer} or other
astrophysical sources \cite{clumps}. In particular, PAMELA
experiment has reported \cite{pamela} (confirming previous
experiments \cite{experiments}) an unexpected rise of $\ps$ flux
fraction for values of the $e^+$ energy, $E_{\ps}$, in the range
$(10 - 100)~\GeV$, in contrast to the power-law falling
background. Moreover, data by the ATIC experiment \cite{atic}
shows an excess in the total $\ps$ and $\el$ flux for $300\leq
E_{\ps}/\GeV\leq800$. On the other hand, the very recently
released data from Fermi LAT indicates \cite{Fermi} smaller fluxes
than the ATIC data in the same range of energies. Nevertheless, we
consider (separately) both latter data in our study.

In this paper we reconsider the increase of $\Omx$ within a
\emph{low reheating scenario} (LRS) or a \emph{quintessential
kination scenario} (QKS) in light of the experimental results
above. Namely, we recall (\Sref{sec:nonSC}) comparatively the
salient features of the two non-standard scenaria, solving
numerically the relevant system of equations, reviewing the
cosmological dynamics and imposing a number of observational
constraints. Particularly, in the LRS we consider the late decay
of a massive field which reheats the universe to a low reheating
temperature. In the QKS, we consider the recently implemented
\cite{patra} generation of a KD era (associated with an
oscillatory evolution of the quintessence field) in the context of
tracking quintessential model with a Hubble-induced mass term for
the quintessence field. We then (Sec.~\ref{sec:boltz}) investigate
the enhancement of $\Omx$ w.r.t its value in SC within these
non-standard scenaria. We show that the increase of $\Omx$ depends
on {\sf\small (i)} the reheat temperature and the number of
$\chi$'s produced per decay and unit mass of the decaying field,
in the case of LRS, and {\sf\small (ii)} the proximity between the
freeze-out temperature and the temperature where the evolution of
the quintessence develops extrema, in the case of QKS. We also
present (\Sref{sec:pamela}) the energy spectra of the $e^\pm$-CR,
assuming that $\chi$'s annihilate into $e^+e^-$, $\mu^+\mu^-$ or
$\tau^+\tau^-$ and adopting an isothermal halo profile
\cite{gstrumia}. Although \emph{Cosmic Microwave Background} (CMB)
\cite{CMBold, CMB, CMBnew} tightly constrain the relevant
parameters, we achieve rather satisfactory fittings especially to
the combination of PAMELA and Fermi-LAT $e^\pm$-CR data and for
the case where $\chi$'s annihilate into $\mu^+\mu^-$
(\Sref{sec:mxsv}). Fulfilment of \Eref{cdmb} is also possible by
appropriately adjusting the parameters of the LRS or QKS. We end
up with our conclusions in Sec.~\ref{sec:con}.

Throughout the text, brackets are used by applying disjunctive
correspondence, the subscript or superscript $0$ is referred to
present-day values (except for the coefficient $\vVo$) and
$\log~[\ln]$ stands for logarithm with basis $10~[e]$. Besides
\Sref{sec:pamela}, natural units for the Planck's constant,
Boltzmann's constant and the velocity of light ($\hbar=c=k_{\rm
B}=1$) are assumed.

\section{Non-Standard Cosmological Scenaria}\label{sec:nonSC}

In this section we present comparatively the main features of the
two non-standard cosmological scenaria under consideration.
Namely, in \Sref{Beqs0} we expose the basic assumptions of each
scenario with reference to the SC and introduce notation.
\Sref{sec:lr} and \Sref{sec:quint} are devoted to a review of the
LRS and QKS respectively. Despite the fact that the displayed
scenaria have been already analyzed in \cref{rhodos, patra} we
prefer to briefly recall and update our results for completeness
and clarity.

\subsection{The General Set-up}
\label{Beqs0}

According to SC, primordial inflation is followed by a RD era. The
$\chi$ species {\sf\small (i)} are produced through thermal
scatterings in the plasma, {\sf\small (ii)} reach chemical
equilibrium with plasma and {\sf\small (iii)} decouple from the
cosmic fluid at a temperature $T_{\rm F}\sim (10-20) ~{\rm GeV}$
during the RD era. The assumptions above fix the form of the
relevant Boltzmann equation, the required strength of the $\chi$
interactions for \emph{thermal production} (TP) and lead to an
isentropic cosmological evolution during the $\chi$ decoupling:
The Hubble parameter is $H\propto T^2$ with temperature $T\propto
R^{-1}$ where $R$ is the scale factor of the universe. In this
context, the $\Omx$ calculation depends only on two parameters:
The $\chi$ mass, $m_\chi$ and the thermal-averaged cross section
of $\chi$ times velocity, $\sigv$ -- see \Tref{tab1}. Although
$\sigv$ can be derived from $m_\chi$ and the residual (s)particle
spectrum once a low energy theory is adopted (see, e.g.,
\cref{prof}), we treat $\mx$ and $\sigv$ as unrelated input
parameters in order to keep our approach as general as possible
(see, e.g., \cref{jcapa, gstrumia}). Also, to be in harmony with
the assumptions employed in the derivation of the restrictions
mentioned in \Sref{sec:sv}, we consider throughout constant
$\sigv$'s, i.e., independent of $T$.

The modern cosmo-particles theories, however, are abundant in
scalar massive particles (moduli) which can decay out of
equilibrium when $H$ becomes equal to their mass creating episodes
of reheating. In the LRS, we assume that such a scalar particle
$\phi$, with mass $m_\phi$, decays with a rate $\Gamma_\phi$ into
radiation, producing an average number $N_{\chi}$ of $\chi$'s. The
key point in this case is that the reheating process is not
instantaneous \cite{moroi, rhodos}. During its realization, the
maximal temperature, $T_{\rm max}$, is much larger than the
so-called reheat temperature, $T_{\rm RH}$, which can be taken to
be lower than $T_{\rm F}$. Also, for $T>T_{\rm RH}$, $H\propto
T^4$ with $T \propto R^{-3/8}$ and an entropy production occurs
(in contrast with the SC). The $\chi$ species {\sf\small
(i$^\prime$)} decouple during the decaying-$\phi$ dominated era
{\sf\small (ii$^\prime$)} do or do not reach chemical equilibrium
with the thermal bath {\sf\small (iii$^\prime$)} are produced by
thermal scatterings and directly by the $\phi$ decay (which
naturally arises even without direct coupling \cite{moroi}). As a
consequence, the $\Omx$ calculation depends also on $T_{\rm RH}$,
$m_\phi$ and $N_\chi$ -- see \Tref{tab1}.

Another role that a scalar field could play when it does not
couple to matter (contrary to $\phi$) is this of quintessence. In
our QKS, such a scalar field $q$ (not to be confused with the
deceleration parameter \cite{wmapl}) is supposed to roll down its
inverse power-law potential (with exponent $a$ and a mass scale
$M$) supplemented with a Hubble-induced mass term (with
coefficient $b$) motivated mainly by non-canonical K\"ahler
potential \cite{mas, Hq} -- c.f. \cref{Tq}. A mild tuning of $b$
and of the initial conditions at an initial temperature $\Ti$  --
$H_{\rm I}=H(\Ti)$ and $q_{\rm I}=q(\Ti)$ -- ensures the
coexistence of an early modified KD phase with the tracking
\cite{attr, trak} solutions and offers the desirable property of
the insensitivity to the initial conditions \cite{mas, patra}.
Since the $q$ kinetic energy, which decreases as $T^6$ (except for
isolated points) dominates we get $H\propto T^3$ with $T\propto
R^{-1}$. If the $\chi$-decoupling occurs during this KD phase --
the assumptions {\sf\small (i)} and {\sf\small (ii)} are
maintained -- the $\Omx$ calculation depends also on $a, b, M,
H_{\rm I}, \Ti$ and $q_{\rm I}$ in this scenario -- see
\Tref{tab1}.

\begin{table}[!t]
\begin{center}
\begin{tabular}{|c|c|c|} \hline
{\bf  SC} & {\bf  LRS} & {\bf QKS}
\\ \hline \hline
$\brhoq=\brhof=0$&$\brhofi\gg\brhoRi,~\brhoq=0$&
$\brhoqi\gg\brhoRi,~\brhof=0$\\
$H\propto T^2$& $H\propto T^4$&$H\propto T^3$\\
$T\propto R^{-1}$&$T\propto R^{-3/8}$&$T\propto R^{-1}$\\
$sR^3={\rm cst}$&$sR^3\neq{\rm cst}$&$sR^3={\rm
cst}$\\
$N_\chi=0$&$N_\chi\neq0$&$N_\chi=0$\\ \hline
\multicolumn{3}{|c|}{\sc Free Parameters of the $\Omx$
Calculation}\\\hline
$\mx,~\sigv$&$\mx,~\sigv,$&$\mx,~\sigv,$\\
&$\Trh,~m_\phi,~N_\chi$&$a,~b,~M,~H_{\rm I},~q_{\rm I},~\Ti$\\
\hline
\end{tabular}
\end{center}\vspace*{-.155in}
\caption{\sl\ftn Comparing the SC with the LRS and the QKS (the
various symbols are explained in \Sref{Beqs0}, the subscript I is
referred to the onset of each scenario and ``cst'' stands for
``constant'').\label{tab1}}
\end{table}

In the two non-standard scenaria under consideration, $H$ is given
by
\begin{equation}\label{rhoqi}
H^2={1\over3m^2_{\rm P}}\left\{\matrix{
\left(\rho_\phi +\rho_\chi +\rho_{{\rm R}}\right)~~&\mbox{for the
LRS}, \hfill \cr
\left(\rho_q +\rho_{{\rm R}}+ \rho_{{\rm M}}\right)~~&\mbox{for
the QKS}, \hfill \cr}\right.\eeq
where $\rho_i$ with $i=\phi, q$ and $\chi$ is the energy density
of $\phi$, $q$ and $\chi$ respectively and $m_{{\rm P}}=M_{\rm
P}/\sqrt{8\pi}$ where $M_{\rm P}=1.22\cdot10^{19}~{\rm GeV}$ is
the Planck mass. The energy density of radiation, $\rho_{{\rm
R}}$, and the entropy density, $s$, can be evaluated as a function
of $T$, whilst the energy density of matter, $\rho_{{\rm M}}$,
with reference to its present-day value:
\beq\label{rhos} \rho_{{\rm
R}}=\frac{\pi^2}{30}g_{\rho*}T^4,~s=\frac{2\pi^2}{45}g_{s*}T^3~~\mbox{and}~~\rho_{{\rm
M}}R^3=\rho_{{\rm M0}}R_0^3
\end{equation}
where $g_{\rho*}(T)~[g_{s*}(T)]$ is the energy [entropy] effective
number of degrees of freedom at temperature $T$. Their precise
numerical values are evaluated by using the tables included in
public packages \cite{micro} and assuming the particle spectrum of
the Minimal Supersymmetric Standard Model.

The initial value of $H$, $H_{\rm I}$, in both non-standard
scenaria can be restricted, assuming that a primordial phase of
inflation (driven by a scalar field different from $\phi$ or $q$,
in general) is responsible for the generation of the power
spectrum of the curvature scalar $P_{\rm s}$ and tensor $P_{\rm
t}$ perturbations. Indeed, imposing the conservative restriction
$r=P_{\rm t}/P_{\rm s}\lesssim1$ and using the observational
\cite{wmap} normalization of $P_{\rm s}$, an upper bound on
$H_{\rm I}$ can be found as follows:
\beq H_{\rm I} \lesssim{\pi\over\sqrt{2}}\mP P^{1/2}_{\rm s*}
~~\Rightarrow~~H_{\rm I}\lesssim2.65\cdot10^{14}~\GeV,
\label{para}\eeq
where $*$ means that  $P_{\rm s*}$ is measured at the pivot scale
$k_*=0.002/{\rm Mpc}$.

Let us, finally, introduce a set of normalized quantities which
simplify significantly the relevant formulas. In particular we
define
\numparts\bea && \label{bar1}\bar \rho_i={\rho_i/\rho_{\rm
c0}},~\mbox{with}~i={\rm R},~{\rm M},~\phi~\mbox{and}~q,~\bar
J={J/\rho_{\rm c0}^{3/4}}~\mbox{with}~J=n_\chi,~n^{\rm
eq}_{\chi}~\mbox{and}~n_\phi,\\
&& \label{bar2} \bar m_i={m_i/\rho_{\rm
c0}^{1/4}}~\mbox{with}~i=\chi~\mbox{and}~\phi,\bar J={J/
H_0}~\mbox{with}~J=H~\mbox{and}~\Gm{\phi}~\mbox{and}~\vsv=\sqrt{3}m_{\rm
P}\rho_{\rm c0}^{1/4}\sigv~~~~~~~.\eea
\endnumparts
\hspace{-.14cm}where $n_i$ with $i=\phi$ and $\chi$ is the number
density of $\chi$ and $\phi$ respectively. Note that
$\rho_\chi=m_\chi\, n_\chi$ and $\rho_\phi=m_\phi\,\Delta_\phi
n_\phi$ where $\Delta_\phi=(m_\phi-N_{\chi}m_{\chi})/m_\phi$. In
our numerical calculation, we use the values:
\beq \rho_{\rm
c0}\simeq8.1\cdot10^{-47}h^2~\GeV^4,~~\mbox{with}~~h=0.72,~~\vrho_{{\rm
M0}}=0.26 ~~\mbox{and}~~T_0=2.35\cdot 10^{-13}~{\rm GeV}.\eeq
We have also $H_0=2.13\cdot10^{-42}h~{\rm GeV}$ and from
Eq.~(\ref{rhos}) we get $\vrho_{{\rm R0}}=8.04\cdot10^{-5}$.

\subsection{The Low Reheating Scenario}
\label{sec:lr}

We below (\Sref{Beqs11}) present the system of equations which
governs the cosmological evolution in the LRS, summarize
(\Sref{reqlr}) the various observational restrictions that have to
be imposed and sketch (\Sref{dynlr}) the basics of the relevant
dynamics.

\subsubsection{Relevant Equations.} \label{Beqs11}
Under the assumption that the decay products of $\phi$ are rapidly
thermalized (see below) the energy densities $\rho_\phi$ and
$\rho_{\rm R}$ obey the following Boltzmann equations
\numparts
\begin{eqnarray}
&& \dot \rho_\phi+3H\rho_\phi+\Gamma_\phi \rho_\phi=0,\label{rf}\\
&& \dot \rho_{{\rm R}}+4H\rho_{{\rm R}}-\Gamma_\phi
\rho_\phi-2m_{\chi}\sigv \left( n_{\chi}^2 - n_{\chi}^{\rm
eq2}\right)=0, \label{rR}
\end{eqnarray}
\endnumparts
\hspace{-.14cm}where dot stands for derivative w.r.t the cosmic
time, $t$ and $n_{\chi}^{\rm eq}$ is the equilibrium number
density of $\chi$, which obeys the Maxwell-Boltzmann statistics:
\begin{equation} \label{neq}
n_{\chi}^{\rm eq}(x)=\frac{g}{(2\pi)^{3/2}}
m_{\chi}^3\>x^{3/2}\>e^{-1/x}P_2\left({1\over
x}\right),~~\mbox{where}~~x={T\over
m_{\chi}}~~\mbox{and}~~P_n(z)=1+{(4n^2-1)\over8z}
\end{equation}
is obtained by expanding the modified Bessel function of the 2nd
kind of order $n$ for $x\ll1$. Assuming that ${\chi}$'s are
Majorana fermions, we set $g=2$ for their number of degrees of
freedom. Note that although in our numerical program \eqs{rf}{rR}
are resolved together with \Eref{nx} -- see \Sref{Beq}--, we here
prefer to present just the two first equations since the influence
of $n_\chi$ to the dynamics of reheating via the last term of the
left hand side in \Eref{rR} is in general negligible. Moreover,
$\Gamma_{\phi}$ can be replaced by $T_{\rm RH}$ through the
relation \cite{rhodos}:
\begin{equation}
\Gamma_\phi = 5\sqrt{\frac{\pi^3 g_{\rho*}(\Trh)}{45}}
\frac{\Trh^2}{M_{{\rm P}}}=\sqrt{\frac{5\pi^2
g_{\rho*}(\Trh)}{72}} \frac{\Trh^2}{m_{{\rm P}}} \cdot\label{GTrh}
\end{equation}
Note that the adopted prefactor, which is slightly different than
those used in the bibliography \cite{riotto, gondolo}, assists us
to approach accurately the numerical solution of $\rho_\phi(T_{\rm
RH})=\rho_{\rm R}(T_{\rm RH})$.

The numerical integration of \eqs{rf}{rR} is facilitated by
absorbing the dilution terms. To this end, we find it convenient
to define the following variables \cite{riotto, rhodos}:
\begin{equation} \label{fdef}
f_\phi=\brhof R^3,~f_{\rm R}=\brhoR R^4,~ f_{\chi}=\vn_{\chi}
R^3~~\mbox{and}~~ f_{\chi}^{\rm eq}=\vn^{\rm eq}_{\chi} R^3
\end{equation}
and convert the time derivatives to derivatives w.r.t the
logarithmic time \cite{rhodos}:
\beq \btau=\ln\left(R/R_{\rm
I}\right)~\Rightarrow~R^\prime=R~~\mbox{and}~~R=R_{\rm I
}e^{\btauf}\label{btau} \eeq
where prime in this section denotes derivation w.r.t $\btau$ and
the value of $R_{\rm I}$ in this definition can be conveniently
selected so as the resolution of the system is numerically stable.
After realize the modifications above, \eqs{rf}{rR} become:
\beq \mbox{\sf\small (a)}~~\vH f^\prime_\phi=-\vGm{\phi}
f_\phi~~\mbox{and}~~\mbox{\sf\small (b)}~~\vH R^2 f^\prime_{\rm
R}=\vGm{\phi} f_\phi R^3+2 \bar m_{\chi} \vsv \left(f_{\chi}^2 -
f_{\chi}^{\rm eq2}\right), \label{fffr} \eeq
where $H$ and $T$ can be expressed correspondingly, in terms of
the variables in \Eref{fdef}, as follows:
\begin{equation} \label{H2exp}
\mbox{\sf\small (a)}~~\vH=R^{-3/2}\sqrt{f_\phi +f_{\rm R}/R
}~~\mbox{and}~~\mbox{\sf\small (b)}~~T=\left(\frac{30\ f_{\rm
R}}{\pi^2 g_{\rho\ast} R^4}\rho_{\rm
c0}\right)^{1/4}\cdot\end{equation}
The system of \Eref{fffr} can be solved from 0 to $\btau_{\rm
f}\sim 50$, imposing the following initial conditions (recall that
the subscript I is referred to quantities defined at $\btau=0$):
\begin{equation}
H_{\rm
I}=m_\phi~\Rightarrow~\brhofi=m_\phi^2/H_0^2~~\mbox{and}~~\brhoRi=\bar\rho_{\chi_{\rm
I}}=0. \label{init}
\end{equation}
However, the results on $\Omx$ do not depend on the explicit value
of $\brhofi$ as long as $T_{\rm RH}<T_{\rm F}<T_{\rm max}$, and
are invariant \cite{rhodos, gondolo} for fixed
$N_{\chi}m^{-1}_\phi$ (and $T_\phi,~m_\chi,~\sigv$). Therefore,
for presentation purposes, it is convenient to define the
following quantity \cite{gondolo}:
\beq\label{cxf}\cxf= N_\chi\;{100~\TeV\over\mff}\cdot\eeq

\subsubsection{Imposed Requirements.} \label{reqlr}
We impose on the LRS the following requirements:

\begin{list}{}{\setlength{\rightmargin=0cm}{\leftmargin=0.7cm}}

\item[$\bullet$] The BBN Constraint. The presence of $\rho_\phi$
should not jeopardize the successful predictions of BBN which
commences at about $T_{\rm BBN}=1~{\rm MeV}$ \cite{oliven}.
Namely, we require:
\beq\Trh\geq1~\MeV~~\mbox{($95\%$ c.l.).} \label{nuc1}\eeq
Given that $\phi$ decays mostly through gravitational
interactions, $\Gm{\phi}$ and consequently $\Trh$ -- see
\Eref{GTrh} -- are highly suppressed. Therefore \cite{gondolo}
fulfilment of BBN constraint with more or less natural coupling
constants requires $m_\phi\geq100~\TeV$.

\item[$\bullet$] Constraints on the range of $m_\phi$. \Eref{para}
assists us to impose an upper bound on $m_\phi$ due to our initial
condition in \Eref{init}. On the other hand, $m_\phi$ can be
bounded from below, too, demanding the decay of $\phi$ to a pair
of $\chi$'s (with mass $\mx$) to be kinematically allowed. All in
all we require:
\beq 2\mx\leq m_\phi\lesssim2.65\cdot10^{14}~{\GeV}.
\label{mfb}\eeq
Note that the upper bound of \Eref{mfb} assures also the rapid
thermalization of the $\phi$-decay products. Indeed, the latter
condition, which is crucial for \eqs{rf}{rR} to be applicable, is
satisfied \cite{sarkar} for $m_\phi\lesssim8\cdot10^{14}~{\rm
GeV}$.

\item[$\bullet$] Constraint on the range of $N_{\chi}$. Depending
on the coupling between $\phi$ and $\chi$ in a specific theory, a
variety of $N_{\chi}$'s is possible \cite{moroi, gondolo, koichi}.
In our approach we conservatively set the upper bound
$N_{\chi}\leq 1$.

\end{list}

Let us finally mention that, on quite general ground, any modulus
$\phi$ has an unsuppressed coupling to gravitino, $\Gr$. Possible
decay of $\phi$ to $\Gr$ creates the co-called moduli-induced
$\Gr$ problem \cite{koichi}. To avoid these complications, we are
obliged to assume that the masses of $\Gr$ and $\phi$ are of the
same order of magnitude.

\subsubsection{Dynamics of Reheating.} \label{dynlr}

\begin{figure}[t]\vspace*{-.1in}
\hspace*{-.25in}
\begin{minipage}{8in}
\epsfig{file=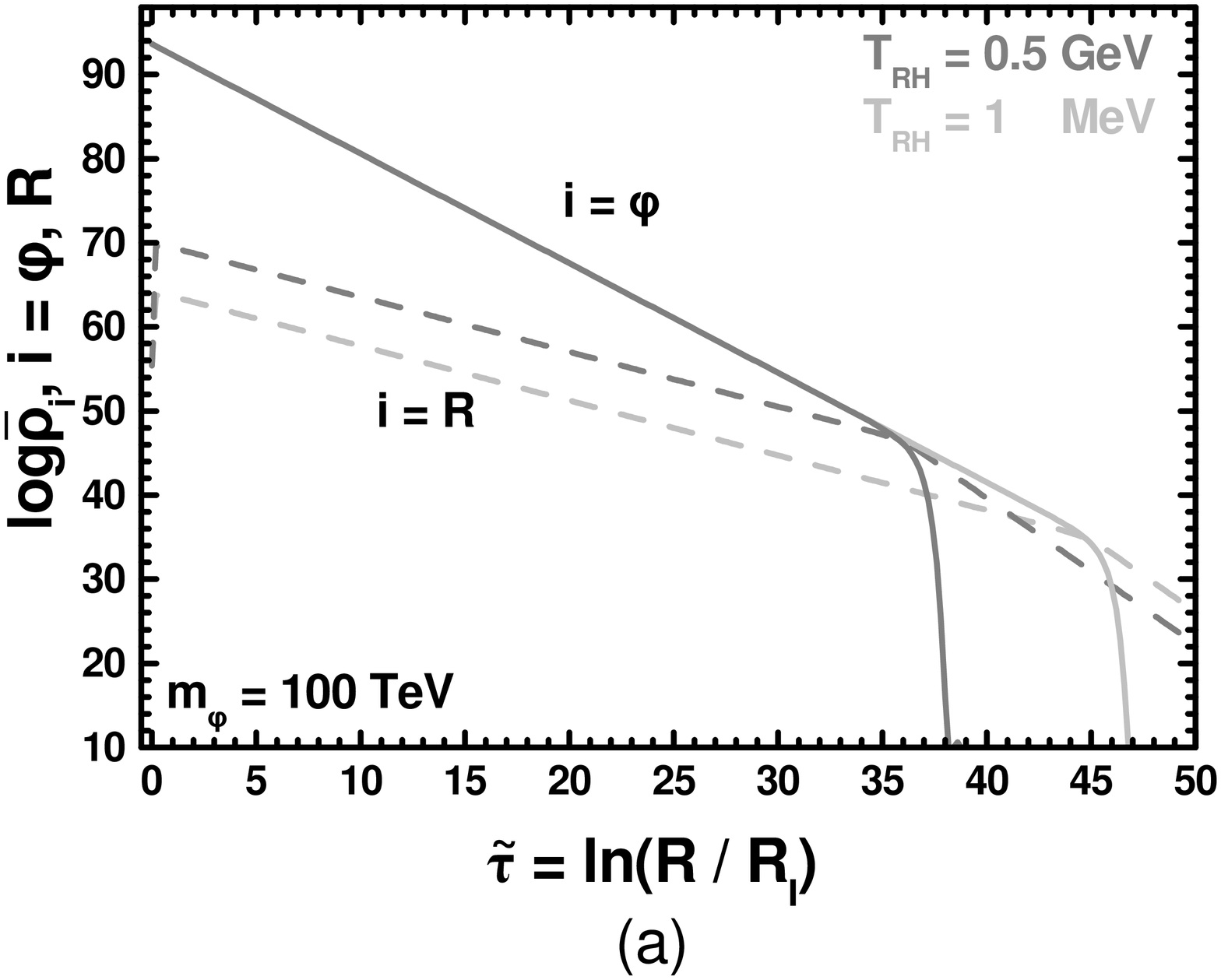,height=3.55in,angle=-90}
\hspace*{-1.37 cm}
\epsfig{file=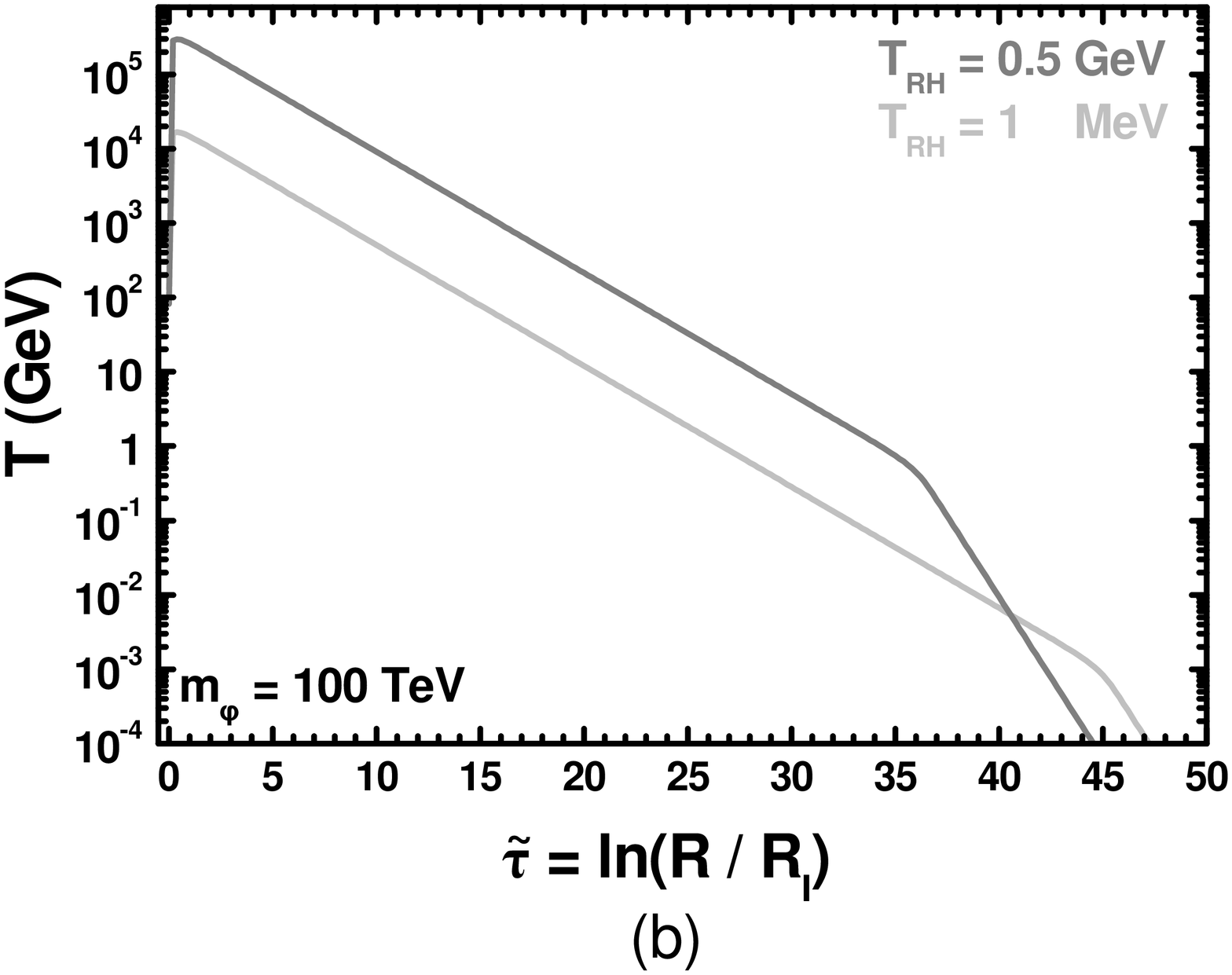,height=3.55in,angle=-90} \hfill
\end{minipage}\vspace*{-.01in}
\hfill \caption[]{\sl\ftn The evolution of {\sf\ssz (a)}
$\log\vrho_i$ with $i=\phi$ (solid lines) and  R (dashed lines)
and {\sf\ssz (b)} $T$ as a function of $\btau=\ln(R/R_{\rm I})$
for $\mff=100~\TeV$ and $(N_\chi, \Trh)=(1, 0.5~\GeV)$ (gray
lines) or $(N_\chi, \Trh)=(7.5\cdot10^{-5}, 1~\MeV)$ (light gray
lines). In both cases, we take $\Omx=0.11$ for $\mx=0.5~\TeV$ and
$\sigv=3\cdot 10^{-7}~\GeV^{-2}$. } \label{figrh}
\end{figure}

The cosmological evolution of the various quantities involved in
the LRS as a function of $\btau$ is illustrated in \Fref{figrh}
for $\mff=100~\TeV$ and $(N_\chi, \Trh)=(1, 0.5~\GeV)$ (gray
lines) or $(N_\chi, \Trh)=(7.5\cdot10^{-5}, 1~\MeV)$ (light gray
lines). In particular, we design $\log\vrho_i$ with $i=\phi$
(solid lines) and $i=$R (dashed lines) [$T$] versus $\btau$ in
\sFref{figw}{a} [\sFref{figw}{b}]. The quantities $\vrho_i$ with
$i=\phi$ and R [$T$] are computed by substituting the numerical
solution of \Eref{fffr} into \Eref{fdef} [\sEref{H2exp}{b}]. From
\Fref{figrh} we can understand the dynamics of the universe during
the two distinct phases \cite{riotto, rhodos}:

\begin{list}{}{\setlength{\rightmargin=0cm}{\leftmargin=0.7cm}}

\item[$\bullet$] For $T\gg T_{\rm RH}$, we have
$\rho_\phi\gg\rho_{\rm R}$. Consequently, inserting
$\vH\simeq\sqrt{\brhof}$ into \Eref{fffr} we extract:
\beq \label{rrh}\mbox{\sf\small (a)}~~\vrho_\phi=\brhofi
e^{-3\btauf}~~\mbox{and}~~\mbox{\sf\small (b)}~~\vrho_{{\rm
R}}=\frac{2}{5}\vGamma_\phi\brhofi^{1/2}
\left(e^{-3\btauf/2}-e^{-4\btauf}\right).\eeq
The function $\vrho_{{\rm R}}(\btau)$ reaches at $\btau_{{\rm
max}}\simeq\ln(1.48)=0.39$ a maximum value $\vrho_{\rm
Rmax}\simeq0.14\sqrt{\brhofi}\vGamma_\phi$ corresponding to a
$T=T_{\rm max}$ derived through \Eref{rhos}. The completion of the
reheating is realized at $\btau=\btrhp$, such that:
\beq \label{trh} \rho_{{\rm
R}}(\btrhp)=\rho_\phi(\btrhp)~\Rightarrow~\btrhp\simeq-
\frac{2}{3}\ln\frac{2}{5}\vGamma_\phi\brhofi^{-1/2}.\eeq
where a corner is observed on the curves of \sFref{figrh}{b}.

\item[$\bullet$] For $T< T_{\rm RH}$, we get $\rho_{{\rm
R}}\gg\rho_\phi$, and so, $\vH\simeq\sqrt{\vrho_{{\rm R}}}$.
Plugging it into \Eref{fffr} we can obtain approximately the
following expressions:
\beq\vrho_\phi=\vrho_\phi(\btrhp)\
\exp\left(-3(\btau-\btrhp)-\frac{5}{4}\left(e^{2(\btau-\btrhp)}-1\right)\right)~~\mbox{and}~~\vrho_{{\rm
R}}=\vrho_{{\rm R}}(\btrhp)\ e^{-4(\btau-\btrhp)}.\label{rfrp}
\eeq
\end{list}

\subsection{The Quintessential Kination Scenario}
\label{sec:quint}

We present below (\Sref{Beqs1}) the equations which govern the
cosmological evolution in the QKS and then enumerate
(Sec.~\ref{reqq}) the various observational restrictions that have
to be imposed. We also highlight the $q$ dynamics (\Sref{Qev}) and
describe the allowed parameter space (\Sref{ap}).

\subsubsection{Relevant Equations.\label{Beqs1}} Under the assumption that
$q$ is spatially homogeneous, it obeys the following Klein-Gordon
equation:
\beq \ddot q+3H\dot
q+V_{,q}=0,~~\mbox{where}~~V=V_a+V_b~~\mbox{with}~~V_a={M^{4+a}\over
q^a}~~\mbox{and}~~V_b={b\over2}H^2q^2,\label{qeq} \eeq
is the adopted potential for the field $q$ with $M$ a mass-scale
and $,q$ stands for derivative w.r.t $q$. In our approach $V_b$ is
present throughout the cosmological evolution of $q$. The induced
coupling between $q$ and CDM during the matter dominated era is
too suppressed to have any observational consequence. Nonetheless,
we have checked that our results remain intact even if we switch
off this term after the onset of the matter domination. The
numerical integration of \Eref{qeq} is facilitated by converting
the time derivatives to derivatives w.r.t the logarithmic time
\cite{jcapa} which is defined as a function of the redshift $z$:
\beq \vtau=\ln\left(R/R_0\right)=-\ln (1+z).\label{dtau} \eeq
Changing the differentiation and introducing the following
quantities:
\beq \label{vrhos}\vV_a=V_a/\rho_{\rm c0},~~f_q=\dot q
R^3/\sqrt{\rho_{\rm c0}}~~\mbox{and}~~\vq=q/\sqrt{3}m_{{\rm
P}},\eeq
\Eref{qeq} turns out to be equivalent to the system of two
first-order equations:
\bea &f_q=\vH\vq^\prime R^3 ~~&\mbox{and}~~\vH
f_q^\prime/R^3+b\,\vH^2\vq+b\vH\vH_{,\vq}\;\vq^2+
\vV_{a,\bar q}=0,~~\label{vH1}\\
&\mbox{where}&\vH^2={1\over1-b\vq^2/2}\;\left({1\over2}f_q^2/R^6+\vV_a+\vrho_{{\rm
R}}+\vrho_{{\rm M}}\right), \label{vH} \eea
prime in this section denotes derivative w.r.t. $\vtau$ and $M$
can be found from the dimensionless quantities as follows:
\beq M=\left(\left(\sqrt{3}\mP\right)^a\vVo\rho_{\rm
c0}\right)^{1/(4+a)}~~\mbox{with}~~\vV_a={\vVo/\vq^a}.\label{Mq}\eeq

Eq. (\ref{vH1}) can be resolved numerically if two initial
conditions are specified at an initial $\vtau$, $\vti$
corresponding to a temperature $\Ti$, which is defined as the
maximal $T$ after the end of primordial inflation, assuming
instantaneous reheating. We take $\vq(\vti)=10^{-2}$ throughout
our investigation, without any loss of generality (see below) and
let as free parameter $\vHi=\bar H(\Ti)$.

\subsubsection{Imposed Requirements.} \label{reqq}
Our QKS can be \cite{patra} consistent with the following
restrictions:

\begin{list}{}{\setlength{\rightmargin=0cm}{\leftmargin=0.7cm}}

\item[$\bullet$] Constraints on the initial conditions. We focus
on the initial conditions which assure a complete initial
domination of kination consistently with \Eref{para}, i.e.,
\bea &&\mbox{\sf\small (a)}~~\Omega^{\rm
I}_q=\Omega_q(\Ti)=1~~\mbox{and}~~\mbox{\sf\small
(b)}~~\vHi\lesssim1.72\cdot10^{56}\label{domk} \\
&&\mbox{with}~~\Omega_q=\rho_q/(\rho_q+\rho_{{\rm R}}+\rho_{\rm
M})~~\mbox{where}~~\rho_q=\dot q/2+V \label{rq}.\eea
\item[$\bullet$] The BBN Constraint. The presence of $\rho_q$ has
not to spoil the successful predictions of BBN which commences at
about $\vtns=-22.5$ ($T_{\rm BBN}=1~\MeV$) \cite{oliven}. Namely,
we require:
\beq\Omqns=\Omega_q(\vtns)\leq0.21~~\mbox{($95\%$ c.l.)}
\label{nuc}\eeq
where 0.21 corresponds to additional effective neutrinos species
$\delta N_\nu<1.6$ \cite{oliven}.

\item[$\bullet$] DE-Density and Coincidence Constraint. These two
constraints can be addressed if (i) the present value of $\rho_q$,
$\rho_{q0}$, is compatible with the abundance of DE in the
universe \cite{wmap} and (ii) $\rho_q$ has already reached the
tracking behavior. In other words we have to demand \cite{mas}
\beq \mbox{\sf\small
(a)}~~\Omega_{q0}=\vrho_{q0}=0.74~~\mbox{and}~~\mbox{\sf\small
(b)}~~d^2V(\vtau=0)/dq^2\simeq H^2_0,\label{rhoq0}\eeq
where we restrict ourselves to the central experimental value of
$\Omega_{q0}$, since, this choice does not affect crucially our
results on the CDM abundance.

\item[$\bullet$] Acceleration Constraint. Any successful
quintessential model has to account for the present-day
acceleration of the universe, i.e., \cite{wmap} (see also
\cref{snae})
\beq -1.12\leq w_q(0)\leq-0.86~~\mbox{($95\%$
c.l.)}~~\mbox{with}~~w_q=(\dot q^2/2-V)/(\dot q^2/2+V).
\label{wqd}\eeq
\end{list}

\subsubsection{Quintessential Dynamics.} \label{Qev}

\begin{figure}[t]\vspace*{-.1in}
\hspace*{-.25in}
\begin{minipage}{8in}
\epsfig{file=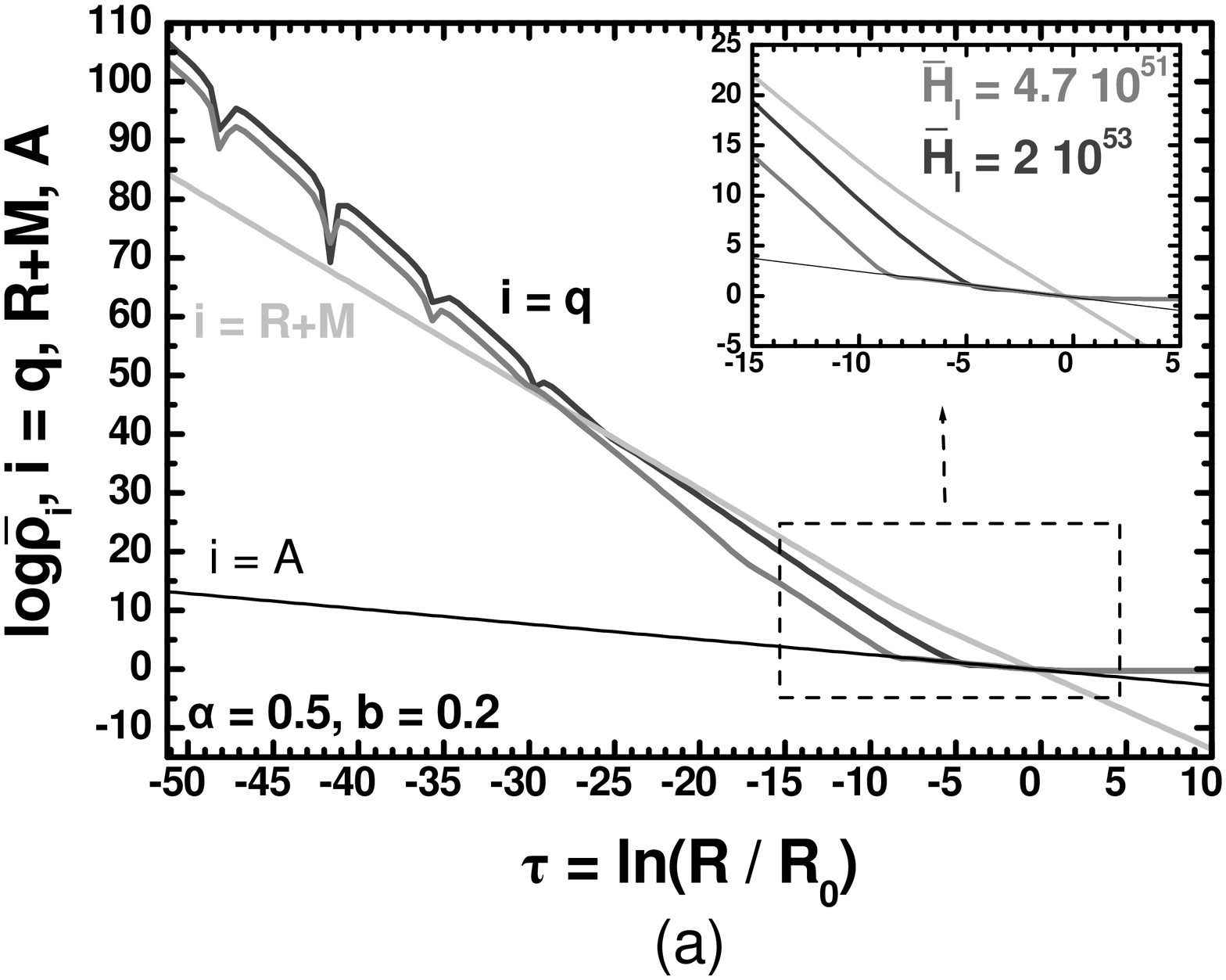,height=3.55in,angle=-90}
\hspace*{-1.37 cm}
\epsfig{file=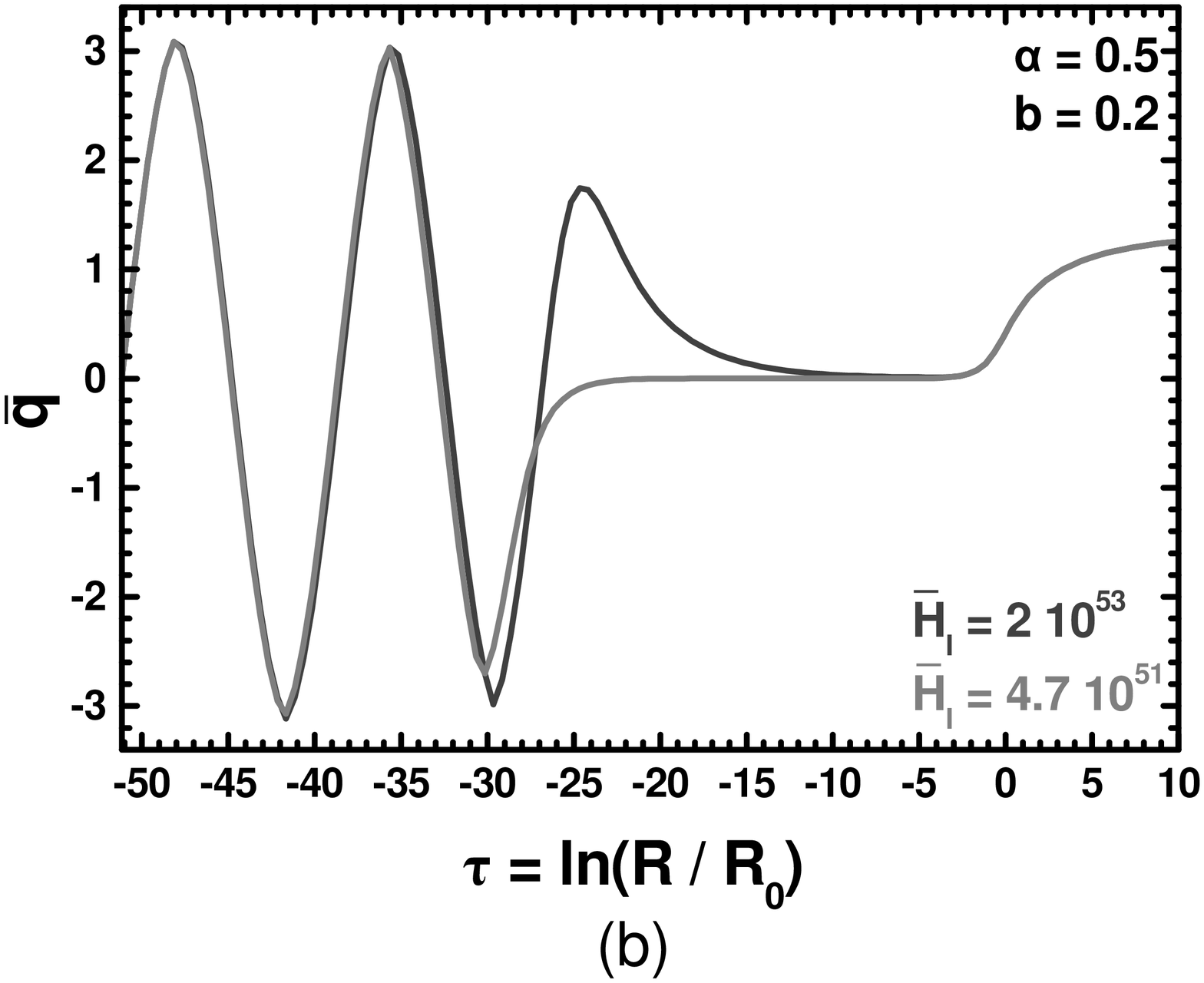,height=3.55in,angle=-90} \hfill
\end{minipage}\vspace*{-.01in}
\hfill \caption[]{\sl\ftn The evolution of {\sf\ssz (a)}
$\log\vrho_i$ with $i=q$ (gray [dark gray] line), R+M (light gray
line) and A (thin line); and {\sf\ssz (b)} $\vq$ (gray [dark gray]
line) as a function of $\vtau=\ln(R/R_0)$ for $\vqi=0.01$,
$a=0.5$, $b=0.2$, $\Ti=10^9~\GeV$ ($\vti\simeq-51.16$) $M=4.8~{\rm
eV}$ and $\vHi=4.7\cdot10^{51}$ ($\Omqns=0.0001$)
[$\vHi=2\cdot10^{53}$ ($\Omqns=0.21$)].} \label{figw}
\end{figure}

The cosmological evolution of the various quantities involved in
our QKS as a function of $\vtau$ is illustrated in \Fref{figw} for
$\vqi=0.01$, $a=0.5$, $b=0.2$, $\Ti=10^9~\GeV$ ($\vti\simeq-51.2$)
$M=4.8~{\rm eV}$ and $\vHi=2\cdot10^{53}$ ($\Omqns=0.21$ dark gray
lines) or $\vHi=2\cdot10^{53}$ ($\Omqns=0.0001$, gray lines).
Particularly, in \sFref{figw}{a} [\sFref{figw}{b}] we draw
$\log\vrho_i$ [$\vq$] versus $\vtau$. For $i=q$ (gray and dark
gray lines), $\vrho_q$ is computed by inserting in the last
equation of \Eref{rq} the numerical solution of \Eref{vH1}. For
$i={\rm R+M}$ (light gray line), we show $\vrho_{{\rm
R+M}}=\vrho_{{\rm R}}+\vrho_{{\rm M}}$ given by Eq.~(\ref{rhos}).
For $i={\rm A}$ (thin black line), $\vrho_{{\rm A}}$ is the
dimensionless energy density of the attractor solution (see
below).

From \Fref{figw} we can conclude that $q$ undergoes four phases
during its cosmological evolution \cite{mas, patra, binetruy}:

\begin{list}{}{\setlength{\rightmargin=0cm}{\leftmargin=0.7cm}}

\item[$\bullet$] The kinetic-energy dominated phase during which
the evolution of both the universe and $q$ is dominated by $\dot
q/2\gg V$. Therefore $\vH\simeq\vH\vq'/\sqrt{2-b\vq^2}$ and
integrating it we obtain \cite{patra}
\beq \label{rhok} \vq\simeq\sqrt{2\over
b}\sin\sqrt{b}\,(\vtau-\vti). \eeq
Obviously, for $b>0$, $q$ is set in harmonic oscillations during
the KD era. In particular, $\vq$ develops extrema at
\beq \label{tmax}\vtp\simeq(2k+1)\sqrt{1\over
b}{\pi\over2}+\vti~~\mbox{with}~~k=0,1,2,...\eeq
On the other hand, $\dot\vq=\vH\vq'$ almost vanish for
$\vtau=\vtp$. Therefore, at $\vtau\simeq\vtp$, $\vrhoR$ dominates
instantaneously over $\dot q/2$. As a consequence, the $\vq$
oscillations become anharmonic. This phase terminates for
$\vtau=\vtkr$ where $\rho_q=\rho_{\rm R}$. For the inputs of
\Fref{figw} we get $\vtkr=-25.6$ [$\vtkr=-28.2$] (or
$\Tkr=0.02~\GeV$ [$\Tkr=0.21~\GeV$]) for $\vHi=2\cdot10^{53}$
[$\vHi=4.7\cdot10^{51}$]. We observe that the lowest $\Tkr$
corresponds to the largest $\vHi$ (and $\Omqns$). From
\sFref{figw}{b} we also remark that the height of the fifth peak
of $\vq$ decreases with $\vHi$. In fact, for
$\vHi<4.7\cdot10^{51}$ we take $\vq_0<0$ and so, $q$ can not serve
as quintessence (see below).

\item[$\bullet$] The frozen-field dominated phase, where the
universe becomes RD and $\rho_q$ is dominated initially by $\dot
q/2$ and subsequently by $V$ and $\vq$ is stabilized to a constant
value -- see \sFref{figw}{b}.

\item[$\bullet$] The attractor dominated phase, where
$\rho_q\simeq V$ and $\rho_{\rm M}$ dominates the evolution of the
universe. The system in \Eref{qeq} admits \cite{binetruy} a
tracking solution since the energy density of the attractor:
\beq \label{rA} \vrho_{\rm A}\simeq\vrho_{\rm
Af}\exp\left(-3(1+w_q^{\rm fp})(\vtau-\vtau_{\rm
Af})\right)~~\mbox{with}~~w_q^{\rm fp}=-\frac{2}{a+2} \eeq
tracks $\vrho_{\rm M}$ until $\vtau=\vtau_{\rm Af}$ where the
tracking regime terminates and $\vrho_{\rm M}\simeq\vrho_{\rm A}$.
For both $\vHi$'s used in \Fref{figw}, we get $\vtau_{\rm
Af}=-0.4$ whereas the onset of this phase occurs at $\vtau_{\rm
Ai}=-4.8$ [$\vtau_{\rm Ai}=-8.6$] for $\vHi=2\cdot10^{53}$
[$\vHi=4.7\cdot10^{51}$]. We observe that although the used
$\vHi$'s differ by two orders of magnitude, both $\vrho_q$'s reach
$\vrho_{\rm A}$ highlighting thereby the insensitivity of our QKS
to the initial conditions.

\item[$\bullet$] Vacuum Dominated Phase. For $\vtau>\vtau_{\rm
Af}$, the evolution of the universe is dominated by $V$. For the
parameters used in \Fref{figw} we get $w_q(0)\simeq-0.88$ and
$\Omega_q(0)\simeq0.74$.
\end{list}

\subsubsection{Allowed Parameter Space.} \label{ap}

\begin{figure}[!t]\vspace*{-.1in}
\hspace*{-.25in}
\begin{minipage}{8in}
\epsfig{file=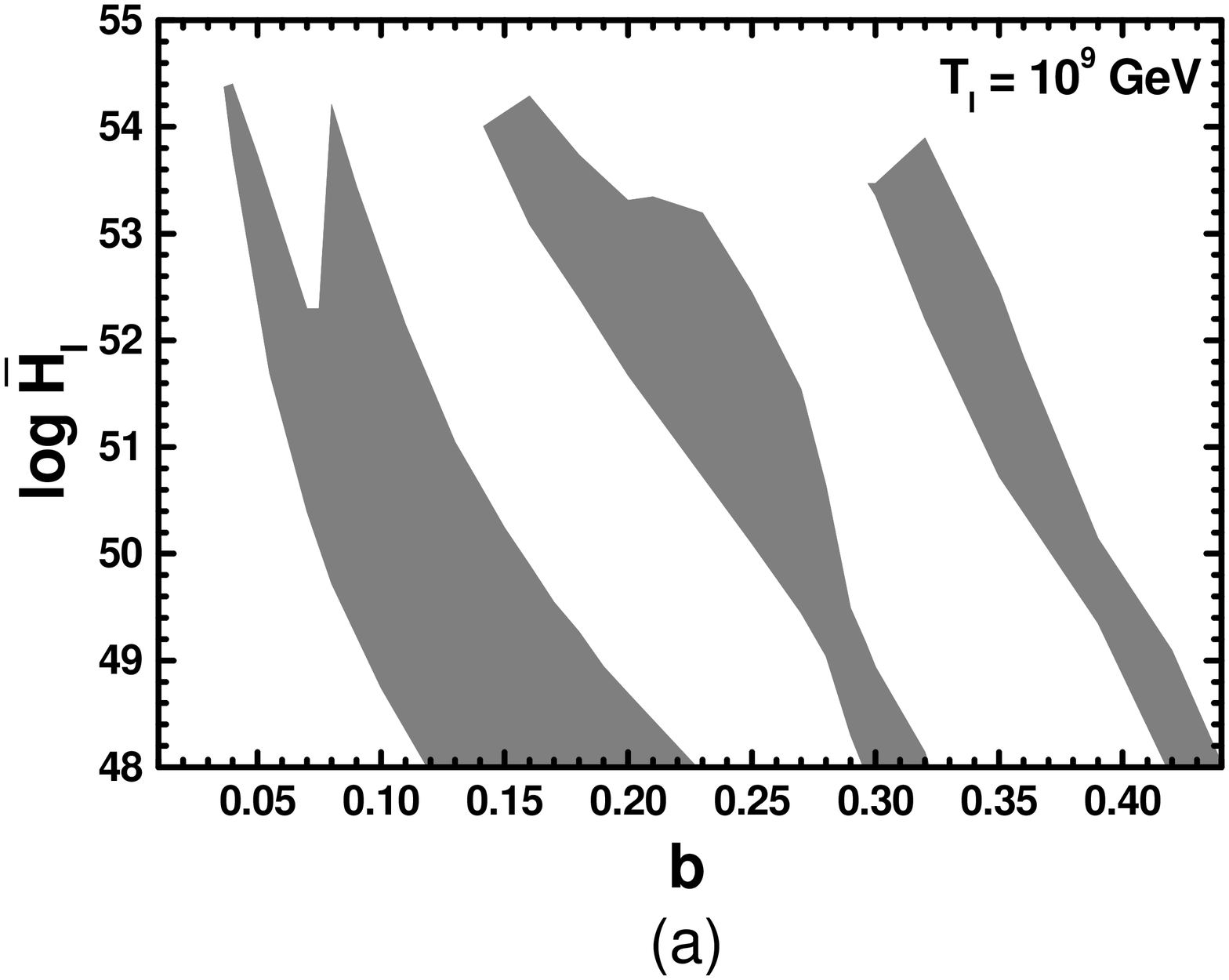,height=3.55in,angle=-90}
\hspace*{-1.37 cm}
\epsfig{file=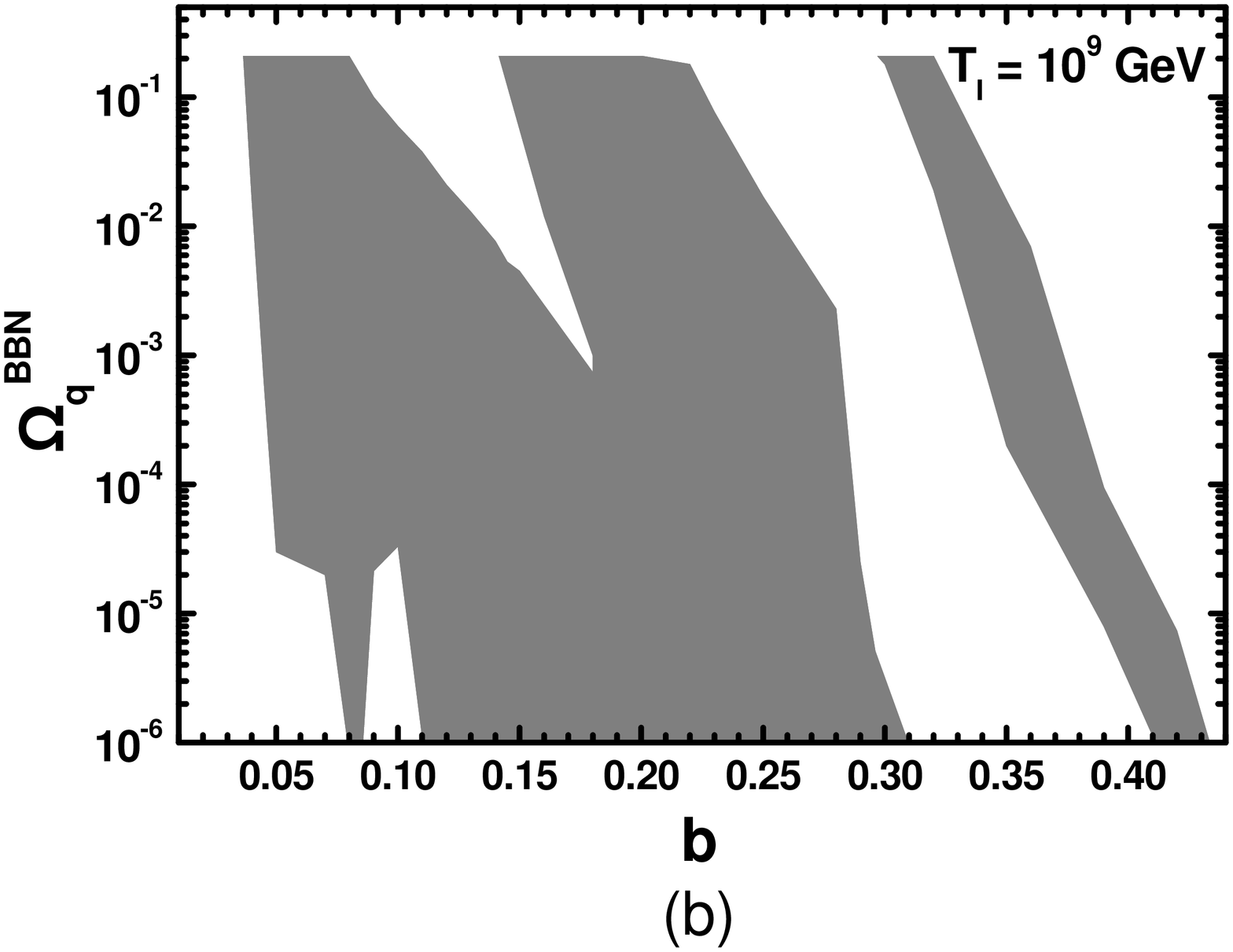,height=3.55in,angle=-90} \hfill
\end{minipage}
\hfill \caption[]{\sl\ftn Allowed (gray shaded) region by
Eqs.~(\ref{domk}) -- (\ref{wqd}) in the {\sf\ssz (a)} $b-\vHi$ and
{\ssz\sf (b)} $b-\Omqns$ plane for $a=0.5,~\vqi=0.01$ and
$\Ti=10^9~\GeV$.} \label{OmT}
\end{figure}

The free parameters of our QKS listed in Table~\ref{tab1} can be
restricted using the criteria presented in \Sref{reqq}. Agreement
with Eq.~(\ref{wqd}) entails $0<a\lesssim0.6$ (compare also with
\cref{bacci}, where less restrictive upper bound on $w_q(0)$ is
imposed). The parameter $M$ can be determined for every $a$
through \Eref{Mq} so that \sEref{rhoq0}{a} is satisfied. The
determination of $a$ and $M$ is independent of $\vti,~\vqi$ and
$\vHi$ provided that the tracking solution is reached in time. To
reduce somehow the parameter space of our investigation we fix
$\Ti=10^9~\GeV$ (or $\vti=-51.16$). This choice is motivated by
the majority of the inflationary models (see, e.g.,
\cref{susyhybrid}). We thereby focus on the two residual free
parameters of our model and we design in \sFref{OmT}{a}
[\ref{OmT}-{\sf\small (b)}] the allowed areas in the $b-\log\vHi$
[$b-\log\Omqns$] plane. In the gray shaded areas Eqs.~(\ref{domk})
- (\ref{wqd}) are fulfilled. Obviously our model possesses an
allowed parameter space with a band structure. The upper boundary
curves of the allowed bands come from \Eref{nuc}. Note, however,
that saturation of \Eref{nuc} is not possible for $0.08<b<0.16$.

For any $(b,\log\vHi)$, which is consistent with \Eref{nuc} and
belongs in a white [gray] band, the resulting $\vq$ after the
oscillatory phase turns out to be negative [positive] and so, it
cannot [can] serve as quintessence. E.g., let us fix $b=0.2$. For
$51.7\lesssim\log\vHi\lesssim53.3$, $\vq$ develops five extrema
during its evolution -- which is of the type shown in -- resulting
to $\vq_0>0$. Actually in \sFref{figw}{b} we display the evolution
of $\vq$ as a function of $\vtau$ for the two bounds of this band.
As $\log\vHi$ decreases below $53.3$ (where the bound of
\Eref{nuc} is saturated), the amplitude of the fifth peak, which
appears in the $\vq$-evolution (at about $\vtau\simeq-24.5$)
eventually decreases and finally this peak disappears at
$\log\vHi\simeq51.7$ where the first allowed band terminates. For
$48.7\lesssim\log\vHi\lesssim51.7$, $\vq$ develops four extremes
during its evolution resulting to $\vq_0<0$. As $\log\vHi$
decreases below $51.7$ the amplitude of the forth peak which
appears in the $\vq$-evolution (at about $\vtau\simeq-30$)
decreases and finally this peak disappears at $\log\vHi\simeq48.7$
where the second allowed band commences. Note that, in the first
allowed band, $\Omqns$ increases with $\vHi$ but this is not a
generic rule.

Variation of $\Ti$ (or equivalently $\vti$) modifies the range of
the obtainable $\Omqns$ and changes somehow the position of the
several bands of our parameter space but do not alter the
essential features of our results. E.g., for $b=0.2$ and
$\Ti=10^{10}~\GeV$ [$\Ti=10^{8}~\GeV$] (or $\vti=-53.5$
[$\vti=-48.9$]) the margin of the first allowed band is
$53.6\lesssim\log\vHi\lesssim56.4$
[$49.6\lesssim\log\vHi\lesssim51.1$] and the second allowed band
commences at $\log\vHi\simeq50.7$ [$\log\vHi\simeq46.75$].

\section{The WIMP Relic Density}\label{sec:boltz}

We turn to the calculation of the relic density, $\Omx$, of a
WIMP-CDM candidate, $\chi$. Employing the symbols defined in
\Eref{fdef}, $\Omx$ can be found from the well-know formula:
\begin{equation}
\label{omh} \Omega_{\chi}={\rho_{\chi0}\over\rho_{\rm c0}}=
{s_0\over\rho_{\rm c0}}\left.{m_{\chi} n_\chi\over
s}\right|_{\tau_{{\rm f}}} ~\Rightarrow~\Omx=4.533
\cdot10^{-27}~\GeV^2\ \left.{m_{\chi}f_\chi \over
sR^3}\right|_{\tau_{{\rm f}}},
\end{equation}
where $s_0\,h^2 /\rho_{\rm c0}^{1/4}= 4.533 \cdot10^{-27}~\GeV^2$
and $\tau_{{\rm f}}$ is a value of $\btau$ [$\vtau$] for the LRS
[QKS] -- see \Eref{btau} [\Eref{dtau}] -- large enough so as
$f_\chi$ is stabilized to its present constant value, $f_{\chi0}$.
Recall that $sR^3$ in the denominator of \Eref{omh} remains
constant only for $\btau>\btrhp$ in the LRS but for every $\vtau$
in the QKS -- see Table~\ref{tab1}. The evolution of $f_\chi$
obeys a Boltzmann equation. In \Sref{Beq} we present and solve
this Boltzmann equation and in \Sref{sec:wimp} we investigate how
we can achieve within our scenaria an enhancement of $\Omx$ w.r.t
its value in the SC.

\subsection{The Boltzmann Equation}
\label{Beq}

Since $\chi$'s are in kinetic equilibrium with the cosmic fluid,
their number density, $n_{\chi}$, evolves according to the
Boltzmann equation:
\beq \label{nx}\dot n_\chi+3Hn_\chi+\sigv \left(n_\chi^2 -
n_\chi^{\rm eq2}\right)=\left\{\matrix{
\Gamma_\phi N_{\chi} n_\phi~~\hfill &\mbox{for the LRS}, \hfill
\cr
0~~\hfill &\mbox{for the QKS}, \hfill \cr}
\right.\eeq
where $H$ is found from \Eref{rhoqi}. In order to find a precise
numerical solution to our problem, we have to solve \Eref{nx}
together with \Eref{qeq} [\eqs{rf}{rR}] for the QKS [LRS]. To this
end, we rewrite Eq.~(\ref{nx}) in terms of the quantities defined
in \Eref{fdef} as follows
\beq \vH R^3f'_{\chi}+\vsv\left(f_{\chi}^2- f_{\chi}^{\rm
eq2}\right)=\left\{\matrix{
\vGamma_\phi N_{\chi} \vn_\phi R^6\hfill &\mbox{for the LRS},
\hfill \cr
0~~\hfill&\mbox{for the QKS}, \hfill \cr}
\right. \label{fn} \eeq
where $\vH$ given by \sEref{H2exp}{a} [\Eref{vH}] and prime in
this section denotes derivation w.r.t $\tau=\btau$ [$\tau=\vtau$]
for the LRS [QKS]. \Eref{fn} can be solved numerically in
conjunction with \Eref{fffr} [\Eref{vH1}] for the LRS [QKS]. In
the LRS, we use the initial conditions in \Eref{init} and we
integrate from $\btau=0$ to $\btau_{{\rm f}}\simeq 50$ (with $g$'s
fixed to their values at $\Trh$). In the QKS, we impose the
initial condition $\vn_\chi(\vtau_\chi)=\vn_{\chi}^{\rm
eq}(\vtau_\chi)$, where $\vtx$ corresponds to the beginning
($x=1$) of the Boltzmann suppression of $\vn_{\chi}^{\rm eq}$ in
\Eref{neq}. The integration of \Eref{fn} is realized from $\vti$
down to $\vtns\simeq-22.5$ (an integration until to $0$ gives also
the same result).

Based on the semi-analytical expressions of \Sref{sec:nonSC}, we
can proceed to an approximate computation, which facilitates the
understanding of the problem and gives, in most cases, accurate
results. In particular, we can set -- see \Eref{vH}
[\eqs{rrh}{trh}] for the QKS [LRS]:
\bea  \nonumber &&\vH\simeq\sqrt{\vrhoR g_{\rm
C}}~~\mbox{where}~~g _{\rm C}\simeq\left\{\matrix{
1+T^4/\Trh^4\hfill &\mbox{for}~~ \btau\ll\btrhp,\cr
1\hfill & \mbox{for}~~\btau\gg\btrhp\hfill \cr}
\right.~~\mbox{for the LRS, or}\\
&& g_{\rm C}\simeq{1\over(1-b\vq^2/2)}\left\{\matrix{
1+f_q^2/2R^6\vrhoR\hfill &\mbox{for}~~ \vtau\ll\vtkr,\cr
1\hfill & \mbox{for}~~\vtau\gg\vtkr\hfill \cr}
\right.~~\mbox{for the QKS.}~~\label{Hgc}\eea

Introducing the notion of freeze-out temperature, $T_{\rm
F}=T(\tf)=x_{{\rm F}}m_{\chi}$ (see, e.g., \cref{rhodos, jcapa}
and references therein) we are able to study \Eref{fn} in the two
extreme regimes:

\begin{list}{}{\setlength{\rightmargin=0cm}{\leftmargin=0.7cm}}

\item[$\bullet$] For $\tau\ll \tf$, $f_\chi\simeq f_\chi^{\rm
eq}$. So, it is more convenient to rewrite \Eref{fn} in terms of
the variable $\Delta(\tau)=f_\chi(\tau)-f_\chi^{\rm eq}(\tau)$ as
follows:
\beq \label{deltaBE} \vH R^3\left(\Delta^{\prime}+{f_\chi^{\rm
eq}}^{\prime}\right)+\vsv \Delta\left(\Delta+2f_\chi^{\rm
eq}\right)=\left\{\matrix{
\vGm{\phi} N_{\chi} \vn_\phi R^6 \hfill&\mbox{for the LRS},  \cr
0 \hfill&\mbox{for the QKS}. \cr}
\right.\eeq
The freeze-out point $\vtf$ can be defined by
$\Delta(\tf)=\delta_{\rm F}f_\chi^{\rm eq}(\tf)$ where
$\delta_{\rm F}$ is a constant of order unity, determined by
comparing the exact numerical solution of \Eref{fn} with the
approximate under consideration one. Inserting this definition
into \Eref{deltaBE}, we obtain the equation:
\bea &&(\delta_{\rm F}+1)f_\chi^{\rm eq}(\tf)\vH R^3\Big(\ln
f_\chi^{\rm eq}\Big)^\prime(\tf) +\delta_{\rm F} (\delta_{\rm
F}+2)\vsv f_\chi^{\rm eq2}(\tf)=\left\{\matrix{
\vGm{\phi} N_{\chi} \vn_\phi R^6  \hfill&\mbox{for the LRS}, \cr
0 \hfill&\mbox{for the QKS},\cr}
\right.\nonumber \\ \label{xf} &&\mbox{with}~~\Big(\ln f_\chi^{\rm
eq}\Big)^{\prime}(\tau)=3+x^\prime\frac{(16+3x)(18+25x)}{2x^2(8+15x)}
\eea
which can be solved w.r.t $\tf$ iteratively. The $\vtau-x$
[$\btau-x$] dependence can be derived by combining \Eref{neq} and
\Eref{rhos} [\Eref{H2exp}] for the QKS [LRS].

\item[$\bullet$] For $\tau\gg\tf$, $f_\chi\gg f_\chi^{\rm eq}$ and
so, we can set $f_\chi^2-f_\chi^{\rm eq2}\simeq f_\chi^2$ in
\Eref{fn}. Let us analyze this case for each scenario separately:

\begin{list}{}{\setlength{\rightmargin=0cm}{\leftmargin=0.4cm}}

\item[$\blacklozenge$] In the LRS and for the range of the
parameters under consideration -- see \Sref{sec:wimp} -- we single
out two cases:

\begin{list}{}{\setlength{\rightmargin=0cm}{\leftmargin=0.4cm}}

\item[{\boldmath $\ast$}] Dominant \emph{non-thermal production}
(non-TP). In this case, which is mainly applicable for very low
$\Trh$'s, $f^2_\chi\vsv\ll\vGm{\phi}N_\chi\vn_\phi R^6$.
Therefore, \Eref{fn} can be integrated analytically inserting into
it \sEref{rrh}{a} as follows:
\beq \label{fxntp} f_{\chi
0}\simeq{2\over3}\sqrt{\brhofi}\,\vGm{\phi} N_\chi\Delta^{-1}_\phi
\bar m^{-1}_\phi \left(e^{3\btrhp/2}-e^{3\btf/2}\right).\eeq
Since $f_{\chi 0}$ takes its main contribution close to
$\btrhp\gg\btf$ our result is more or less independent of $\btf$.

\item[{\boldmath $\ast$}] Equipartition between TP and non-TP.
Besides $f_\chi^{\rm eq}$ none of the other terms of \Eref{fn} can
be neglected in this case. A rather precise result for $f_{\chi
0}$ can be obtained by numerically integrating \Eref{fn}
subsequently from {\sf\small (i)} $\btau=\btf$ until
$\btau=\btrhp$ with initial condition $f_\chi(\btf)=(\delta_{\rm
F}+1)f^{\rm eq }_\chi(\btf)$ and using $\vH$ [$\brhof$] given by
\Eref{Hgc} [\sEref{rrh}{a}]; {\sf\small (ii)} $\btau=\btrhp$ until
$\btau=\btau_{\rm f}\simeq50$, employing the estimates of
\Eref{rfrp} for $\brhof$ and $\vrhoR$. A less accurate result for
this case can be derived \cite{moroi, riotto} by equating the
annihilation rate $\Gamma_\chi=n_\chi\sigv$ to the expansion rate
$H$ at the completion of reheating. Combining \sEref{rrh}{a} and
\Eref{trh} we obtain
$\vH(\btrhp)=2\vrho_\phi(\btrhp)=2\sqrt{2}\vGamma_\phi/5$ and so,
we arrive at:
\beq \label{nstp} {n_\chi\over
s}\simeq{9\sqrt{2}\over\pi^2}{\Gamma_\phi\over\sigv\Trh^3}\eeq
In general, this result underestimates the numerical one by a
factor of unity. However, the method applied reveals the presence
of the phenomenon of reannihilation \cite{masiero} in this case,
i.e., the occurrence of a secondary (for $\btau\gg\btf$) $\chi$
decoupling -- see \Sref{wimp1}.

\end{list}

\item[{\boldmath $\blacklozenge$}] In the QKS, we can integrate
numerically \Eref{fn} from $\vtf$ down to 0, as follows:
\beq \label{BEsol} f_{\chi0} = \left(f_{\chi\rm F}^{-1}+J_{\rm F}
\right)^{-1},~~\mbox{where}~~J_{\rm F}= \int_{\vtauf_{{\rm
F}}}^{0} d\vtau\ \frac{\vsv}{\vH R^3}~~\mbox{and}~~f_{\chi\rm F}
=(\delta_{\rm F} +1)\> f_\chi^{\rm eq}(\vtf).\eeq
Although not crucial, a  choice $\delta_{\rm F}=1.2\mp0.2$ assists
us to approach better the precise numerical solution of \Eref{fn}.

\end{list}

\end{list}

\subsection{The Enhancement of $\Omx$}\label{sec:wimp}

As we explain in \Sref{sec:pamela} the interpretation of the
$e^\pm$-CR anomalies favors
$10^{-7}\lesssim\sigv/\GeV^{-2}\lesssim10^{-6}$ which results to
$0.0025\gtrsim\Omxsc\gtrsim0.00027$ for $0.1\leq\mx/\TeV\leq3$,
where $\Omxsc$ denotes $\Omx$ within the SC. Clearly, the
resulting $\Omxsc$ lie much lower than the range of \Eref{cdmb}.
However, the proposed non-standard scenaria can increase $\Omx$
w.r.t $\Omxsc$. The resulting enhancement can be quantified, by
defining the quantity:
\beq\label{dom}\Domx=
\left(\Omx-\Omxsc\right)\left/\right.\Omxsc.\eeq
We below analyze the behavior of $\Domx$ as a function of the free
parameters of each non-standard scenario separately.

\subsubsection{The LRS. \label{wimp1}} 
\begin{figure}[!t]\vspace*{-.25in}\begin{tabular}[!t]{cc}{\begin{minipage}[t]{8in}
\hspace*{-.25in}\epsfig{file=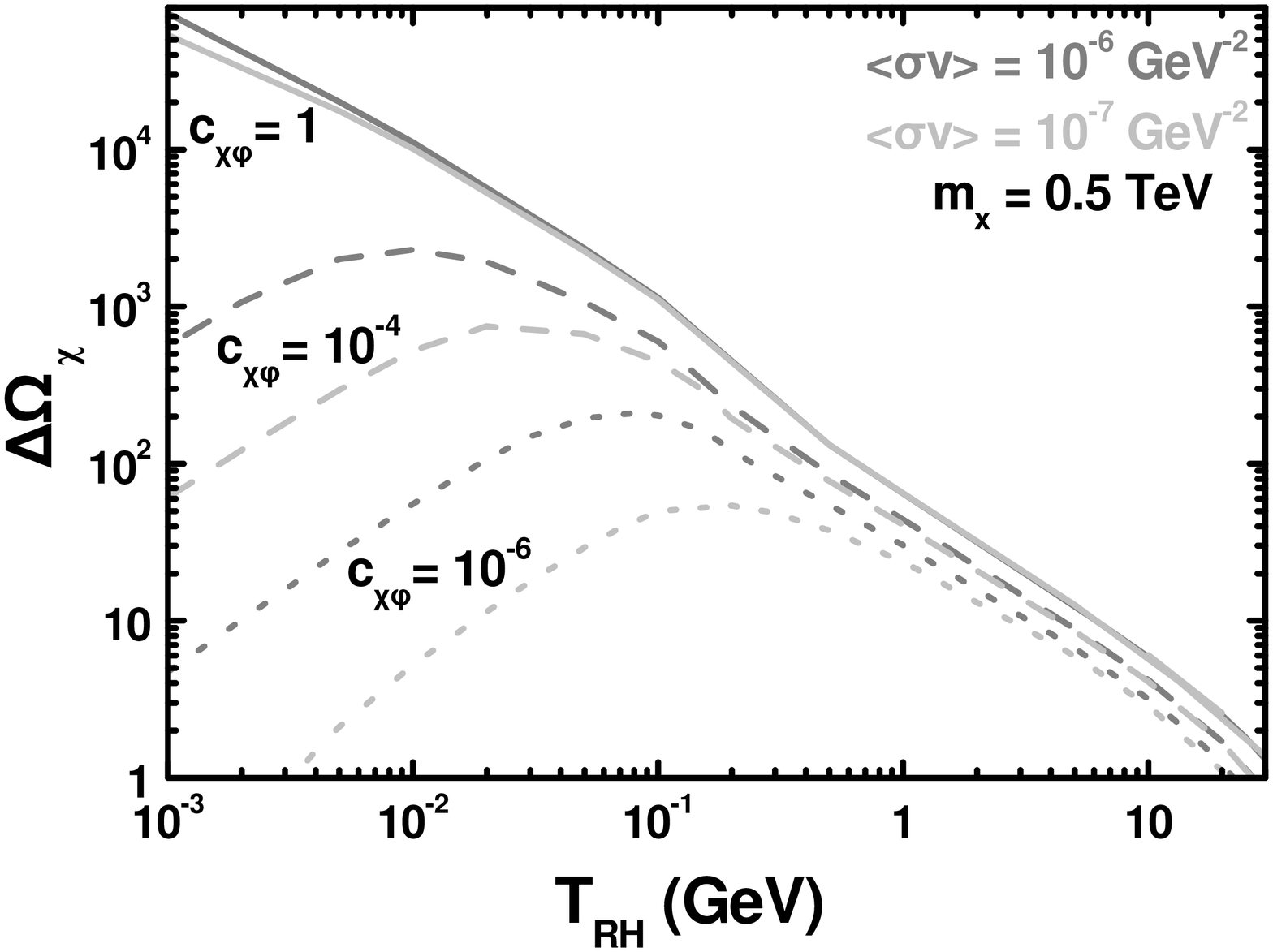,height=3.55in,angle=-90}\end{minipage}}
&\begin{minipage}[h]{2in}\hspace{-4.8in}{\vspace*{-2.15in}\begin{tabular}{|c|c|c|}
\hline
$\cxf$&$\Trh~(\GeV)$&$\chi$ {\sc Production}\\
\hline \hline
$10^{-6}$&$0.001-0.2$&{\sc non-TP}\\
&$0.2-30$&{\sc non-TP + TP}\\\hline
$10^{-4}$&$0.001-0.02$&{\sc non-TP}\\
&$0.02-30$&{\sc non-TP + TP}\\\hline
$1$&$0.001 - 30$&{\sc non-TP+TP}\\\hline
\end{tabular}} \hfill
\end{minipage}
\end{tabular} \hfill \vspace*{-.13in}\caption[]{\sl \ftn $\Domx$ versus $\Trh$ for
the LRS with $m_\chi=0.5~\TeV$, $\langle\sigma
v\rangle=10^{-6}~\GeV^{-2}~[\langle\sigma
v\rangle=10^{-7}~\GeV^{-2}]$ (gray [light gray] lines) and
$\cxf=1$ (solid lines), $\cxf=10^{-4}$ (dashed lines) and
$\cxf=10^{-6}$ (dotted lines). In the table we also show the type
of the $\chi$ production for each $\cxf$ and the various ranges of
$\Trh$.}\label{om}
\end{figure}

Let us initially clarify that in the LRS,
both signs of $\Domx$ are possible, as emphasized in \cref{rhodos,
gondolo}. However, we here confine ourselves to the combination of
parameters which assures the favored from the $e^\pm$-CR data
$\Domx>0$. The dependence of $\Domx$ on the free parameters of the
LRS can be inferred from \Fref{om}, where we depict $\Domx$ versus
$\Trh$ for $m_\chi=0.5~\TeV$,
$\sigv=10^{-6}~\GeV^{-2}~[\sigv=10^{-7}~\GeV^{-2}]$ (gray [light
gray] lines) and $\cxf=1$ (solid lines), $\cxf=10^{-4}$ (dashed
lines) and $\cxf=10^{-6}$ (dotted lines). The ranges of parameters
where each production mechanism is activated are also shown in the
table included. Note that the exposed ranges depend very weakly on
the employed $\mx$'s and $\sigv$'s. We observe that $\Domx$
increases with $\Trh$ when we have non-TP as expected from
\Eref{fxntp}, but it decreases as $\Trh$ increases when we have
equipartition between non-TP and TP, as anticipated in
\Eref{nstp}. The former mechanism is dominant mainly for very low
$\Trh$'s whereas the latter is present for higher $\Trh$'s. The
accuracy of the corresponding empirical expressions in
\Eref{fxntp} [\Eref{nstp}] increases as $\Trh$ decreases
[increases] and as $\sigv$ decreases [increases]. It is remarkable
that for $\cxf$'s where both production mechanisms are possible
(e.g., $\cxf=10^{-6}$ or $10^{-4}$) we can obtain the same $\Domx$
for two values of $\Trh$. In general, $\Domx$ increases with
$\Nx$. Augmentation of $\mx$ increases $\Omx$, too, but does not
alter the dependence of $\Omx$ on $\Trh$ and the ranges where the
$\chi$-production mechanisms are activated.

\renewcommand{\arraystretch}{1.1}
\renewcommand{\arraystretch}{1.0}

\begin{figure}[!t]\vspace*{-.1in}
\hspace*{-.25in}
\begin{minipage}{8in}
\epsfig{file=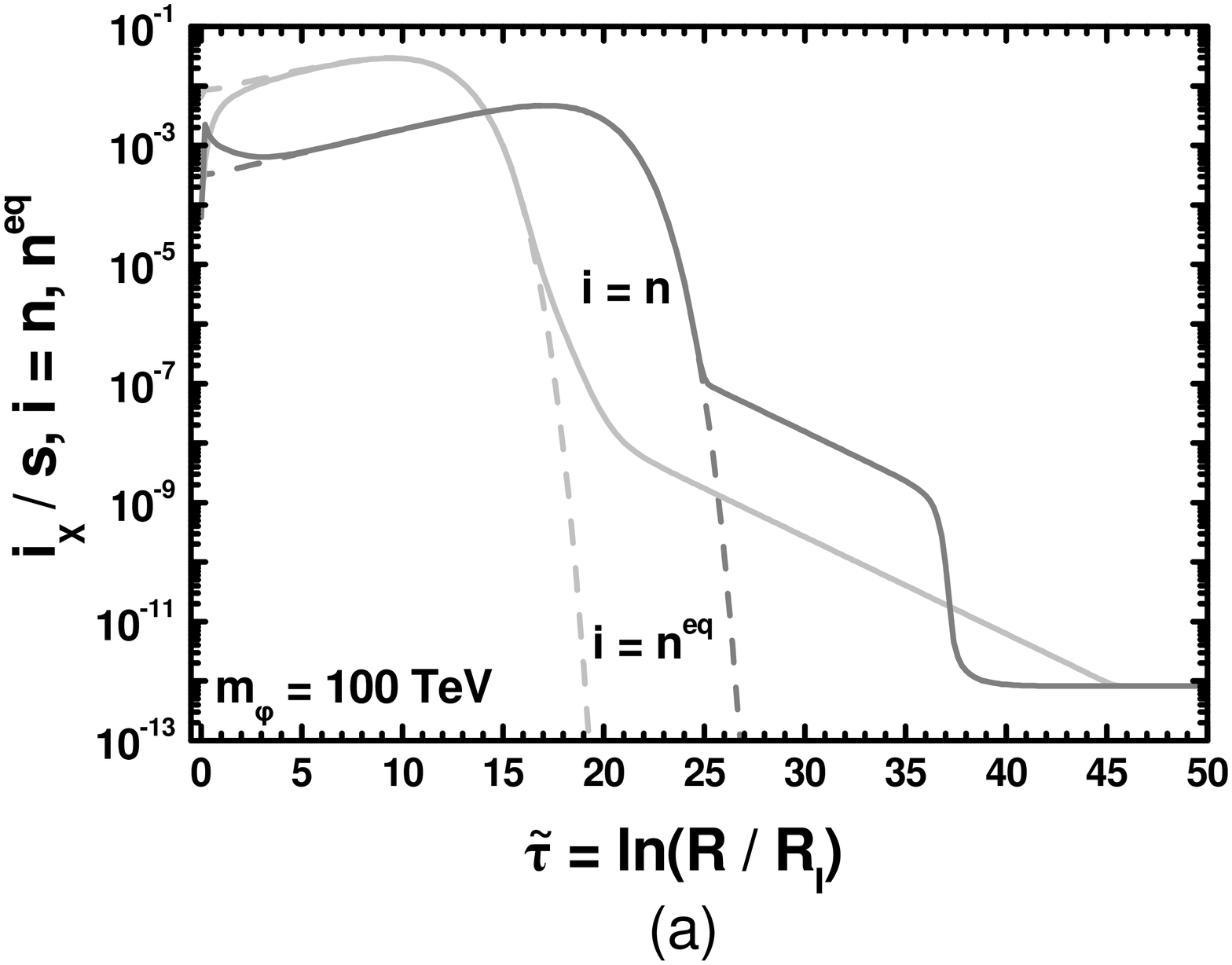,height=3.55in,angle=-90}
\hspace*{-1.37 cm}
\epsfig{file=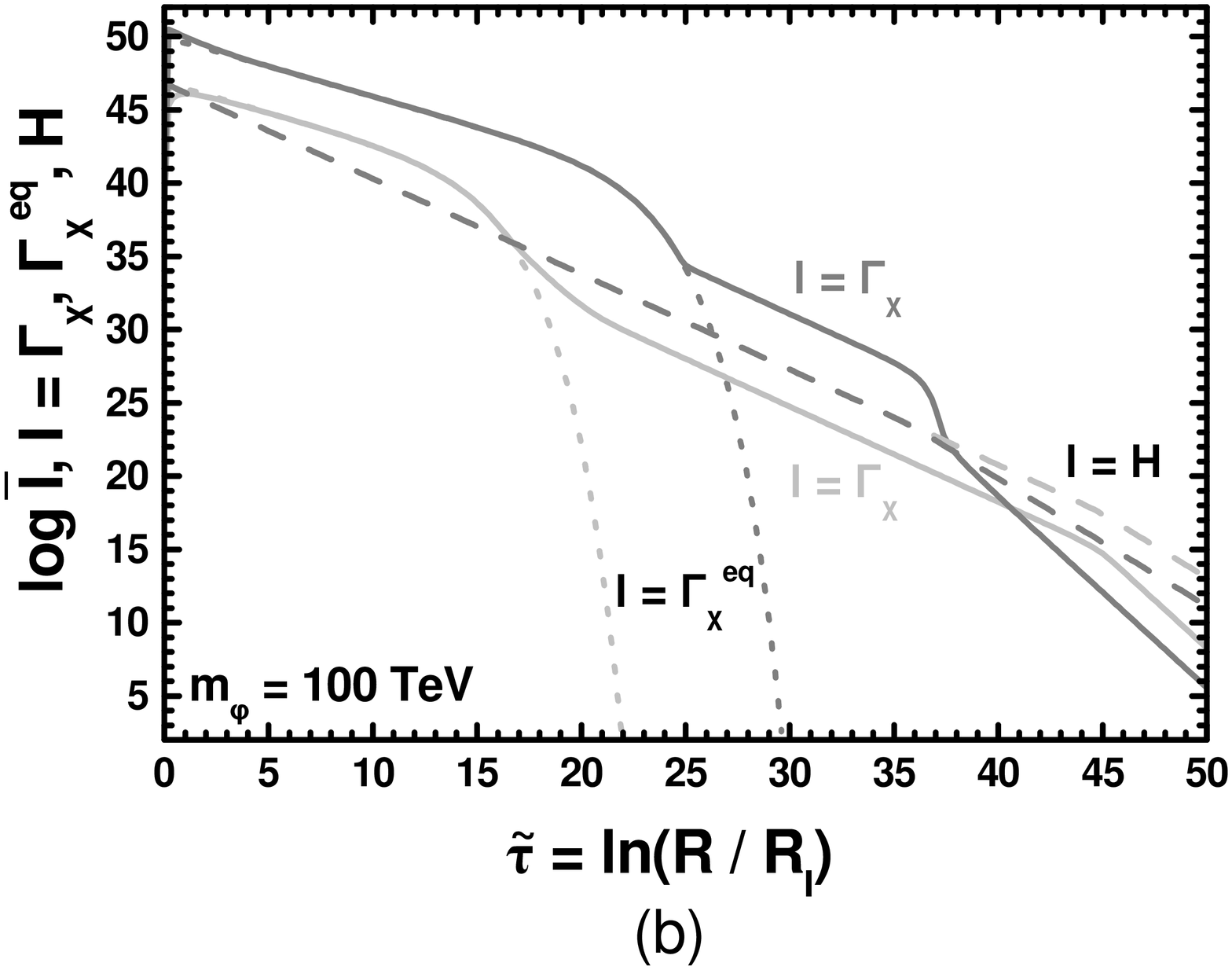,height=3.55in,angle=-90} \hfill
\end{minipage}\vspace*{-.1in}
\hfill\begin{center}
\epsfig{file=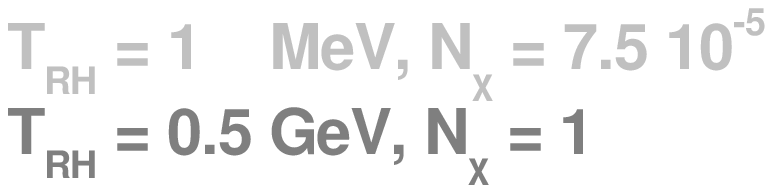,height=1.6in,angle=-90}
\end{center}\vspace*{-.1in}
\hfill \caption[]{\sl\ftn {\sf\ssz (a)} $n_\chi/s$ (solid lines)
and $n^{\rm eq}_\chi/s$ (dashed lines) versus $\btau$ and {\sf\ssz
(b)} $\vGm{\chi}$ (solid lines), $\vGm{\chi}^{\rm eq}$ (dotted
lines) and $\vH$ (dashed lines) versus $\btau$ for the LRS with
$m_\chi=0.5~\TeV$, $\mff=100~\TeV$, $\langle\sigma
v\rangle=3\cdot10^{-7}~\GeV^{-2}$ and $(N_\chi,
\Trh)=(1,0.5~\GeV)$ [$(N_\chi, \Trh)=(7.5\cdot10^{-5}, 1~\MeV)$]
(gray [light gray] lines) -- in both cases, we get
$\Omx=0.11$.}\label{om1}
\end{figure}

The operation of the two types of $\chi$ production encountered in
\Fref{om} is visualized in \ssFref{om1}{a}{b}. In these, we
display the actual $\chi$ yield, $n_\chi/s$ (solid lines) and its
equilibrium value, $n^{\rm eq}_\chi/s$ (dashed lines) the
dimensionless actual interaction rate of $\chi$,
$\vGm{\chi}=\vn_\chi\vsv$ (solid lines), its equilibrium value,
$\vGm{\chi}^{\rm eq}=\vn^{\rm eq}_\chi\vsv$ (dotted lines) and the
dimensionless expansion rate $\vH$ (dashed lines) -- given by
\sEref{H2exp}{a} -- versus $\btau$. In both figures we use
$m_\chi=0.5~\TeV$, $\mff=100~\TeV$, $\langle\sigma
v\rangle=3\cdot10^{-7}~\GeV^{-2}$ and $(N_\chi,
\Trh)=(1,0.5~\GeV)$ [$(N_\chi, \Trh)=(7.5\cdot10^{-5}, 1~\MeV)$]
(gray [light gray] lines). For the selected parameters, the
evolution of the background energy densities ($\log\brhof$ and
$\log\brhoR$) and $T$ is presented in \ssFref{figrh}{a}{b}. The
completion of reheating occurs at $\btrhp\simeq36.1$
[$\btrhp=45.1$] for $\Trh=0.5~\GeV$ [$\Trh=1~\MeV$].

From \sFref{om1}{a} we can deduce that $n_\chi/s$ takes its
present value close to [clearly above] $\btrhp$ for $\Trh=1~\MeV$
[$\Trh=0.5~\GeV$]. For this reason, the integration of \Eref{fn}
until $\btrhp$ for non-TP is sufficient for an accurate result --
see \Eref{fxntp} --, but insufficient when non-TP and TP
interplay. The $\chi$ reannihilation takes place along the almost
vertical part of the gray line for $\btau$ a little lower than
$\btrhp$. It is notable that in \cref{rhodos, gondolo} which focus
on lower $\sigv$'s than the ones considered here, the phenomenon
of reannihilation is not stressed.

This effect is further analyzed, following the approach of the
first paper in \cref{masiero}, in \sFref{om1}{b}. From this we
infer that for $\Trh=1~\MeV$, where non-TP outstrips, $\chi$
decouples from plasma only once at $\btf=16.5$ where
$\Gm{\chi}=\Gm{\chi}^{\rm eq}=H$ and an intersection of the light
gray lines is observed (note that the light gray and gray dashed
lines coincide at that region). On the contrary, for
$\Trh=0.5~\GeV$, where non-TP and TP coexist, we observe that
$\chi$ decouples from plasma initially at about $\btf=26.3$ where
$\Gm{\chi}^{\rm eq}=H$ but also at $\btau_{\rm Fr}=37.9>\btrhp$
where $\Gm{\chi}=H$. In other words, we observe two intersections
between the two pairs of the three gray lines. This effect
signalizes the existence of a period of $\chi$ reannihilation
similar to that noticed in \cref{masiero}. Contrary to that
situation, in our case {\sf\small (i)} $\Gm{\chi}$ remains larger
than $H$ after the first $\chi$ decoupling and drops sharply below
$H$ after reannihilation, and {\sf\small (ii)} $H$ smoothly
evolves from its behavior during LRS to that within RD era.

\vspace*{-0.2cm}\subsubsection{The QKS. \label{wimp2}} 
\begin{figure}[!t]{\vspace*{-.25in}\begin{tabular}[!t]{cc}{\begin{minipage}[t]{8in}
\hspace*{-.25in}\epsfig{file=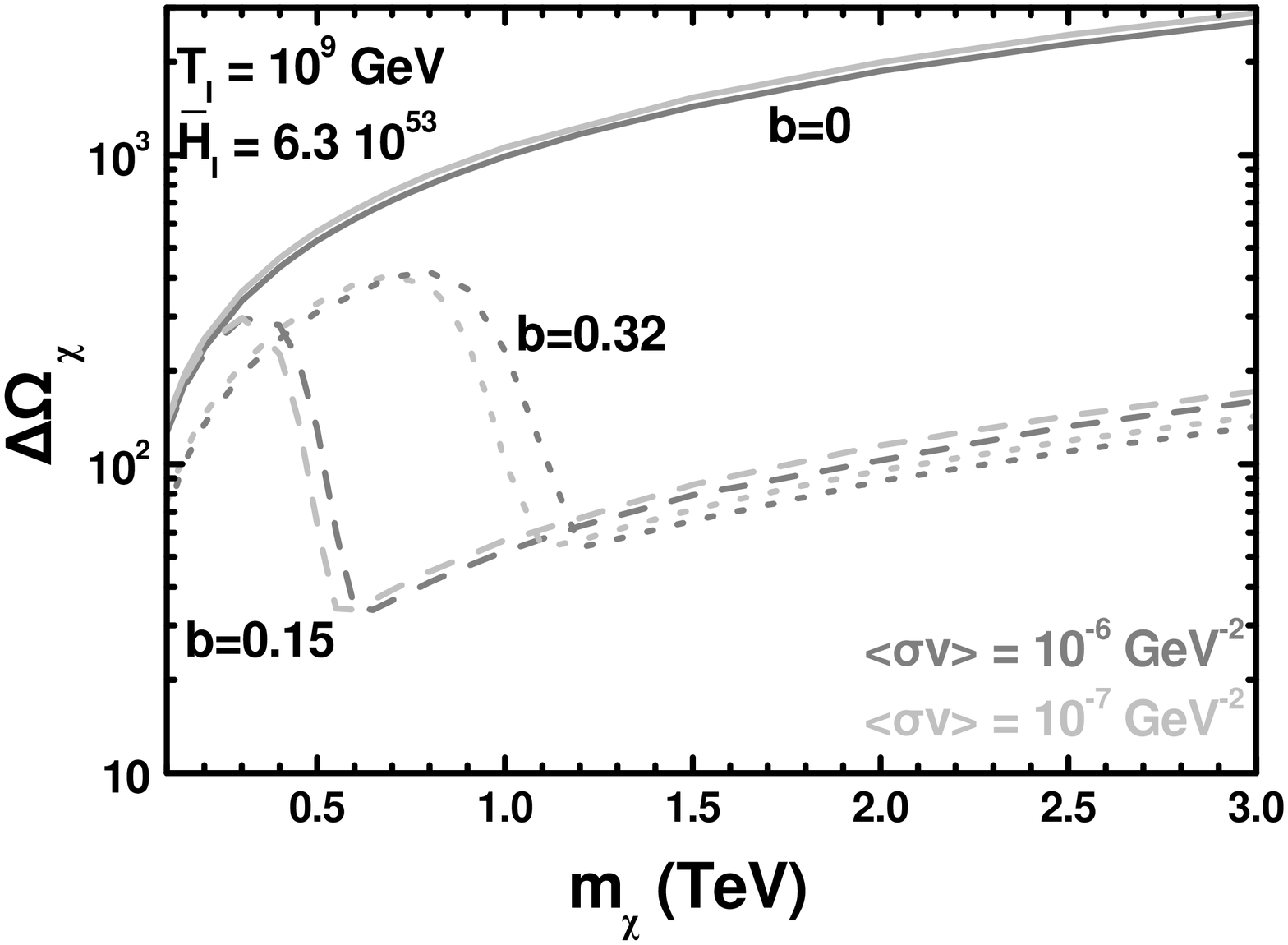,height=3.55in,angle=-90}\end{minipage}}
&\begin{minipage}[h]{2in}\hspace{-4.8in}{\vspace*{-2.15in}\begin{tabular}{|c|c|c|}
\hline
$\sigv$&$10^{-7}$&$10^{-6}$\\
$\left(\GeV^{-2}\right)$&&\\\hline \hline
$-\vtf$&$31.7-35.2$&$31.6-35.1$\\\hline\hline
&\multicolumn{2}{|c|}{$b=0.15$, $\vtp\simeq-33.1$}\\\cline{2-3}
$\vtau_{\rm F}^{\rm min}$&$-33.3$&$-33.3$\\\hline\hline
&\multicolumn{2}{|c|}{$b=0.32$, $\vtp\simeq-33.8$}\\\cline{2-3}
$\vtau_{\rm F}^{\rm min}$&$-34.1$&$-34.0$\\\hline
\end{tabular}} \hfill
\end{minipage}
\end{tabular}} \hfill \vspace*{-.13in}\caption[]{\sl \ftn
$\Domx$ versus $m_\chi$ for the QKS with $a=0.5$,
$\vHi=6.3\cdot10^{53}$, $\vTi=10^9~\GeV$, $\langle\sigma
v\rangle=10^{-6}~\GeV^{-2}~[\langle\sigma
v\rangle=10^{-7}~\GeV^{-2}]$ (gray [light gray] lines) and $b=0$
(solid lines), $b=0.15$ (dashed lines) and $b=0.32$ (dotted
lines). In the table we also show the range of $\vtf$ for
$0.1\leq\mx/\TeV\leq3$, the freeze-out logarithmic time
$\vtau_{\rm F}^{\rm min}$ at which the minima of $\Domx$ occur and
the closest to $\vtau_{\rm F}^{\rm min}$'s $\vtp$'s for the
selected $b$'s.}\label{om2}
\end{figure}

The
presence of $g_{\rm C}>1$ in \Eref{xf} and, mainly, in
\Eref{BEsol} reduces $J_{\rm F}$ w.r.t its value in the SC
generating, thereby, $\Domx>0$ within the QKS. The mechanism of
the $\chi$ decoupling in this case, for both $b=0$ and $b\neq0$,
is pretty known -- see \cref{Kam, salati, jcapa}. However, a
peculiar effect emerges in the dependence of $\Domx$ on $\mx$ for
$b\neq0$ which can be inferred from \Fref{om2}, where we display
$\Domx$ versus $\mx$ for $a=0.5$, $\vHi=6.3\cdot10^{53}$,
$\langle\sigma v\rangle=10^{-6}~\GeV^{-2}~[\langle\sigma
v\rangle=10^{-7}~\GeV^{-2}]$ (gray [light gray] lines) and $b=0$
(solid lines), $b=0.15$ (dashed lines) and $b=0.32$ (dotted
lines). The chosen $\vHi$'s result to $\Omqns\simeq0.01,~0.068$ or
$0.19$ for $b=0, 0.15$ or $0.32$ correspondingly.

Obviously, for $b=0$ we get a pure KD era and our results reduce
to those presented in \cref{jcapa}, i.e., $\Domx$ increases when
$\mx$ increases or $\sigv$ decreases. On the contrary, for
$b\neq0$, $\Domx$ depends crucially on the hierarchy between
$\vtf$ and $\vtp$ found from \eqs{xf}{tmax} respectively. Given
that $J_{\rm F}$ takes its main contribution from $g_{\rm C}$ for
$\vtau\sim\vtf$, $J_{\rm F}$ is enhanced -- see \Eref{BEsol} -- if
$\vtf$ is lower than $\vtp$ and close to it, since $g_{\rm C}$ is
suppressed ($g_{\rm C}\simeq1$) for $\vtau\simeq\vtp$. As a
consequence -- see \eqs{omh}{BEsol} -- $\Domx$ diminishes. This
argument is highlighted in the table of \Fref{om2}. There, we list
the range of $\vtf$ for $0.1\leq\mx/\TeV\leq3$ and
$\sigv=10^{-7}~\GeV^{-2}$ or $\sigv=10^{-6}~\GeV^{-2}$ and the
logarithmic time $\vtp$ at which the closest to $\vtf$'s peak in
the $q$ evolution takes place for $b=0.15$ or $b=0.32$ and
$\vHi=6.3\cdot10^{53}$. Clearly $\vtf$ [$\vtp$] is independent of
$b$ and $\vHi$ [$\mx$ and $\sigv$]. As $\mx$ increases above
$0.1~\TeV$, $\vtf$ moves closer to $\vtp$ and $\Domx$ decreases
with its minimum $\left.\Domx\right|_{\rm min}$ occurring at
$\vtau_{\rm F}^{\rm min}\simeq\vtp$. The small deviation of
$\vtau_{\rm F}^{\rm min}$ from $\vtp$ can be attributed to the
presence of $f_{\chi\rm F}$ in \Eref{BEsol}. The appearance of the
minima can be avoided if $\vtf$'s happen to remain constantly
lower than $\vtp$'s -- see, e.g., \sFref{svmx}{c$_2$}. Variation
of $\Ti$ or $\vHi$  leads to a displacement of $\vtp$'s -- see
\Eref{tmax} -- relocating, thereby, the minima of $\Domx$ in
\Fref{om2}. However, our conclusions on the behavior of $\Domx$
remain intact.

\renewcommand{\arraystretch}{1.1}
\renewcommand{\arraystretch}{1.0}

\section{PAMELA, ATIC and Fermi-LAT Anomalies}\label{sec:pamela}

The aforementioned $\Domx$ obtained within the LRS or QKS assists
us to explain the experimental data on the $e^\pm$-CRs
consistently with \Eref{cdmb}. Indeed, the observed anomalies on
the CR $e^\pm$ fluxes can be attributed to the annihilation of
$\chi$'s in the galaxy provided that $\mx$ and $\sigv$ are chosen
appropriately. In \Sref{sec:form} we outline the basic formalism
that we employ in order to estimate the observable quantities as a
function of these parameters and in \Sref{sec:fits} we display our
fittings.

\subsection{Cosmic Rays from Annihilation of WIMPs}\label{sec:form}

After being produced in the Milky Way halo, charged CRs propagate
in the galaxy and its vicinity in a rather complicated way before
reaching the earth. Their propagation is commonly evaluated by
solving a diffusion equation \cite{edjo, matsumoto} with static
cylindrical boundary conditions. The solution can be casted into
the following semi-analytical form \cite{edjo, matsumoto, strumia}
which yields the $\ps$ flux per energy -- in units ${\rm GeV^{-1}
cm^{-2} s^{-1} sr^{-1}}$ -- at earth from the $\chi$ annihilation:
\beq \Phi^{\nt\nt}_{e^+} (E) = {1\over2}\frac{v_{e^+}}{4\pi b(E)}
\, \left( \frac{\rho_{\odot} }{m_{\nt}} \right )^2 \, \langle
\sigma v \rangle \; \int_E^{m_{\nt} }\; d E' \,
I\left(\lambdaup_{D}(E,E') \right)\, \frac{d N_{e^+}}{ d E'_{e^+}
} \;, \label{semi} \eeq
where $v_{e^+}$ is the velocity of $\ps$ which is practically
equal to the one of the light, the pre-factor of 1/2 arises from
our assumption that $\chi$ is a Majorana particle,
$\rho_{\odot}=0.3~\GeV/{\rm cm}^3$ is the local CDM density,
$b(E)=E^2/(\GeV\,t_E)$ with $t_E = 10^{16}~{\rm s}$ is the energy
loss rate function and $dN_{e^+}/dE_{e^+}$ denotes the energy
distribution of $\ps$'s per $\chi$ annihilation. Motivated by the
highly restrictive PAMELA data \cite{pamelap} on the anti-proton
mode of the CDM annihilation, we consider a purely leptophilic
$\chi$ (however, see also \cref{winos}). In particular, we
incorporate in our investigation the following annihilation modes:
$\chi\chi\to e^-e^+$, $\chi\chi\to \mu^-\mu^+$ or
$\chi\chi\rightarrow \tau^-\tau^+$. In the first case,
$dN_{e^+}/dE_{e^+}$ is given by a simple analytic expression
\cite{peskin} whereas in both latter cases, we use analytic
parametrizations (presented in \cref{chi2,carena}) of
$dN_{e^+}/dE_{e^+}$ which reproduce quite accurately the numerical
outputs of the package {\sc Pythia} \cite{pythia}. In all cases
the effect of final state radiation \cite{fsr} is taken into
account. Namely, we take
\beq {dN_{e^+}\over dE_{e^+}} =\left\{\matrix{
{\delta(E_{e^+}-\mx)+\left(\alpha_{\rm
em}/2\pi\right)}\left[3\delta(E_{e^+}-\mx)/2+\right. \hfill &
\hfill \cr
\left.\left(1+y^2\right)\ln{\left(4\mx^2/m_e^2\right)/\left(1-y\right)\mx}
\right]~~~\hfill \mbox{for} & \chi\chi\to \ps\el, \hfill \cr
({\alpha_{\rm em}/\pi}) A\exp[-(A_1y + A_2y^2)] + B_1 + B_2y
~~~~\hfill \mbox{for} & \chi\chi\to\mu^+\mu^-, \hfill \cr
({1/\mx}) \left(\exp[A^-_0+A^-_1y +A^-_2y^2+A^-_3y^3]+\right.
\hfill & \hfill \cr
\left.\exp[A^+_0+A^+_1y
+A^+_2y^2+A^+_3y^3+A^+_4y^4+A^+_5y^5]\right) ~~~\hfill \mbox{for}
& \chi\chi\to \tau^+\tau^-, \hfill \cr
}
\right. \label{dNE}\eeq
where $\alpha_{\rm em}$ is the fine-structure constant computed at
a scale equal to $2\mx$, $m_e=0.511~{\rm MeV}$ is the $\el$ mass
and $0<y=E_{e^+}/\mx\leq1$. The infrared singularity encountered
for $\chi\chi\to \ps\el$ and $y=1$ is handled as described in
\cref{peskin}. For the $\chi\chi\to \mu^+\mu^-$ mode, we take
$J=\tilde J\,(\mx/ 0.5~\TeV)^{\delta_J}$ with
$J=A,~A_1,~A_2,~B_1,~B_2$ and
\bea && (\tilde A, \tilde A_1, \tilde A_2, \tilde B_1, \tilde B_2
)= (-0.296635, 2.65121,
14.8445, 0.0042505, -0.00427157),\nonumber\\
&& (\delta_{A} , \delta_{A_1} , \delta_{A_2}, \delta_{B_1},
\delta_{B_2})=(-1.01424, 0.017198, -0.0107585, -0.999819,
-0.999819).\nonumber\eea
We checked that the parametrization above gives results quite
similar to those obtained using the simpler parametrization
suggested in \cref{carena}. For the $\chi\chi\to \tau^+\tau^-$
mode, we take \cite{carena}
\bea && (A^-_0, A^-_1, A^-_2, A^-_3)= (0.951, -18.083, 15.79, -15.575),\nonumber\\
&& (A^+_0, A^+_1, A^+_2, A^+_3, A^+_4, A^+_5)=(2.783, -22.942,
82.595, -193.748, 223.389, -97.716).\nonumber\eea

Also, $I(\lambda_D)$ is the dimensionless halo function which
fully encodes the galactic astrophysics with $\lambda_D(E,E')$ the
diffusion length from energy $E$ to energy $E'$ which is given by
\begin{equation} \label{lD}
 \lambdaup^2_D = 4 K_0 t_E \left[ \frac{ (E'/{\rm GeV} )^{\delta - 1}
 -(E/{\rm GeV} )^{\delta - 1} }{\delta-1}\right ] \;\cdot
\end{equation}
To compute $I(\lambdaup_D)$ we employ the semi-empirical function
proposed in \cref{strumia, korea}. Namely,
\begin{equation}\label{hf}
  I(\lambdaup_D) = a_0 + a_1 \tanh \left( \frac{ b_1 - l}{c_1} \right )
\, \left[ a_2 \exp\left(- \frac{(l - b_2)^2}{c_2}\right) + a_3
\right ]~~\mbox{with}~~l = \log\left( {\lambdaup_D\over{\rm kpc}}
\right)\cdot
\end{equation}
The involved in \eqs{lD}{hf} constants \cite{strumia} depend on
the CDM distribution and the propagation model that we consider.
As we emphasize in \Sref{grays}, the constraint from the $\gamma$
CRs enforces us to adopt the isothermal halo profile \cite{isoT}
which weakens the relative restrictions. Note, however, that our
results on $\Phi^{\nt\nt}_{e^+}$ are quite close to those that we
would had obtained if we had used the NFW halo \cite{NFW} profile
-- c.f. \cref{gstrumia, strumia}. We also use the MED propagation
model for $\chi\chi\to e^-e^+$ and $\chi\chi\to \mu^-\mu^+$ and
the MIN (M2) model for $\chi\chi\rightarrow \tau^-\tau^+$. These
choices provide the bets fits to the combined experimental data -
c.f. \cref{korea, moroiNS}. Note that only these two propagation
models are consistent \cite{matsumoto} with the observed
boron-to-carbon ratio in the CR flux. Therefore, we use
\cite{strumia}
\bea \nonumber &&(a_0, a_1, a_2, a_3, b_1, b_2, c_1, c_2, K_0,
\delta)=\\ &&\left\{\matrix{
(0.495,0.629,0.137,0.784,0.766,0.55,0.193,0.296, 0.0112~{\rm
kpc^2/My}, 0.7) ~~~~\hfill \mbox{(MED)}, \cr
(0.5,0.903,-0.449,0.557,0.096,192.8,0.210,33.91,0.00595~{\rm
kpc^2/My}, 0.55)~~~\hfill \mbox{(MIN)}.  \cr
}
\right. ~~~~~~~~~~~~\eea
We explicitly verified that the numerically fitted function in
\Eref{hf} reproduces quite accurately and fast enough the results
obtained by performing numerically the relevant integrations
presented in the earlier formulae of \cref{matsumoto}. Moreover,
the formalism of \cref{strumia} overcomes successfully the
mismatching problem (in the numerical integration) which is
mentioned in \cref{matcing}.

In order to calculate the total fluxes, we also have to estimate
the background $e^\pm$ fluxes. In our study, we take into account
the fluxes of {\sf\small (i)} secondary $\ps$
$\left(\Phi_{e^+}^{\rm sec}\right)$ produced by collisions between
primary protons and interstellar medium in our galaxy {\sf\small
(ii)} primary $\el$ $\left(\Phi_{e^-}^{\rm prim}\right)$
presumably produced in supernova remnants and {\sf\small  (iii)}
secondary $\el$ $\left(\Phi_{e^-}^{\rm sec}\right)$ produced by
spallation of CRs in the interstellar medium. These fluxes are
commonly parameterized as \cite{edjo}
\numparts
\begin{eqnarray}
\Phi_{e^+}^{\rm sec} &=& \frac{ 4.5\, (E/\GeV)^{0.7} }{ 1 +
650\, (E/\GeV)^{2.3} + 1500\,  (E/\GeV)^{4.2} },\\
\Phi_{e^-}^{\rm prim} &=& \frac{ 0.16\,  (E/\GeV)^{-1.1} }{
1 + 11\, (E/\GeV)^{0.9} + 3.2 (E/\GeV)^{2.15} } ,\\
\Phi_{e^-}^{\rm sec}&=&\frac{ 0.7\, (E/\GeV)^{0.7} }{ 1 + 110\,
(E/\GeV)^{1.5} + 600\, (E/\GeV)^{2.9}+ 580\, (E/\GeV)^{4.2} },
\end{eqnarray}\endnumparts
\hspace{-.14cm}in units ${\rm GeV^{-1} cm^{-2} s^{-1} sr^{-1}}$.
With these backgrounds, the total $e^\pm$ fluxes read
\beq \label{bfluxes}\Phi_{e^+}=\Phi_{e^+}^{\nt\nt}+\Phi_{e^+}^{\rm
sec}
~~\mbox{and}~~\Phi_{e^-}=\Phi_{e^-}^{\nt\nt}+c_{\el}\,\Phi_{e^-}^{\rm
prim}+\Phi_{e^-}^{\rm sec} \eeq
where $\Phi^{\nt\nt}_{e^-}=\Phi^{\nt\nt}_{e^+}$ and
$c_{\el}\simeq(0.6-0.8)$ is a normalization factor. We take
$c_{\el}=0.6~[c_{\el}=0.7]$ so that our flux calculation is
consistent with the ATIC [Fermi-LAT] data in the low energy range
of $(20-70)~\GeV$ \cite{korea}.

\subsection{Fitting the PAMELA and ATIC or Fermi-LAT Data}\label{sec:fits}

Using the fluxes defined above, we can evaluate the observable
quantities and compare them with the experimental outputs. In
order to qualify our fittings to the experimental data, we perform
a $\chiup^2$ analysis. In particular, we define the $\chiup^2$
variables as \cite{chi2, korea, matcing}
\beq\label{x2} \chiup^2_A =\sum_{i=1}^{N_A} {\left(F_{Ai}^{\rm
obs} - F_{Ai}^{\rm th}\right)^2 \over \left(\Delta F_{Ai}^{\rm
obs}\right)^2 },~~\mbox{with}~~F_A=\left\{\matrix{
\Phi_{e^+}/\left(\Phi_{e^+}+\Phi_{e^-}\right)\hfill
&\mbox{and}~~N_A=7~~ \mbox{for}~~A=1, \hfill \cr
E^3_{e^+}\left(\Phi_{e^+}+\Phi_{e^-}\right)~\hfill
&\mbox{and}~~N_A=\left\{\matrix{
21\hfill &\mbox{for}  & A=2 , \hfill \cr
26 \hfill &\mbox{for} &A=3 , \hfill \cr}
\right. \hfill \cr}
\right. \eeq
where $A=1,2,3$ stands for the PAMELA \cite{pamela}, ATIC
\cite{atic} or Fermi-LAT \cite{Fermi} data respectively which are
considered as independent sets. The index $i$ runs over the data
points of each experiment $A$, the superscript ``obs'' [``th'']
refers to measured [theoretically predicted] quantities whereas
$\Delta F^{\rm obs}$ means error in the experimentally observable
$F$. $N_A$ is the number of data points considered from the
experiment $A$. Note that, from the PAMELA data-set, we use
\cite{korea, chi2} only the 7 data points above $9.1~\GeV$ where
the effect of solar modulation is expected to be small. In our
analysis we take into account only the vertical errors. We also
conservatively combine, independently for each data-point, in
quadrature statistical and systematic errors released from Fermi
LAT \cite{Fermi}.

\begin{figure}[!t]\vspace*{-.09in}
\hspace*{-.25in}
\begin{minipage}{8in}
\epsfig{file=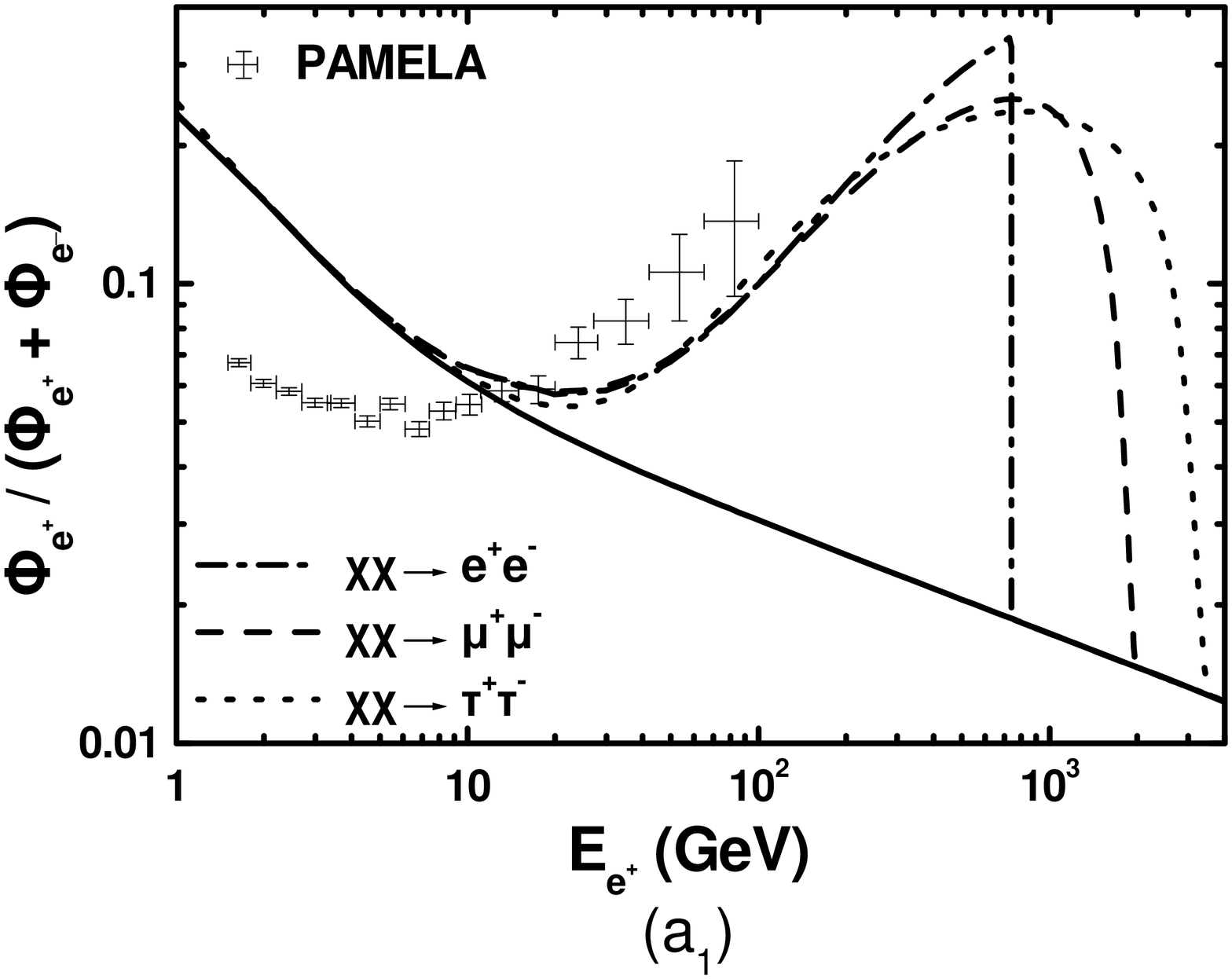,height=3.55in,angle=-90}
\hspace*{-1.37 cm}
\epsfig{file=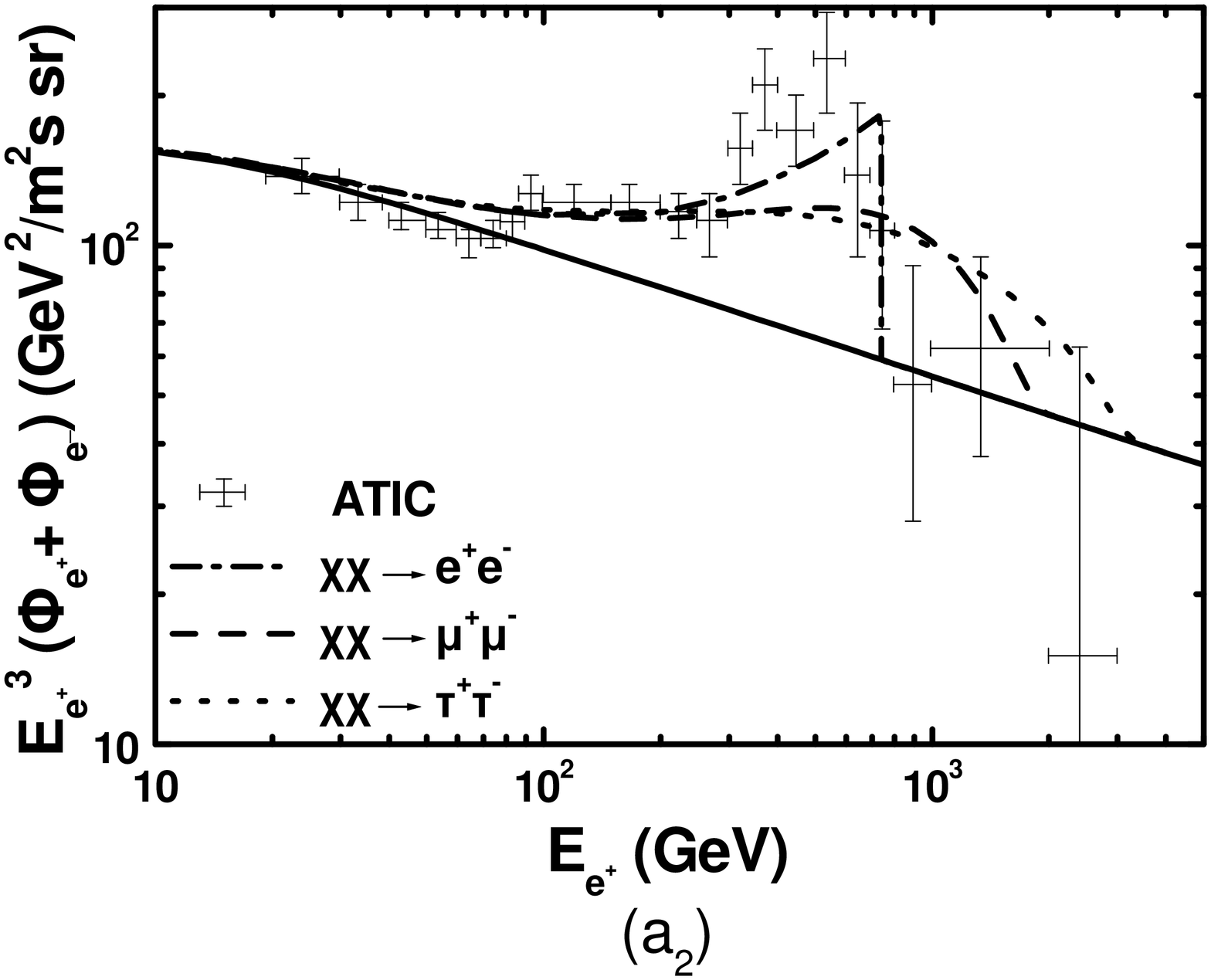,height=3.55in,angle=-90} \hfill
\end{minipage}\vspace*{-.01in}
\hfill \hspace*{-.25in}
\begin{minipage}{8in}
\epsfig{file=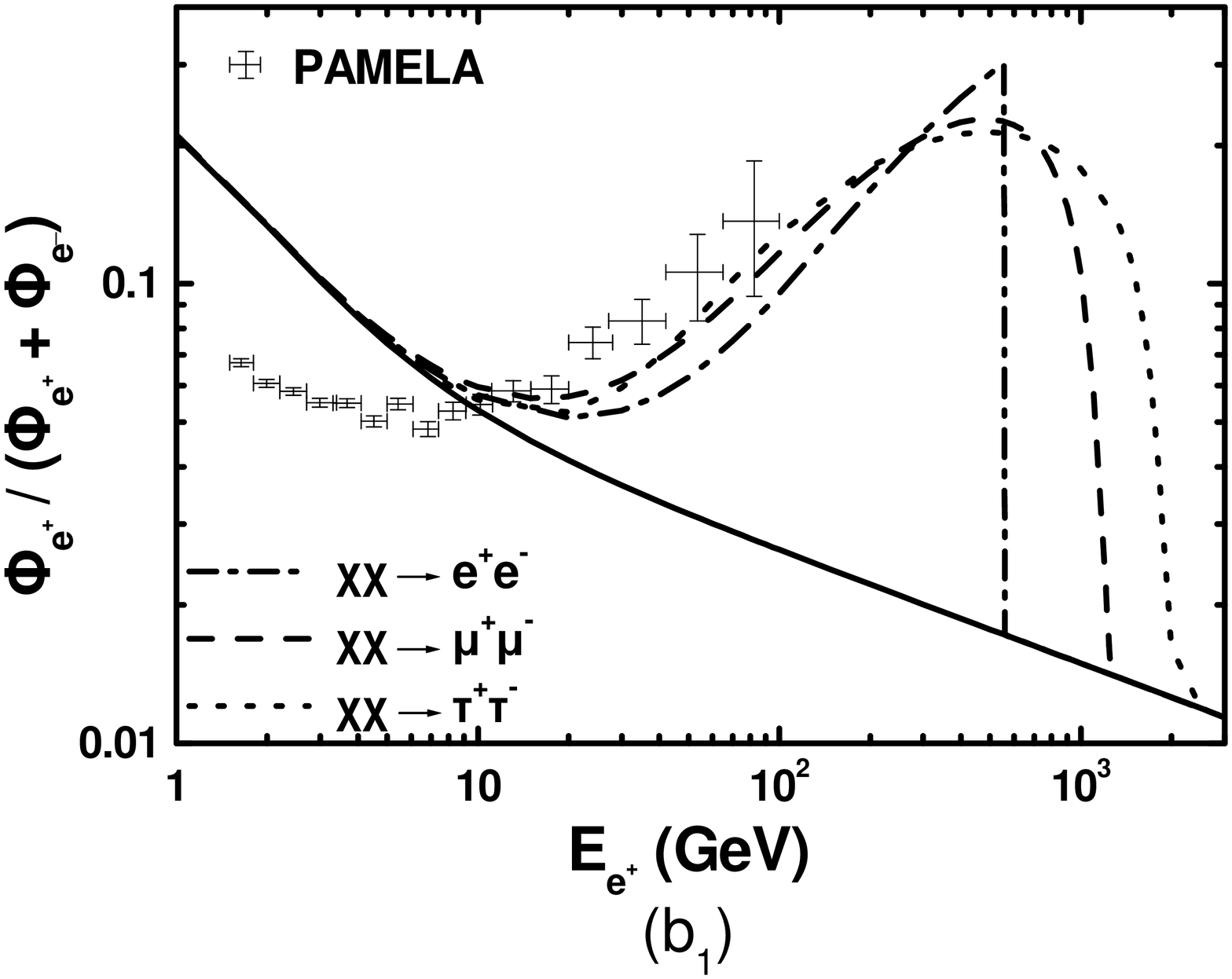,height=3.55in,angle=-90}
\hspace*{-1.37 cm}
\epsfig{file=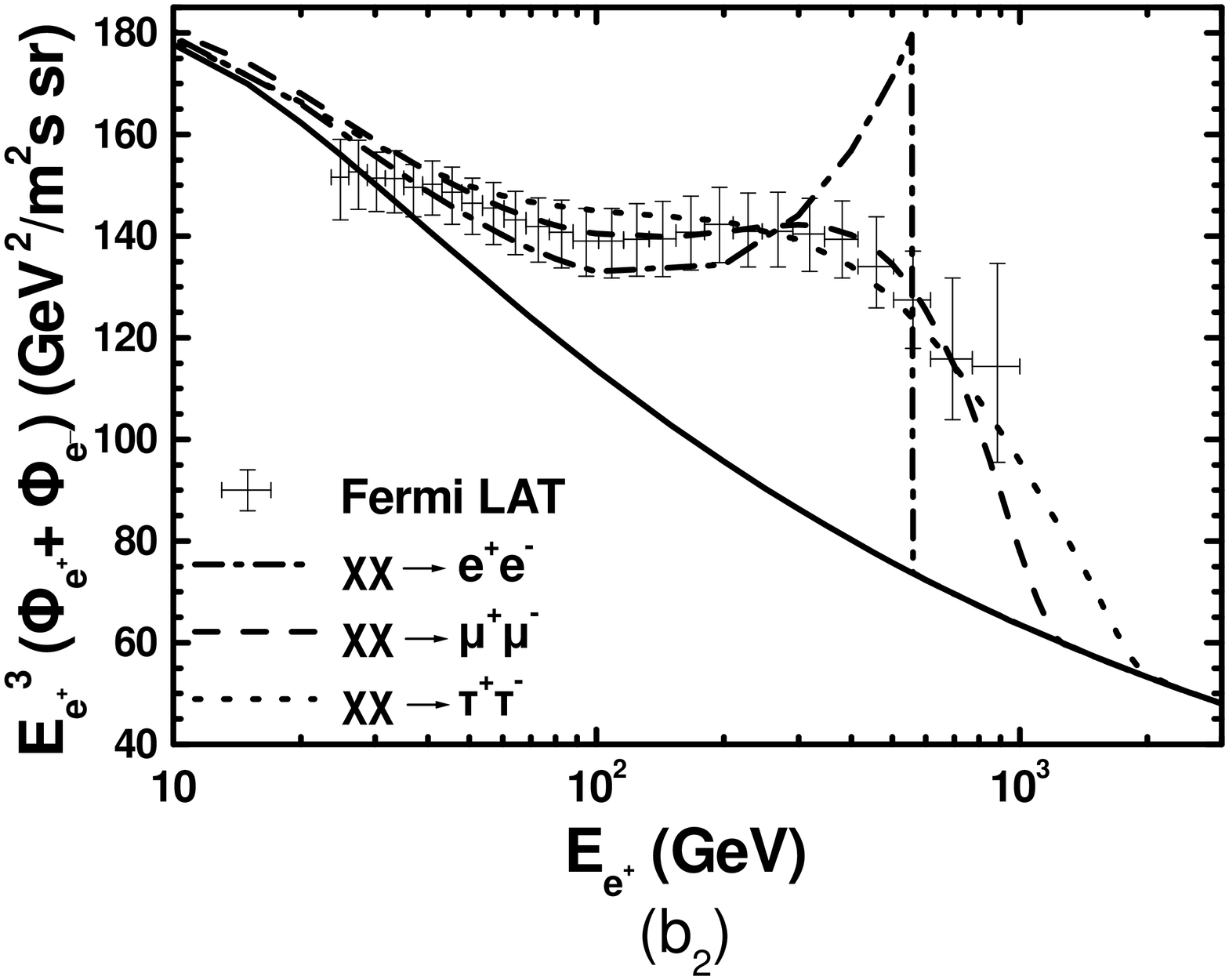,height=3.55in,angle=-90} \hfill
\end{minipage}
\hfill \caption[]{\sl\ftn The $e^+$-flux fraction ({\sf\ssz a$_1$}
and {\sf\ssz  b$_1$}) and the total $e^-$ and $e^+$ flux times
$E^3_{e^+}$ ({\sf\ssz a$_2$} and {\sf\ssz  b$_2$}) as a function
of $E_{e^+}$, with $E_{e^+}$ being the ${e^+}$ energy. We use the
best-fit points ($m_{\nt},~\sigv$), indicated in Table~\ref{Tfit},
for the PAMELA and ATIC data ({\sf\ssz  a$_1$} and {\sf\ssz
a$_2$}) or the PAMELA and Fermi-LAT data ({\sf\ssz  b$_1$} and
{\sf\ssz b$_2$}), assuming $\nt$ annihilating into $e^+e^-$
(dot-dashed lines), $\mu^+\mu^-$ (dashed lines) or $\tau^+\tau^-$
(dotted lines). The background fluxes are denoted by solid lines
and are computed for $c_{\el}=0.6$ ({\sf\ssz  a$_1$} and {\sf\ssz
a$_2$}) or $c_{\el}=0.7$ ({\sf\ssz b$_1$} and {\sf\ssz b$_2$}).
The data from PAMELA {\sf\ssz (a$_1$, b$_1$)}, ATIC {\sf\ssz
(a$_2$)} and Fermi-LAT {\sf\ssz  (b$_2$)} experiments are also
shown (an additional uncertainty from the Fermi-LAT energy scale,
which can shift all the points by $5\%$ (up) to $10\%$ (down) is
not shown). Recall that we adopt the isothermal halo profile and
the MED [MIN] propagation model for $\chi\chi\to \ps\el$ and
$\chi\chi\to \mu^+\mu^-$ [$\chi\chi\to \tau^+\tau^-$].}
\label{fit}
\end{figure}

In \Fref{fit} we show the predicted observable quantities compared
to the experimental data as a function of the $\ps$ energy
$E_{\ps}$, assuming $\chi$ annihilating to $\ps\el$ (dot-dashed
lines), $\mu^+\mu^-$ (dashed lines) or $\tau^+\tau^-$ (dotted
lines). We use the best-fit $(\mx, \sigv)$'s obtained from
minimization of $\chiup^2_1+\chiup^2_2$ [$\chiup^2_1+\chiup^2_3$]
in \sFref{fit}{a$_1$} and {\small\sf (a$_2$)} [\sFref{fit}{b$_1$}
and {\small\sf (b$_2$)}]. Since ATIC and Fermi-LAT data are not
consistent with each other, we do not combine them but present
results using only either of the two. The relevant $(\mx,
\sigv)$'s can be read in the Table~\ref{Tfit} together with the
corresponding $\chiup^2/{\rm d.o.f}$, where d.o.f denotes the
number of degrees of freedom involved in our fits which is equal
to $N_1+N_2-2=26$ [$N_1+N_3-2=31$] for PAMELA and ATIC [PAMELA and
Fermi-LAT] data (2 is the number of the fitting variables, $\mx$
and $\sigv$). As can be deduced from Table~\ref{Tfit}, an
exceptionally good fit to PAMELA and Fermi-LAT data arises for
$\chi$ annihilating to $\mu^+\mu^-$, whereas in most other cases
the fits are rather poor since we get just $\chiup^2/{\rm
d.o.f}\simeq2.5-3$ for 28 or 33 data points. Better fits can be
probably attained under the assumption that $\chi$'s both
annihilate and decay, as pointed out in \cref{korea}. Note finally
that, the $\chi$ annihilation into $e^+e^-$ is strongly disfavored
\cite{referee} by Fermi-LAT data since it predicts a spectrum with
a too sharp end-point.


In Table~\ref{Tfit} we also list the maximal $\sigv$, $\sigv_{\rm
max}$, allowed by \Eref{sigCMB} (see \Sref{sCMB}) and the
resulting $\Omxsc$'s. It is remarkable that $\Omxsc$ turns out to
be well below the range of \Eref{cdmb} implied by the CDM
considerations. As a consequence, the SC can not be consistent
with the interpretation of the $e^\pm$-CR anomalies via $\chi$
annihilation, unless we invoke an enhancement mechanism of $\sigv$
at present \cite{pole,sommer}. In other words, some $\Domx$ is
necessitated in order to reconcile the best-fit $(\mx, \sigv)$'s
with \Eref{cdmb}. Moreover, we observe that the bound of
\Eref{sigCMB} (which turns out to be the most restrictive of all
the others presented in \Sref{sec:sv}) is violated in all cases.
This violation is softer [stronger] in the case where $\chi$'s
annihilate to $\mu^+\mu^-$ [$\tau^+\tau^-$]. We observe, also,
that for $\chi\chi\to\tau^+\tau^-$, $\mx$'s and $\sigv$'s are
pushed to larger values than those needed for
$\chi\chi\to\mu^+\mu^-$ and the $\mx$'s and $\sigv$'s used in the
latter case are higher than those used for $\chi\chi\to\ps\el$. As
shown in the next section, best-fit $(\mx, \sigv)$'s consistent
with all the available constraints can be achieved for $\chi$'s
annihilating to $\ps\el$ or $\mu^+\mu^-$.

\renewcommand{\arraystretch}{1.2}
\begin{table}[!t]
\begin{center}
\begin{tabular}{|c|c||c|c|c||c||c|} \hline
{\sc Figure}& {\sc Annihilation}& $\mchi/$ & $\mx/$ & $\sigv/$& $\sigv_{\rm max}/$&$\left.\Omx\right|_{\rm SC}/$\\
&{\sc Mode}&${\rm d.o.f}$&
$\TeV$&$10^{-7}~\GeV^{-2}$&$10^{-7}~\GeV^{-2}$& $10^{-4}$\\
\hline \hline
\ref{fit}-{\sf\small (a$_1$)} and {\small\sf (a$_2$)}&
$\chi\chi\to \ps\el$&$67/26$&$0.74$&$7$&$3.3$&$3.85$\\
\ref{fit}-{\small\sf (a$_1$)} and {\small\sf (a$_2$)}&$\chi\chi\to
\mu^+\mu^-$
&$76/26$&$2$&$28.6$&$26$&$0.97$\\
\ref{fit}-{\small\sf (a$_1$)} and {\small\sf (a$_2$)}&$\chi\chi\to
\tau^+\tau^-$ &$79/26$&$3.5$&$143$&$20$&$0.2$\\\hline
\ref{fit}-{\small\sf (b$_1$)} and {\small\sf (b$_2$)}&
$\chi\chi\to \ps\el$ &$75/31$&$0.5595$&$4.6$&$2.5$&$5.8$\\
\ref{fit}-{\small\sf (b$_1$)} and {\small\sf (b$_2$)}&
$\chi\chi\to \mu^+\mu^-$
&$24/31$&$1.28$&$19.5$&$16.5$&$1.4$\\
\ref{fit}-{\small\sf (a$_1$)} and {\small\sf (a$_2$)}&$\chi\chi\to
\tau^+\tau^-$ &$33/31$&$2.1$&$94.6$&$28.2$&$0.3$\\ \hline
\end{tabular}
\end{center}\vspace*{-.155in}
\renewcommand{\arraystretch}{1.0}
\hfill \caption[]{\sl\ftn The best-fit points ($\mx,~\sigv$) for
the PAMELA and ATIC data (used in Fig.~\ref{fit}-{\sf\ssz (a$_1$)}
and {\sf\ssz (a$_2$)}) or the PAMELA and Fermi-LAT data (used in
Fig.~\ref{fit}-{\sf\ssz (b$_1$)} and {\sf\ssz (b$_2$)}) for each
annihilation channel of $\nt$. Reported are also the corresponding
$\mchi/{\rm d.o.f.}$, the maximal $\sigv$, $\sigv_{\rm max}$,
allowed by \Eref{sigCMB} and the resulting $\Omx$ in the SC,
$\Omxsc$.} \label{Tfit}
\end{table}

\section{Restrictions in the $\mx-\sigv$ Plane}\label{sec:mxsv}

To systematize our approach, we need to delineate in the
$\mx-\sigv$ plane the regions which are favored at $95\%$ c.l. by
the various experimental data on the $e^\pm$-CRs. In particular,
we consider \cite{korea, chi2} regions favored by PAMELA data
only, PAMELA and ATIC data or PAMELA and Fermi LAT data. Since we
have two independent parameters, $\mx$ and $\sigv$ these regions
can be determined imposing the condition \cite{chi2}
\beq
\label{x2min}\chiup^2\lesssim\mchi+6~~\mbox{with}~~\chiup^2=\left\{\matrix{
\chiup^2_1\hfill & \mbox{for PAMELA}, \hfill \cr
\chiup^2_1+\chiup^2_2\hfill & \mbox{for PAMELA and ATIC}, \hfill
\cr
\chiup^2_1+\chiup^2_3~\hfill &\mbox{for PAMELA and Fermi LAT},
\hfill \cr}
\right.\eeq
where $\mchi$ can be extracted numerically by minimization of
$\chiup^2$ w.r.t $\mx$ and $\sigv$. On the other hand, the
interpretation of the data on $e^\pm$-CRs in terms of $\chi$
annihilation can be viable if it can become consistent with a
number of phenomelogical constraints. In \Sref{sec:sv} we
summarize these constraints and in \Sref{sec:res} we examine if
they can be reconciled with the regions favored by the data on
$e^\pm$-CRs.

\subsection{Imposed Constraints}\label{sec:sv}

Though $n_\chi/s$ in \Eref{omh} stays essentially unchanged for
$\tau>\tau_{\rm f}$, residual annihilations of $\chi$'s occur up
to the present with several cosmological consequences besides the
possible interpretation of the data on $e^\pm$-CRs. Recently,
important upper bounds on $\sigv$ have been reported and are
summarized below for the three exemplary $\chi$ annihilation modes
considered in our investigation. The well-known unitarity
constraint is also taken into account.

\subsubsection{Unitarity Constraint.}\label{sigun}
Using partial-wave unitarity \cite{khalil, unitarity} an upper
limit, particularly relevant for $\mx>2~\TeV$, on $\sigv$ can be
derived as a function of $\mx$, i.e.,
\beq\label{sigUn}\sigv\leq8\pi~\GeV^{-2}\;\left({m_{\nt}\over
1~\GeV}\right)^{-2}\cdot \eeq
\subsubsection{BBN Constraint.}\label{sNS}
During BBN, the $\chi$ annihilations inject an amount of energetic
particles which is proportional to $\sigv$ and may strongly alter
\cite{moroiNS, Bi} the abundances of the light elements. Ruining
the successful predictions of the BBN can be avoided if we impose
an upper bound on $\sigv$ which, however, depends on the identity
of the products of the annihilation of $\chi$'s. Taking into
account the most up-to-date analysis of \cref{moroiNS} we demand:
\beq\label{sigNS}\sigv\leq3\cdot10^{-5}~\GeV^{-2}\;{2\mx\over
E_{\rm vis}}{m_{\nt}\over1~\TeV}~~ \mbox{where}~~{E_{\rm vis}\over
\mx} =\left\{\matrix{
2 ~~\hfill \mbox{for}  & \chi\chi\to \ps\el, \hfill \cr
0.7~~\hfill \mbox{for} & \chi\chi\to \mu^+\mu^-, \hfill \cr
0.62~~\hfill \mbox{for} & \chi\chi\to \tau^+\tau^-, \hfill \cr}
\right. \eeq
with $E_{\rm vis}$ being the total visible energy of the produced
particles in the $\chi$ annihilation.

\subsubsection{CMB Constraint.}\label{sCMB}
The $\chi$ annihilations may have \cite{CMBold, CMB} an impact on
the ionization state of the baryonic gas at recombination and
therefore on the CMB angular spectra. Consistency with the WMAP5
data \cite{wmap} dictates \cite{CMB} at 95$\%$ c.l. (see, also,
\cref{CMBnew}):
\beq\label{sigCMB}\sigv\leq{3.1\cdot10^{-7}~\GeV^{-2}\over
f}\;{m_{\nt}\over1~\TeV}\,~~ \mbox{where}~~f \simeq\left\{\matrix{
0.7 ~~\hfill \mbox{for}  & \chi\chi\to \ps\el, \hfill \cr
0.24~~\hfill \mbox{for} & \chi\chi\to \mu^+\mu^-, \hfill \cr
0.23~~\hfill \mbox{for} & \chi\chi\to \tau^+\tau^-, \hfill \cr}
\right. \eeq
is the deposited power fraction which expresses the efficiency of
the coupling between the annihilation products and the
photon-baryon fluid at $z\sim1000$. It is expected that
forthcoming experiments will impose \cite{CMB} even more stringent
bounds on $\sigv$. Note, in passing, that the presence of $q$ in
the QKS does not affect recombination (which occurs at $\vtau_{\rm
rec}\simeq-7$) since $\Omega_q(\vtau_{\rm rec})$ is safely
suppressed provided that \Eref{nuc} is fulfilled.

\subsubsection{Constraint from the $\gamma$-Cosmic
Rays.}\label{grays} The $\chi$ annihilation in the galactic center
yields sizeable amount of $\gamma$-CRs, through the cascade decay
of the annihilation products and/or bremsstrahlung processes.
Comparing the relevant $\gamma$-CR flux with the H.E.S.S
observations \cite{hess} we can further restrict \cite{gstrumia}
$\sigv$ as a function of $\mx$ for the two chosen annihilation
channels. However, this restriction significantly depends on the
CDM halo profile. Adopting the cored isothermal CDM profile
\cite{isoT}, which assures the less restrictive version of this
constraint, we graphically extract the upper bound on $\sigv$ from
the plots of \cref{gstrumia}. To have a feeling of the strength of
this constraint we can give some rough estimations:
\beq\label{sighess}\sigv\lesssim~~\left\{\matrix{
(3-10)\cdot10^{-6}~\GeV^{-2}~~&\mbox{for}&\mx=(0.2-1.5)~\TeV&\mbox{and}&\chi\chi\to
\ps\el, \hfill \cr
(4-25)\cdot10^{-6}~\GeV^{-2}~~&\mbox{for}&\mx=(0.2-3)~\TeV&\mbox{and}&
\chi\chi\to \mu^+\mu^-, \hfill \cr
(1.2-4)\cdot10^{-6}~\GeV^{-2}~~& \mbox{for}
&\mx=(0.5-3)~\TeV&\mbox{and}& \chi\chi\to \tau^+\tau^-. \hfill
\cr}
\right.~~\eeq
Alternatively this constraint can be evaded for every CDM profile,
by allowing the $\chi$-annihilation products to be long lived, as
pointed out in \cref{chi2}.

\paragraph{} Complementary constraints on the $\chi$ annihilation can be imposed
comparing the findings of EGRET satellite \cite{egret} with the
diffuse (secondary) $\gamma$-CR fluxes, which would be produced
\cite{ics} by inverse Compton scatterings on interstellar photons
of the energetic $e^\pm$ generated by the $\chi$ annihilation in
the galactic halo. However, these bounds are expected \cite{ics}
to be weaker than the ones imposed by the high energy $\gamma$-CRs
mentioned in \Sref{grays} and are not included in our analysis.
Similar arguments are \cite{neutrinos} also valid for the
neutrinos generated from the $\chi$ annihilation in the galactic
center, though the dependence on the CDM profile is weaker.

\subsection{Results}\label{sec:res}

Constructing the preferred areas by the various experimental data
and taking into account the constraints quoted in \Sref{sec:sv} we
can check the viability of the interpretation of the anomalies on
$e^\pm$-CR fluxes in terms of the $\chi$ annihilation. In
\Fref{svmx1} we consider the mode $\chi\chi\to\tau^+\tau^-$
whereas in \Fref{svmx}-{\small\sf (a$_1$), (b$_1$)} and {\small\sf
(c$_1$)} [\Fref{svmx}-{\small\sf (a$_2$), (b$_2$)} and {\small\sf
(c$_2$)}] we assume that $\chi$'s annihilate into $\ps\el$
[$\mu^+\mu^-$].

In Fig.~\ref{svmx1} and \ref{svmx} we delineate the regions
preferred at $95\%$ c.l. by PAMELA data (black and red sparse
hatched areas), PAMELA and ATIC data (dense black hatched areas)
and PAMELA and Fermi-LAT data (dense red hatched areas), by
imposing the condition of \Eref{x2min}. In the black [red] hatched
areas the backgrounds fluxes are normalized setting $c_{\el}=0.6$
[$c_{\el}=0.7$] in \Eref{bfluxes}. Evidently, the PAMELA data do
not prefer any $\mx$ since it does not show any peak structure
whereas Fermi-LAT data disfavors the mode $\chi\chi\to e^+e^-$
since the spectrum from such a channel is too peaked to reproduce
data. We also remark that the regions derived by the joint
analysis of two data-sets are rather limited -- c.f.
\cref{gstrumia, CMBnew, chi2, referee}. We consider the latter
results as more reliable, since even when we fit only the PAMELA
data, the data points with low $E_{\ps}$ from ATIC/Fermi LAT are
involved, in order to normalize \cite{chi2} the background fluxes.

In Fig.~\ref{svmx1} and \ref{svmx} drawn is also the upper bound
from \eqss{sigUn}{sigNS}{sigCMB}, denoted by a black dotted, solid
and dashed line respectively and this from the constraint of
\Sref{grays}, depicted by a dot-dashed line. Obviously acceptable
are the regions mainly below the dashed curves, since the bound of
\Eref{sigCMB} is the most restrictive from the others. The bound
of \Eref{sigUn} cuts out some slices of the parameter space for
$\chi\chi\rightarrow \mu^+\mu^-$ and $\chi\chi\to\tau^+\tau^-$ and
large $\mx$'s. We easily conclude that the explanation of the
experimental anomalies via the annihilation mode:

\begin{list}{}{\setlength{\rightmargin=0cm}{\leftmargin=0.7cm}}

\item[$\bullet$] $\chi\chi\to e^+e^-$ is just marginally
consistent with \Eref{sigCMB} -- see \Fref{svmx}-{\small\sf
(a$_1$), (b$_1$)} and {\small\sf (c$_1$)}. Indeed, we observe that
just a minor portion of the area favored by PAMELA at $95\%~{\rm
c.l.}$ is allowed by \Eref{sigCMB} whereas the regions preferred
at $95\%~{\rm c.l.}$ from both combinations of PAMELA and ATIC or
PAMELA and Fermi-LAT data are entirely excluded from the bounds of
\Eref{sigCMB}.

\item[$\bullet$] $\chi\chi\to \mu^+\mu^-$ can be reconciled --
c.f. \cref{khalil, gstrumia, chi2, moroiNS} -- with the various
constraints -- see \Fref{svmx}-{\small\sf (a$_2$), (b$_2$)} and
{\small\sf (c$_2$)}. Namely, we notice that sizable slices of the
regions favored by PAMELA lie below the bound of \Eref{sigCMB}.
Moreover, very close to or even lower than this limit we find
portions of the favored regions at $95\%$ c.l. by the PAMELA and
ATIC or Fermi-LAT data.

\item[$\bullet$] $\chi\chi\to\tau^+\tau^-$ is inconsistent -- c.f.
\cref{gstrumia} -- with both \Eref{sigCMB} and \Eref{sighess} at
$95\%~{\rm c.l.}$ since all the regions favored by the
experimental data lie entirely above the bounds above -- see
\Fref{svmx1}. Violation of the bound of \Eref{sigUn} in a sizable
fraction of these regions is observed too. Because of this fact,
we below concentrate on the other two annihilation modes of
$\chi$'s.

\end{list}

 \begin{center}
\begin{minipage}{0.6\textwidth}
\vspace*{-.1cm}\centerline{\epsfig{file=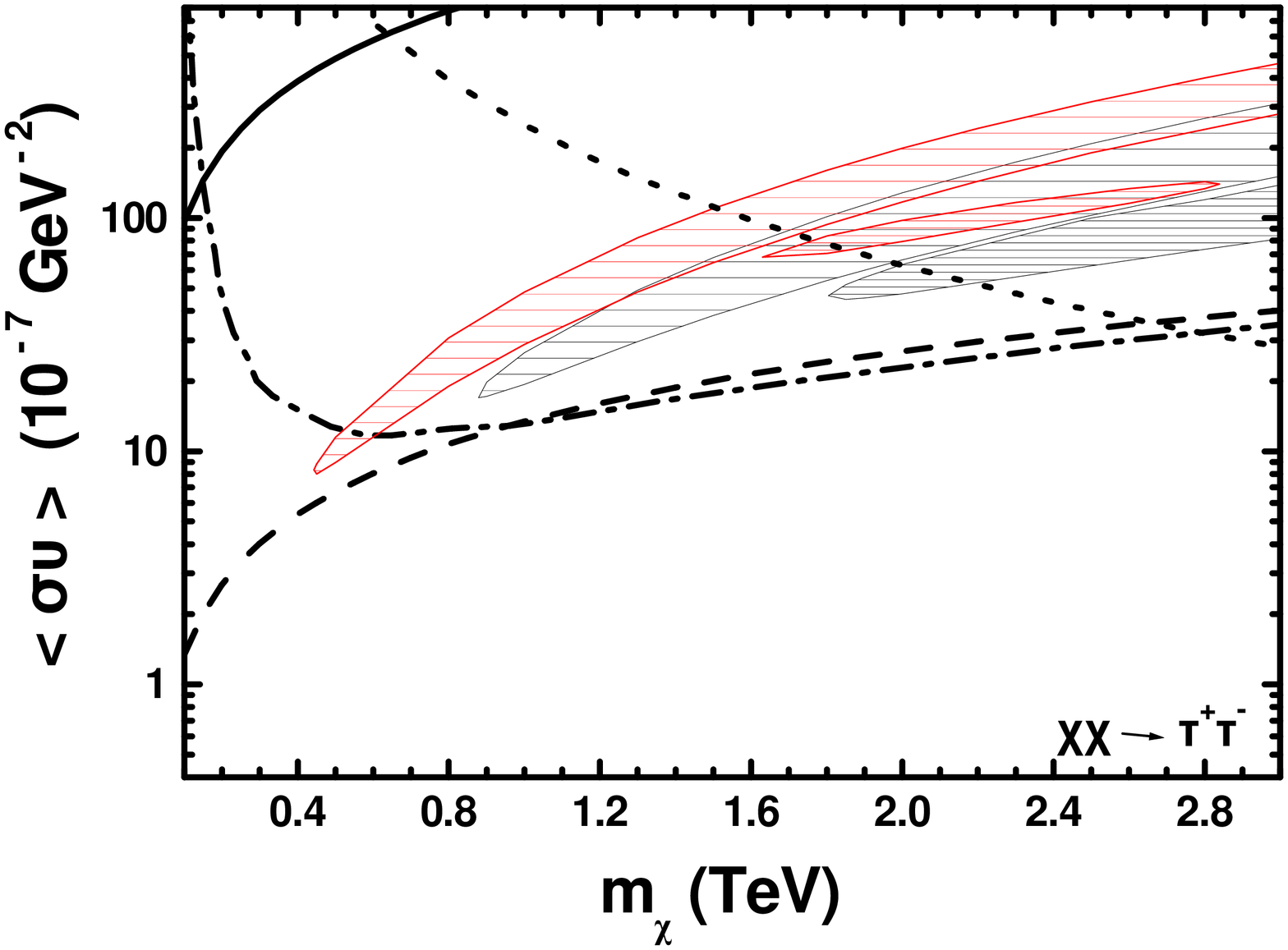,height=3.55in,angle=-90}}
\hfill
\end{minipage}\hfill
\begin{minipage}{0.4\textwidth}
\vspace*{-1.4cm}\captionof{figure}{\sl\ftn Restrictions in the
$\mx-\sigv$ plane for $\chi$'s annihilating into $\tau^+\tau^-$.
The sparse black [red] hatched areas are preferred at $95\%$ c.l.
by the PAMELA data for $c_{\el}=0.6$ [$c_{\el}=0.7$] and the dense
black [red] hatched areas are preferred at $95\%$ c.l. by the
PAMELA and ATIC [PAMELA and Fermi-LAT] data. Regions above the
black solid, dashed, dotted and dot-dashed lines are ruled out by
the upper bounds on $\sigv$ from \Eref{sigNS}, (\ref{sigCMB}),
(\ref{sigUn}) and \Sref{grays} correspondingly.}\label{svmx1}
\end{minipage}
\end{center}

Fixing the parameters related to the LRS [QKS] to some
representative values -- consistent with \Sref{reqlr}
[\Sref{reqq}] --, we can display in the $\mx-\sigv$ plane, as in
\Fref{svmx}, regions (light gray shaded) confronted with
\Eref{cdmb}. The gray dashed [dotted] lines correspond to
\sEref{cdmb}{b} [\sEref{cdmb}{a}], whereas the gray solid lines
are obtained by fixing $\Omx$ to its central value in \Eref{cdmb}.
\ssFref{svmx}{a$_1$}{a$_2$} are devoted to the LRS whereas
\Fref{svmx}-{\small\sf (b$_1$),~(b$_2$), (c$_1$)} and {\small\sf
(c$_2$)} analyze the QKS.

For the LRS, we confine ourselves to some combinations of
parameters which assure a sufficient coexistence of non-TP and TP,
since non-TP alone is obviously -- see \Eref{fxntp} -- $\sigv$
independent and therefore can not be properly depicted in the
$\mx-\sigv$ plane. We take $(\cxf, \Trh)=(2\cdot 10^{-6},
0.1~\GeV)$ and $(\cxf, \Trh)=(1, 1~\GeV)$ [$(\cxf, \Trh)=(10^{-6},
0.1~\GeV)$ and $(\cxf, \Trh)=(1, 0.5~\GeV)$] in
\sFref{svmx}{a$_1$} [\sFref{svmx}{a$_2$}]. For the QKS, we set
throughout $a=0.5,~\vTi=10^9~\GeV$. We also take {\sf\small (i)}
$b=0$ and $\vHi=6.3\cdot 10^{53},~2\cdot 10^{53}$ or
$6.2\cdot10^{52}$ resulting to $\Omqns=0.01,0.001$ or $0.0001$
respectively in \ssFref{svmx}{b$_1$}{b$_2$}; {\sf\small (ii)}
$\vHi=6.3\cdot10^{53}$ and $b=0.15$ [$\vHi=6.2\cdot10^{52}$ and
$b=0.32$] resulting to $\Omqns=0.068$ or [$\Omqns=0.065$]  in
\sFref{svmx}{c$_1$}; {\sf\small (iii)} $\vHi=6.3\cdot10^{53}$ and
$b=0.32$ [$\vHi=2\cdot10^{53}$ and $b=0.2$] yielding $\Omqns=0.19$
or [$\Omqns=0.21$]  in \sFref{svmx}{c$_2$}. Note that in
\ssFref{svmx}{b$_1$}{b$_2$} we present for the sake of comparison
results even for $b=0$, although the tracking behavior of the QKS
is not attained in this case -- see \Sref{ap}.

In all cases, we observe that $\Omx$ decreases as $\sigv$
increases. This is due to the fact that $\Omx\propto1/\sigv$ as
can be deduced from \Eref{omh} and \Eref{nstp} [\Eref{BEsol}] for
the LRS [QKS]. For the LRS , as it is clear from these plots,
there is a minor slice of the allowed region with $\mx<0.35~\TeV$
[$\mx<0.8~\TeV$ ] for $\Trh=0.1~\GeV$ and $\cxf=2\cdot 10^{-6}$
[$\cxf=10^{-6}$] where non-TP is strengthened and our results are
almost $\sigv$ independent. For the QKS, we also observe that for
$\vtf$ far away from $\vtp$ the allowed by \Eref{cdmb} for
$b\neq0$ region reaches the one for $b=0$ -- with fixed $\vHi$.
However, when $\vtf$ reaches $\vtp$, $\Omx$ decreases (as we
explain in \Sref{wimp2}) and so, the required, for obtaining
$\Omx$ in the range of \Eref{cdmb}, $\sigv$ decreases too. As a
consequence, although the allowed by \Eref{cdmb} area in
\sFref{svmx}{c$_1$} [\sFref{svmx}{c$_2$}] for
$\vHi=6.3\cdot10^{53}$ approaches the corresponding area in
\sFref{svmx}{b$_1$} [\sFref{svmx}{b$_2$}] with the same $\vHi$ and
violates the bounds of \Eref{sigCMB} for low $\mx$'s, it becomes
compatible with the latter constraint for larger $\mx$'s. On the
other hand, we observe that there is no such a transition region
in the light gray area of \sFref{svmx}{c$_2$}. This is, because
for $0.1\leq\mx/\TeV\leq3$ we get $31.8\leq-\vtf\leq35.1$ whereas
the closest to $\vtf$'s, $\vtp$ is $\vtp=-35.4$ which remains
constantly lower than $\vtf$. Therefore, no reduction of $\Domx$
occurs for the $\mx$'s used in \sFref{svmx}{c$_2$}.

\newpage

\begin{center}  \vspace*{-1.1cm}\hspace*{-.25in}
\begin{minipage}{8in}
\epsfig{file=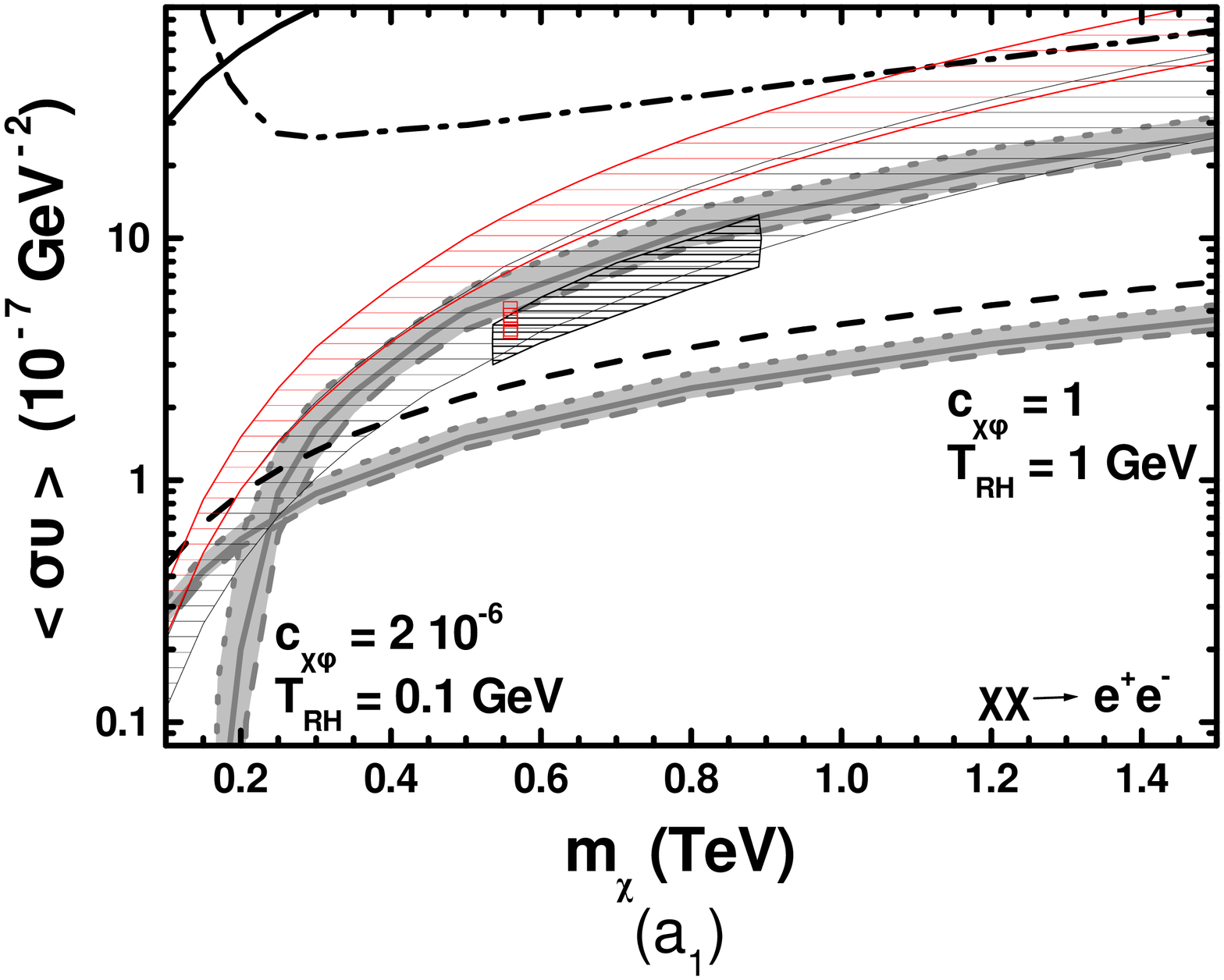,height=3.55in,angle=-90}
\hspace*{-1.37 cm}
\epsfig{file=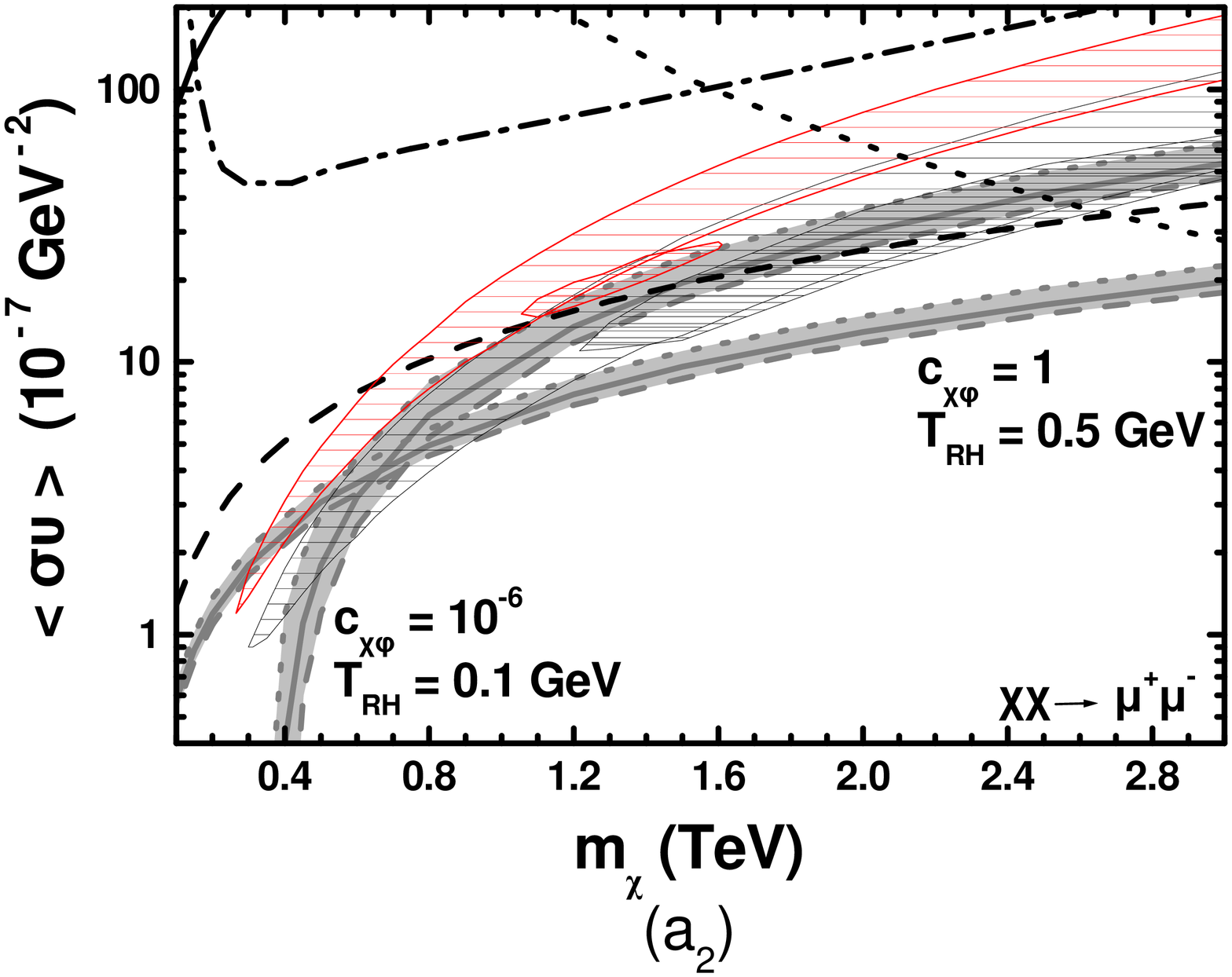,height=3.55in,angle=-90} \hfill
\end{minipage}\vspace*{-.01in}
\hfill \hspace*{-.25in}
\begin{minipage}{8in}
\epsfig{file=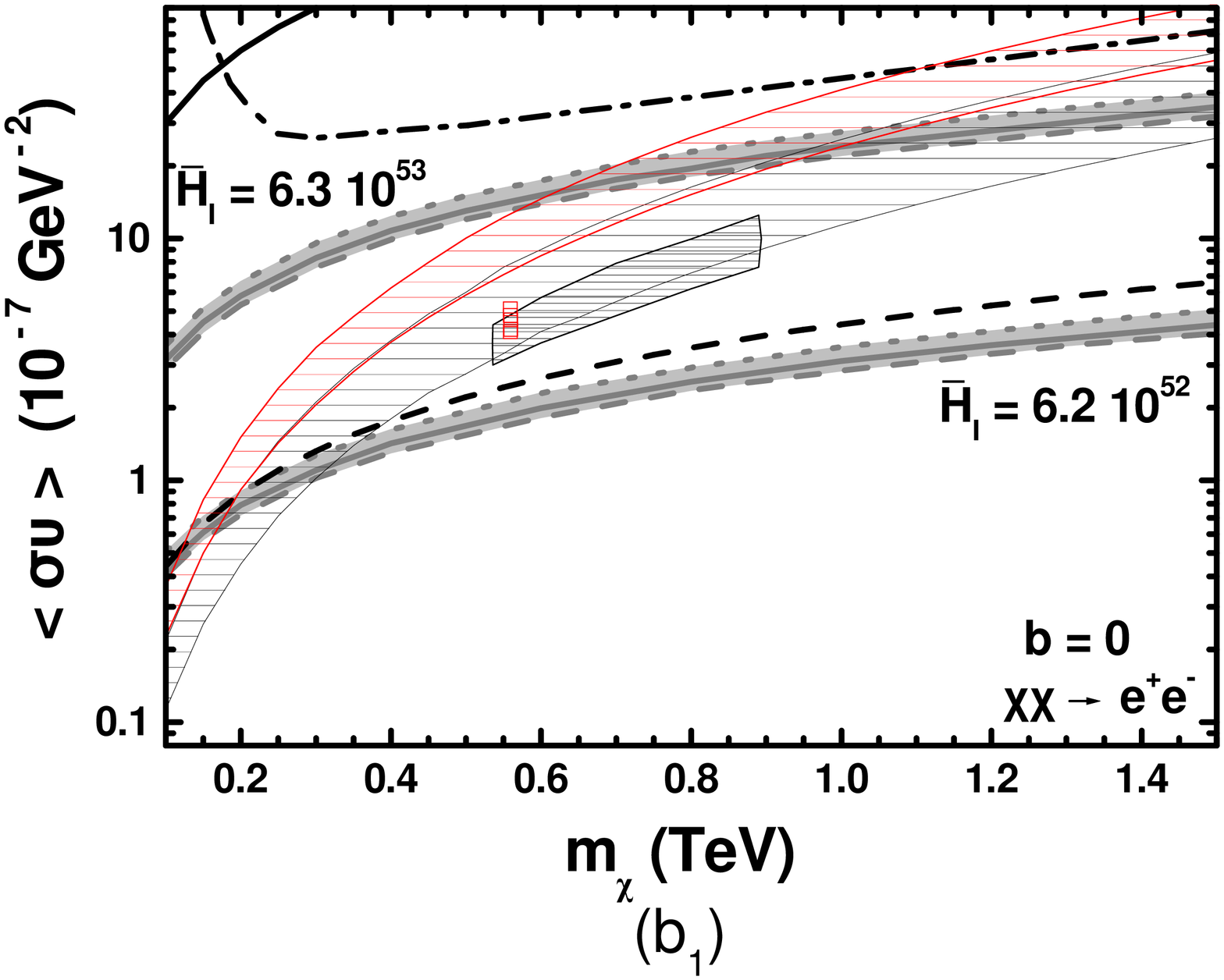,height=3.55in,angle=-90}
\hspace*{-1.37 cm}
\epsfig{file=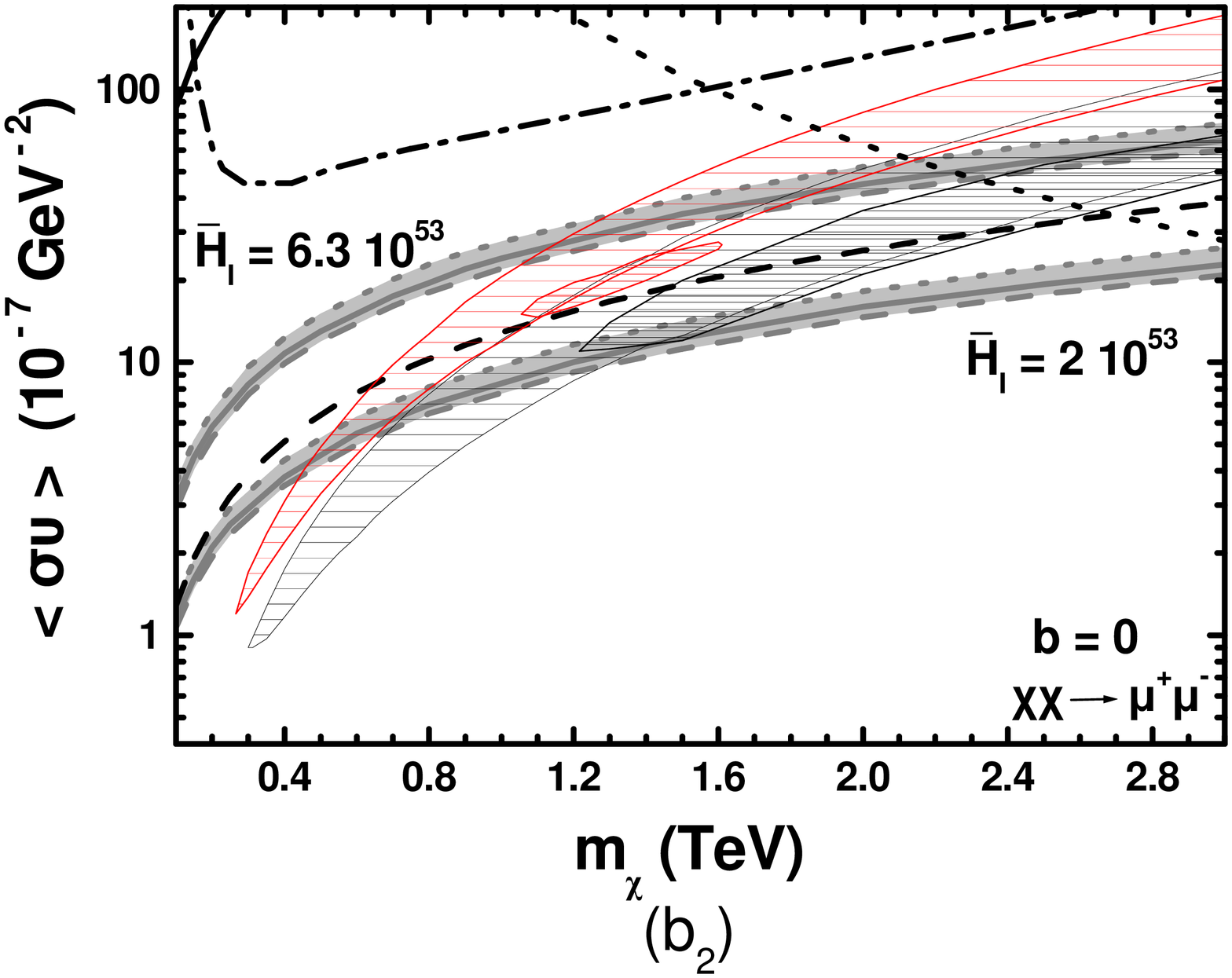,height=3.55in,angle=-90} \hfill
\end{minipage}\vspace*{-.01in}
\hfill \hspace*{-.25in}
\begin{minipage}{8in}
\epsfig{file=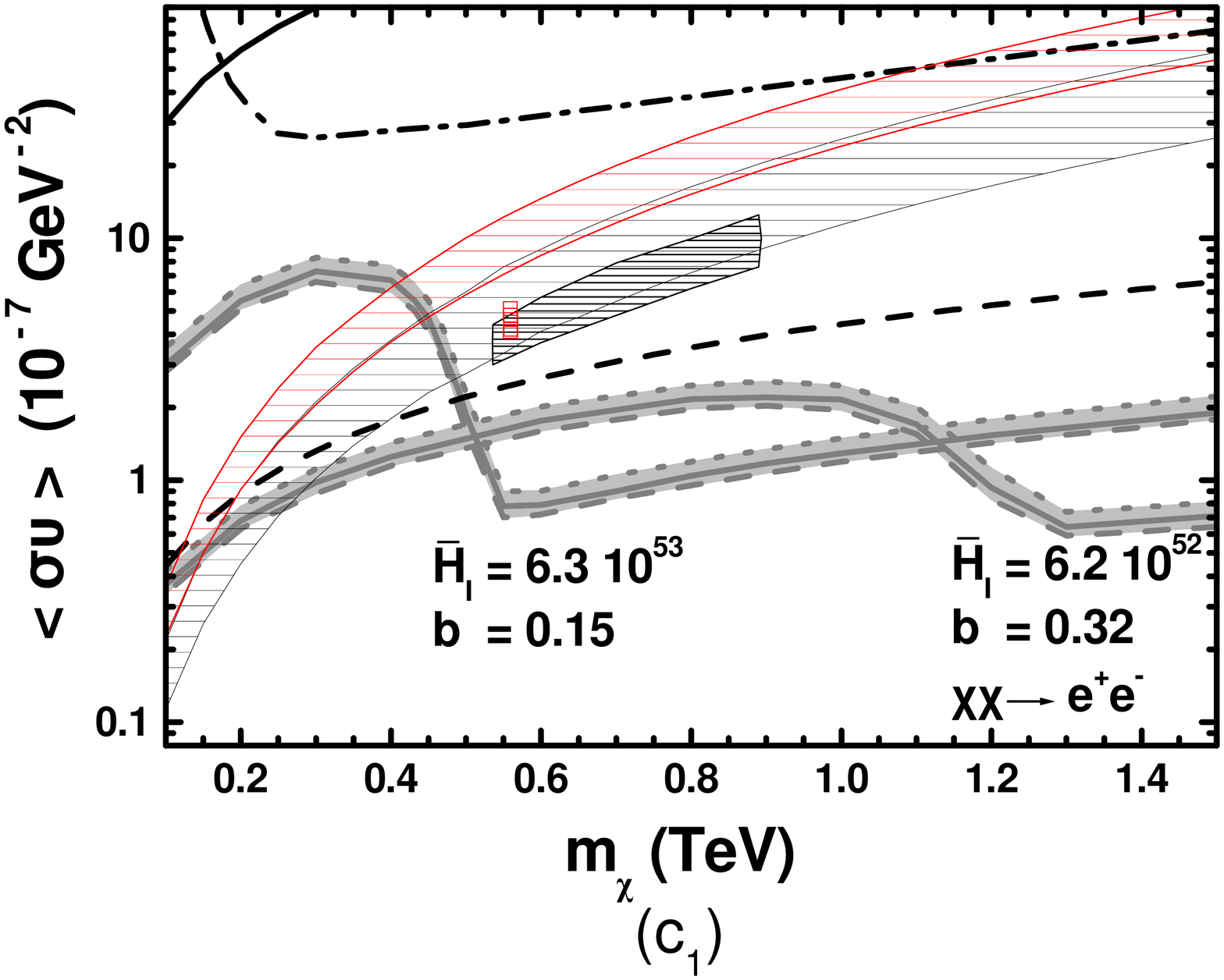,height=3.55in,angle=-90}
\hspace*{-1.37 cm}
\epsfig{file=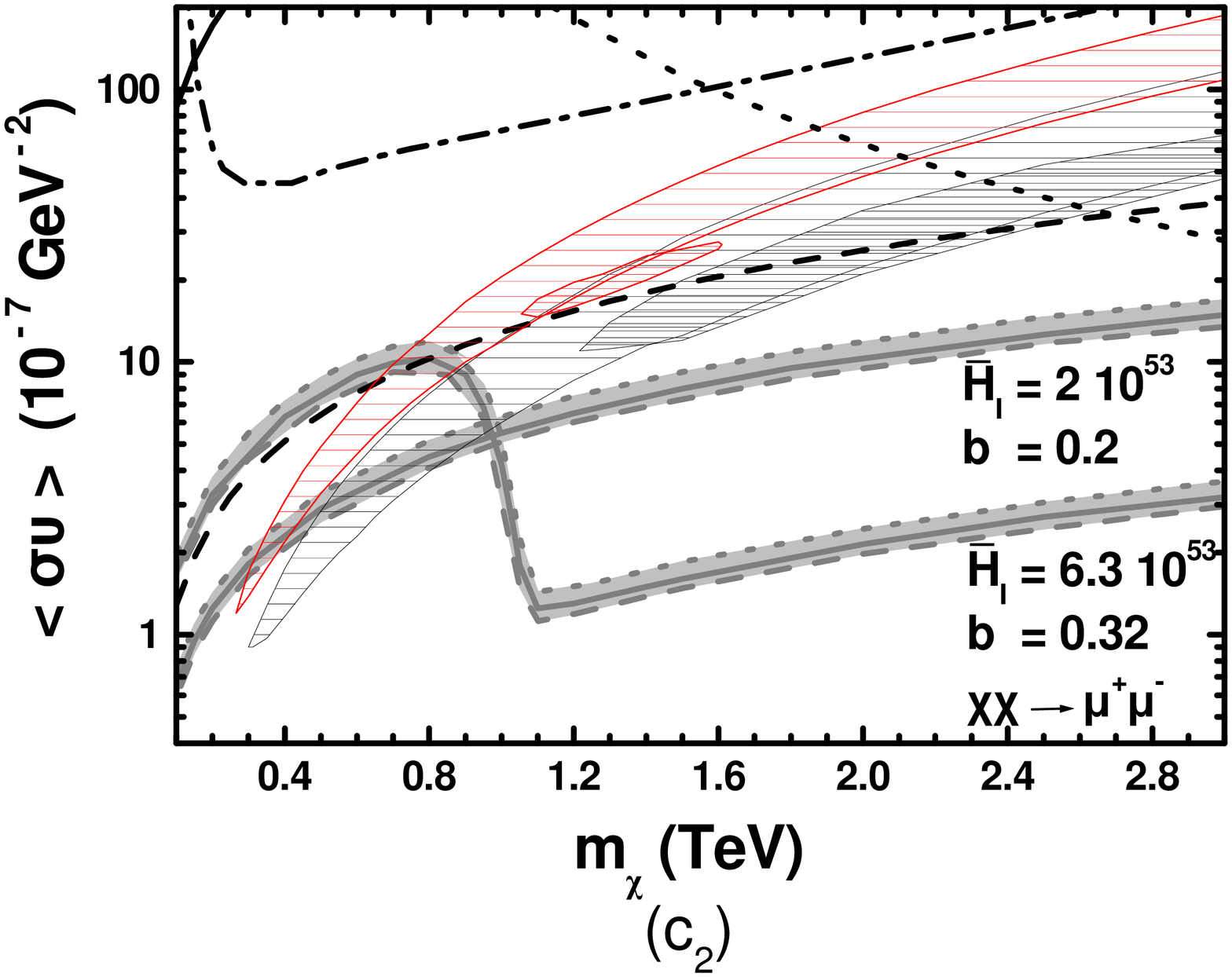,height=3.55in,angle=-90} \hfill
\end{minipage}\vspace*{-.05in}
\begin{minipage}{0.75\textwidth}
\captionof{figure}{\sl\ftn Restrictions in the $\mx-\sigv$ plane
for the LRS {\sf\ssz (a$_1$, a$_2$)} [QKS {\sf\ssz (b$_1$, b$_2$,
c$_1$, c$_2$)} taking $a=0.5,~\vTi=10^9~\GeV$] with various
$\cxf$'s and $\Trh$'s [$b$'s and $\vHi$'s], indicated in the
graphs, and $\chi$'s annihilating into $\ps\el$ {\sf\ssz (a$_1$,
b$_1$, c$_1$)} or $\mu^+\mu^-$ {\sf\ssz (a$_2$, b$_2$, c$_2$)}.
The light gray shaded areas are allowed by \Eref{cdmb}, the sparse
black [red] hatched areas are preferred at $95\%$ c.l. by the
PAMELA data for $c_{\el}=0.6$ [$c_{\el}=0.7$] and the dense black
[red] hatched areas are preferred at $95\%$ c.l. by the PAMELA and
ATIC [PAMELA and Fermi-LAT] data. Regions above the black solid,
dashed, dotted and dot-dashed lines are ruled out by the upper
bounds on $\sigv$ from \Eref{sigNS}, (\ref{sigCMB}), (\ref{sigUn})
and \Sref{grays} correspondingly. The conventions adopted for the
residual lines are also shown.} \label{svmx}\hfill
\end{minipage}\hfill
\begin{minipage}{0.2\textwidth}
\vspace*{-.1in}\centerline{\epsfig{file=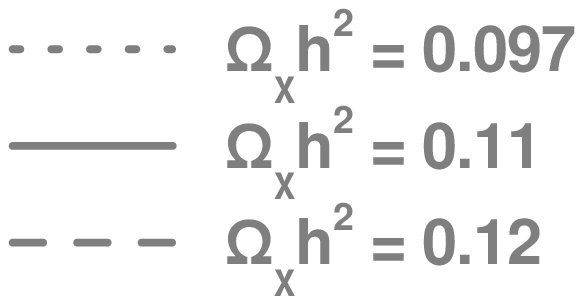,height=1.3in,angle=-90}}
\end{minipage}
\end{center}\vfill

As can be concluded from most of the plots of \Fref{svmx}, a
simultaneous interpretation of the $e^\pm$-CR anomalies
consistently with the requirements of \Sref{sec:sv} can be
achieved in the regions where the gray shaded areas overlap the
lined ones below the dashed lines. To clarify further this
intriguing conclusion of this paper, it would be interesting to
find the best-fit ($\mx$, $\sigv$) (for the various combined
data-sets) which fulfill all the restrictions imposed in
\Sref{sec:sv}. Our results are arranged in Table~\ref{fit1}. The
listed ($\mx$, $\sigv$)'s saturate the bound of \Eref{sigCMB}
which turns out to be essentially the most stringent of the others
-- see \Fref{svmx}. We observe that $\mx\sim0.1~\TeV$ and
$\sigv\sim10^{-7}~\GeV^{-2}$ [$\mx\sim1~\TeV$ and
$\sigv\sim10^{-6}~\GeV^{-2}$] for $\chi$'s annihilating to
$e^+e^-$ [$\mu^+\mu^-$]. From the exposed in Table~\ref{fit1}
$\chiup^2-\mchi$'s, we deduce that all the requirements are met in
a portion of the area favored at $99\%$ c.l. [$68\%$ c.l.] by the
PAMELA and Fermi-LAT [ATIC] data for $\chi\chi\rightarrow
\mu^+\mu^-$, whereas the mode $\chi\chi\to e^+e^-$ can be excluded
at $99\%$ c.l. As regards the quality of the fits, from
Tables~\ref{Tfit} and \ref{fit1}, we can infer that the
$\mu^+\mu^-$ channel gives better fit to the PAMELA and Fermi-LAT
data ($\chiup^2/\dof = 33/31$) than to the PAMELA and ATIC data
($\chiup^2/\dof = 77/26$).

\renewcommand{\arraystretch}{1.2}
\begin{table}[!t]
\begin{center}\begin{tabular}{|c|c|c|c||c|c|c|} \hline
&\multicolumn{6}{|c|}{\sc\bfseries Fits to PAMELA And ATIC
Data}\\\cline{2-7} &\multicolumn{3}{|c||}{$\chi\chi\rightarrow
\ps\el$} &\multicolumn{3}{|c|}{$\chi\chi\rightarrow \mu^+\mu^-$}
\\\hline\hline
$\mx~(\TeV)$&\multicolumn{3}{|c||}{$0.45$}&\multicolumn{3}{|c|}{$2$}\\
$\sigv~\left(\GeV^{-2}\right)$&\multicolumn{3}{|c||}{$1.98\cdot10^{-7}$}
&\multicolumn{3}{|c|}{$2.6\cdot10^{-6}$}\\\hline
$\chiup^2-\mchi$&\multicolumn{3}{|c||}{13.4}&\multicolumn{3}{|c|}{1}
\\\hline\hline
$\left.\Omx\right|_{\rm
SC}$&\multicolumn{3}{|c||}{0.0013}&\multicolumn{3}{|c|}{0.0001}\\\hline\hline
\multicolumn{7}{|c|}{\sc Combinations of Parameters Yielding
$\Omx=0.11$ In The LRS}\\\hline
$\Trh~(\GeV)$&0.001&0.1&0.69&0.001&0.1&0.27\\
$\cxf$&$8.1\cdot10^{-5}$&$1.3\cdot10^{-6}$&$1$&$1.9\cdot10^{-5}$&$9\cdot10^{-7}$
&$1$\\\hline
$\chi$-{\sc Production} &\multicolumn{2}{|c|}{\sc non-TP }& {\sc
non-TP + TP}& {\sc non-TP }& \multicolumn{2}{|c|}{\sc non-TP +
TP}\\\hline\hline
\multicolumn{7}{|c|}{\sc Combinations of Parameters Yielding
$\Omx=0.11$}\\ \multicolumn{7}{|c|}{\sc In The QKS for $a=0.5$ and
$\vTi=10^9~\GeV$}\\\hline
$b$ & 0 & 0.2& 0.32& $~0~$& \multicolumn{2}{|c|}{ $0.08$}\\
$\vHi/10^{53}$& 0.81&1.27& 1& $~3.5~$&\multicolumn{2}{|c|}{
$3.7$}\\\hline
$\Tkr~(\GeV)$ & $0.06$ & $0.03$& $0.038$& $0.017$ &
\multicolumn{2}{|c|}{$0.005$}\\\hline
\hline &\multicolumn{6}{|c|}{\sc\bfseries Fits to PAMELA And
Fermi-LAT Data}\\\cline{2-7}
&\multicolumn{3}{|c||}{$\chi\chi\rightarrow \ps\el$}
&\multicolumn{3}{|c|}{$\chi\chi\rightarrow \mu^+\mu^-$}
\\\hline\hline
$\mx~(\TeV)$&\multicolumn{3}{|c||}{0.383}&\multicolumn{3}{|c|}{1.12}\\
$\sigv~\left(\GeV^{-2}\right)$&\multicolumn{3}{|c||}{$1.69\cdot10^{-7}$}
&\multicolumn{3}{|c|}{$1.44\cdot10^{-6}$}\\\hline
$\chiup^2-\mchi$&\multicolumn{3}{|c||}{84}&\multicolumn{3}{|c|}{9}\\\hline\hline
$\left.\Omx\right|_{\rm
SC}$&\multicolumn{3}{|c||}{0.0015}&\multicolumn{3}{|c|}{0.00019}\\\hline\hline
\multicolumn{7}{|c|}{\sc Combinations of Parameters Yielding
$\Omx=0.11$ In The LRS}\\\hline
$\Trh~(\GeV)$&0.001&0.1&0.68&0.001&0.1&0.27\\
$\cxf $&$9.5\cdot10^{-5}$&$1.5\cdot10^{-6}$
&$1$&$3.3\cdot10^{-5}$&$1.4\cdot10^{-6}$&$1$\\\hline
$\chi$-{\sc Production} &\multicolumn{2}{|c|}{\sc non-TP }& {\sc
non-TP + TP}& {\sc non-TP }& \multicolumn{2}{|c|}{\sc non-TP +
TP}\\\hline\hline
\multicolumn{7}{|c|}{\sc Combinations of Parameters Yielding
$\Omx=0.11$}\\ \multicolumn{7}{|c|}{\sc In The QKS for $a=0.5$ and
$\vTi=10^9~\GeV$}\\\hline
$b$& $0$& $0.2$& $0.32$ & $0$& $0.08$&$0.18$\\
$\vHi/10^{53}$ & $0.79$ & $1.3$& $0.99$& $3.1$ & $3.4$&
$4.7$\\\hline
$\Tkr~(\GeV)$ & $0.07$ & $0.03$& $0.04$& $0.019$ & $0.005$&
$0.009$\\\hline
\end{tabular}
\end{center}\vspace*{-.155in}
\caption{\sl\ftn Best-fit ($\mx$, $\sigv$)'s for the combination
of the PAMELA and ATIC or Fermi-LAT data and the various
annihilation channels, consistently with all the imposed
constraints. Shown are also the resulting $\chiup^2-\mchi$ and
$\Omx$ in the SC, $\left.\Omx\right|_{\rm SC}$, several
combinations of $(b, \vHi)$'s [$(\Trh, \cxf)$'s] leading to
$\Omx\simeq0.11$ and the corresponding $\Tkr$'s  [types of $\chi$
production] within the QKS [LRS].}\label{fit1}
\end{table}
\renewcommand{\arraystretch}{1.0}

From Table~\ref{fit1} we can also appreciate the importance of the
non-SC in boosting $\Omx$ to an acceptable level. Indeed, in this
Table we display $\Omxsc$ for every allowed best-fit ($\mx$,
$\sigv$). We observe that it lies much lower than the range of
\Eref{cdmb} in all cases, i.e., it is insufficient to account for
the present CDM abundance in the universe. However, an appropriate
adjustment (shown also in Table~\ref{fit1}) of the parameters
$\cxf$ and $\Trh$ [$b$ and $\vHi$ (with fixed $a=0.5$ and
$\vTi=10^9~\GeV$)] for the LRS [QKS] -- consistently with the
restrictions of \Sref{reqlr} [\Sref{reqq}] -- elevates adequately
$\Omx$ which can acquire the central experimental value in
\Eref{cdmb} for every best-fit point. In Table~\ref{fit1} we also
expose the type of $\chi$ production for the LRS and the
transition temperature to the RD era for the QKS. We remark that
since the $\sigv$'s required for $\chi$'s annihilating to $e^+e^-$
are lower than those required for the $\mu^+\mu^-$ channel, non-TP
dominates even for $\Trh=0.1~\GeV$. Note that $\Tkr\leq0.04~\GeV$
and the tracking behavior fails for $b=0$ in the case of the QKS.

\section{Conclusions}\label{sec:con}

We presented two non-standard cosmological scenaria which can
increase the relic abundance of a WIMP $\chi$, $\Omx$, w.r.t its
value in the SC due to the generation of a background energy
density steeper than the one of RD era. This increase is
quantified by $\Domx$ defined in \Eref{dom}. According to the
first scenario, termed LRS, a scalar field $\phi$ decays,
reheating the universe to a reheating temperature lower than the
freeze-out temperature of the WIMPs. According to the second
scenario, termed QKS, a scalar field, $q$, rolls down its inverse
power-low potential with a Hubble-induced mass term. In both cases
our approach was both {\sf\small (i)} purely numerical,
integrating the relevant system of the differential equations
{\sf\small (ii)} semi-analytical, producing approximate relations
for the evolution of the various energy densities of the
cosmological background and the $\chi$-number density. We consider
that the exposed semi-analytical findings -- although do not
provide quite accurate results in all cases -- facilitate the
understanding of the cosmological dynamics.

As regards the LRS, we recalled the dynamics of reheating and
showed that $\Domx$ is affected by the two basic types of $\chi$
production which can be discriminated, depending whether non-TP
dominates or equally contributes with TP. The first type is
activated for very low $\Trh$, low $\Nx$ and is more or less
independent of $\sigv$, whereas the latter case requires larger
$\Trh$'s and $\Nx$'s and dependents on $\sigv$. In this last case,
we remarked that a period of $\chi$ reannihilation can emerge.

As regards the QKS, we verified that the included Hubble-induced
mass term ensures the presence of a KD period, which is
characterized by an oscillating evolution of $q$, and allows the
quintessential energy density to join in time a tracker behavior,
alleviating, thereby, the coincidence problem. Observational data
originating from BBN, the present acceleration of the universe,
the inflationary scale and the DE density parameter can be also
met in a sizable fraction of the parameter space of the model.
$\Domx$ crucially depends on the hierarchy between the freeze-out
temperature and the temperature where the evolution of $q$
develops extrema.

Assuming that the WIMP annihilates primarily to $e^+e^-$,
$\mu^+\mu^-$ or $\tau^+\tau^-$ we calculated the induced flux of
$e^\pm$-CRs and fit the current data of PAMELA, ATIC and Fermi LAT
without invoking any ad-hoc boost factor. For simplicity, we did
not include in our fits older experimental results, such as from
PPB-BETS \cite{ppb}, or more uncertain ones, such as from H.E.S.S
\cite{hess1} (however, the latter data may be used for imposing an
upper limit in the $\mx-\sigv$ plane \cite{zah}). Taking into
account the strong bounds originating mostly from CMB, we
concluded that the channel $\chi\chi\to\tau^+\tau^-$ can be
excluded at $95\%$ c.l. and: {\sf \small (i)} large parts of the
regions favored by PAMELA at $95\%$ c.l. for the residual
annihilation modes are ruled out; {\sf \small (ii)} regions
favored by PAMELA and ATIC or Fermi LAT at $99\%$ c.l. for
$\chi\chi\to e^+e^-$ are excluded; {\sf \small (iii)} only a part
of the region favored by PAMELA and ATIC data at $95\%$ c.l. for
$\chi\chi\to \mu^+\mu^-$ can be acceptable. For the latter
annihilation channel we achieved our best fits to PAMELA and
Fermi-LAT data with $\mx\sim1~\TeV$ and
$\sigv\sim10^{-6}~\GeV^{-2}$ which belong within the region
individuated by these data--sets at $99\%$ c.l. In all cases, the
considered $\mx$'s and $\sigv$'s can yield the right amount of
$\Omx$ (entailed by the CDM considerations) by adjusting the
parameters of the QKS or LRS. In both scenaria the required
transition temperature to the conventional RD era turns out to be
lower than about $0.7~\GeV$. In the case of the LRS an appreciable
contribution of non-TP is also necessitated.

As for the prediction of any CDM signal, there are three sources
of uncertainty in our investigation: the CDM distribution, the
propagation of its annihilation products and the role of
astrophysical backgrounds. In our analysis we used {\sf\small (i)}
the isothermal halo profile, to avoid troubles \cite{gstrumia}
with observations on $\gamma$-CRs; {\sf\small (ii)} the MED
propagation model, which provides the best fits to the
combinations of the various data-sets \cite{korea} and {\sf\small
(iii)} commonly assumed background $\ps$ and $\el$ fluxes
\cite{edjo} normalized with the ATIC or Fermi-LAT data. The
uncertainties above in conjunction with the very stringent
constraints from CMB \cite{CMB, CMBnew} may jeopardize the
interpretation of the PAMELA and ATIC or Fermi-LAT anomalies
through the $\chi$ annihilation. Therefore, the proposed scenaria
can be probed in the near future, if a better understanding of the
astrophysical uncertainties becomes available and/or more accurate
experimental data are released.

Our proposal could be supplemented by the construction of a
particle model (see, e.g., \cref{korea, shafi}) with the
appropriate couplings so that $\chi$ annihilates into $\mu^+\mu^-$
with the desired $\sigv$'s derived self-consistently with the
(s)particle spectrum. In a such case, several phenomenological
implications could be examined as in \cref{prof}. Let us finally
mention that another class of non-standard cosmological scenaria
\cite{Kam, khalil, masiero, qcosmology, seto} can be generated
considering modifications to the Friedmann equation due to
corrections to the Einstein gravity. Constraining these
possibilities (as in \cref{nicolo}) through the experimental
results on $e^\pm$-CRs would be another interesting issue.

\newpage
\acknowledgments  This research was funded by the FP6 Marie Curie
Excellence grant MEXT-CT-2004-014297. The author would like to
thank K. Kohri for useful correspondence, A.B. Lahanas for
valuable discussions and N.D.~Vlachos for providing helpful
software.

\end{document}